\documentclass[journal]{IEEEtran}

\usepackage{url}
\usepackage{./bs_macros}
\usepackage[labelformat=simple]{subcaption}

\usepackage{float}
\usepackage[noend]{algpseudocode}

\definecolor{AliceBlue}{rgb}{0.94,0.97,1.00}
\definecolor{AntiqueWhite1}{rgb}{1.00,0.94,0.86}
\definecolor{AntiqueWhite2}{rgb}{0.93,0.87,0.80}
\definecolor{AntiqueWhite3}{rgb}{0.80,0.75,0.69}
\definecolor{AntiqueWhite4}{rgb}{0.55,0.51,0.47}
\definecolor{AntiqueWhite}{rgb}{0.98,0.92,0.84}
\definecolor{BlanchedAlmond}{rgb}{1.00,0.92,0.80}
\definecolor{BlueViolet}{rgb}{0.54,0.17,0.89}
\definecolor{CadetBlue1}{rgb}{0.60,0.96,1.00}
\definecolor{CadetBlue2}{rgb}{0.56,0.90,0.93}
\definecolor{CadetBlue3}{rgb}{0.48,0.77,0.80}
\definecolor{CadetBlue4}{rgb}{0.33,0.53,0.55}
\definecolor{CadetBlue}{rgb}{0.37,0.62,0.63}
\definecolor{CornflowerBlue}{rgb}{0.39,0.58,0.93}
\definecolor{DarkBlue}{rgb}{0.00,0.00,0.55}
\definecolor{DarkCyan}{rgb}{0.00,0.55,0.55}
\definecolor{DarkGoldenrod1}{rgb}{1.00,0.73,0.06}
\definecolor{DarkGoldenrod2}{rgb}{0.93,0.68,0.05}
\definecolor{DarkGoldenrod3}{rgb}{0.80,0.58,0.05}
\definecolor{DarkGoldenrod4}{rgb}{0.55,0.40,0.03}
\definecolor{DarkGoldenrod}{rgb}{0.72,0.53,0.04}
\definecolor{DarkGray}{rgb}{0.66,0.66,0.66}
\definecolor{DarkGreen}{rgb}{0.00,0.39,0.00}
\definecolor{DarkGrey}{rgb}{0.66,0.66,0.66}
\definecolor{DarkKhaki}{rgb}{0.74,0.72,0.42}
\definecolor{DarkMagenta}{rgb}{0.55,0.00,0.55}
\definecolor{DarkOliveGreen1}{rgb}{0.79,1.00,0.44}
\definecolor{DarkOliveGreen2}{rgb}{0.74,0.93,0.41}
\definecolor{DarkOliveGreen3}{rgb}{0.64,0.80,0.35}
\definecolor{DarkOliveGreen4}{rgb}{0.43,0.55,0.24}
\definecolor{DarkOliveGreen}{rgb}{0.33,0.42,0.18}
\definecolor{DarkOrange1}{rgb}{1.00,0.50,0.00}
\definecolor{DarkOrange2}{rgb}{0.93,0.46,0.00}
\definecolor{DarkOrange3}{rgb}{0.80,0.40,0.00}
\definecolor{DarkOrange4}{rgb}{0.55,0.27,0.00}
\definecolor{DarkOrange}{rgb}{1.00,0.55,0.00}
\definecolor{DarkOrchid1}{rgb}{0.75,0.24,1.00}
\definecolor{DarkOrchid2}{rgb}{0.70,0.23,0.93}
\definecolor{DarkOrchid3}{rgb}{0.60,0.20,0.80}
\definecolor{DarkOrchid4}{rgb}{0.41,0.13,0.55}
\definecolor{DarkOrchid}{rgb}{0.60,0.20,0.80}
\definecolor{DarkRed}{rgb}{0.55,0.00,0.00}
\definecolor{DarkSalmon}{rgb}{0.91,0.59,0.48}
\definecolor{DarkSeaGreen1}{rgb}{0.76,1.00,0.76}
\definecolor{DarkSeaGreen2}{rgb}{0.71,0.93,0.71}
\definecolor{DarkSeaGreen3}{rgb}{0.61,0.80,0.61}
\definecolor{DarkSeaGreen4}{rgb}{0.41,0.55,0.41}
\definecolor{DarkSeaGreen}{rgb}{0.56,0.74,0.56}
\definecolor{DarkSlateBlue}{rgb}{0.28,0.24,0.55}
\definecolor{DarkSlateGray1}{rgb}{0.59,1.00,1.00}
\definecolor{DarkSlateGray2}{rgb}{0.55,0.93,0.93}
\definecolor{DarkSlateGray3}{rgb}{0.47,0.80,0.80}
\definecolor{DarkSlateGray4}{rgb}{0.32,0.55,0.55}
\definecolor{DarkSlateGray}{rgb}{0.18,0.31,0.31}
\definecolor{DarkSlateGrey}{rgb}{0.18,0.31,0.31}
\definecolor{DarkTurquoise}{rgb}{0.00,0.81,0.82}
\definecolor{DarkViolet}{rgb}{0.58,0.00,0.83}
\definecolor{DeepPink1}{rgb}{1.00,0.08,0.58}
\definecolor{DeepPink2}{rgb}{0.93,0.07,0.54}
\definecolor{DeepPink3}{rgb}{0.80,0.06,0.46}
\definecolor{DeepPink4}{rgb}{0.55,0.04,0.31}
\definecolor{DeepPink}{rgb}{1.00,0.08,0.58}
\definecolor{DeepSkyBlue1}{rgb}{0.00,0.75,1.00}
\definecolor{DeepSkyBlue2}{rgb}{0.00,0.70,0.93}
\definecolor{DeepSkyBlue3}{rgb}{0.00,0.60,0.80}
\definecolor{DeepSkyBlue4}{rgb}{0.00,0.41,0.55}
\definecolor{DeepSkyBlue}{rgb}{0.00,0.75,1.00}
\definecolor{DimGray}{rgb}{0.41,0.41,0.41}
\definecolor{DimGrey}{rgb}{0.41,0.41,0.41}
\definecolor{DodgerBlue1}{rgb}{0.12,0.56,1.00}
\definecolor{DodgerBlue2}{rgb}{0.11,0.53,0.93}
\definecolor{DodgerBlue3}{rgb}{0.09,0.45,0.80}
\definecolor{DodgerBlue4}{rgb}{0.06,0.31,0.55}
\definecolor{DodgerBlue}{rgb}{0.12,0.56,1.00}
\definecolor{FloralWhite}{rgb}{1.00,0.98,0.94}
\definecolor{ForestGreen}{rgb}{0.13,0.55,0.13}
\definecolor{GhostWhite}{rgb}{0.97,0.97,1.00}
\definecolor{GreenYellow}{rgb}{0.68,1.00,0.18}
\definecolor{HotPink1}{rgb}{1.00,0.43,0.71}
\definecolor{HotPink2}{rgb}{0.93,0.42,0.65}
\definecolor{HotPink3}{rgb}{0.80,0.38,0.56}
\definecolor{HotPink4}{rgb}{0.55,0.23,0.38}
\definecolor{HotPink}{rgb}{1.00,0.41,0.71}
\definecolor{IndianRed1}{rgb}{1.00,0.42,0.42}
\definecolor{IndianRed2}{rgb}{0.93,0.39,0.39}
\definecolor{IndianRed3}{rgb}{0.80,0.33,0.33}
\definecolor{IndianRed4}{rgb}{0.55,0.23,0.23}
\definecolor{IndianRed}{rgb}{0.80,0.36,0.36}
\definecolor{LavenderBlush1}{rgb}{1.00,0.94,0.96}
\definecolor{LavenderBlush2}{rgb}{0.93,0.88,0.90}
\definecolor{LavenderBlush3}{rgb}{0.80,0.76,0.77}
\definecolor{LavenderBlush4}{rgb}{0.55,0.51,0.53}
\definecolor{LavenderBlush}{rgb}{1.00,0.94,0.96}
\definecolor{LawnGreen}{rgb}{0.49,0.99,0.00}
\definecolor{LemonChiffon1}{rgb}{1.00,0.98,0.80}
\definecolor{LemonChiffon2}{rgb}{0.93,0.91,0.75}
\definecolor{LemonChiffon3}{rgb}{0.80,0.79,0.65}
\definecolor{LemonChiffon4}{rgb}{0.55,0.54,0.44}
\definecolor{LemonChiffon}{rgb}{1.00,0.98,0.80}
\definecolor{LightBlue1}{rgb}{0.75,0.94,1.00}
\definecolor{LightBlue2}{rgb}{0.70,0.87,0.93}
\definecolor{LightBlue3}{rgb}{0.60,0.75,0.80}
\definecolor{LightBlue4}{rgb}{0.41,0.51,0.55}
\definecolor{LightBlue}{rgb}{0.68,0.85,0.90}
\definecolor{LightCoral}{rgb}{0.94,0.50,0.50}
\definecolor{LightCyan1}{rgb}{0.88,1.00,1.00}
\definecolor{LightCyan2}{rgb}{0.82,0.93,0.93}
\definecolor{LightCyan3}{rgb}{0.71,0.80,0.80}
\definecolor{LightCyan4}{rgb}{0.48,0.55,0.55}
\definecolor{LightCyan}{rgb}{0.88,1.00,1.00}
\definecolor{LightGoldenrod1}{rgb}{1.00,0.93,0.55}
\definecolor{LightGoldenrod2}{rgb}{0.93,0.86,0.51}
\definecolor{LightGoldenrod3}{rgb}{0.80,0.75,0.44}
\definecolor{LightGoldenrod4}{rgb}{0.55,0.51,0.30}
\definecolor{LightGoldenrodYellow}{rgb}{0.98,0.98,0.82}
\definecolor{LightGoldenrod}{rgb}{0.93,0.87,0.51}
\definecolor{LightGray}{rgb}{0.83,0.83,0.83}
\definecolor{LightGreen}{rgb}{0.56,0.93,0.56}
\definecolor{LightGrey}{rgb}{0.83,0.83,0.83}
\definecolor{LightPink1}{rgb}{1.00,0.68,0.73}
\definecolor{LightPink2}{rgb}{0.93,0.64,0.68}
\definecolor{LightPink3}{rgb}{0.80,0.55,0.58}
\definecolor{LightPink4}{rgb}{0.55,0.37,0.40}
\definecolor{LightPink}{rgb}{1.00,0.71,0.76}
\definecolor{LightSalmon1}{rgb}{1.00,0.63,0.48}
\definecolor{LightSalmon2}{rgb}{0.93,0.58,0.45}
\definecolor{LightSalmon3}{rgb}{0.80,0.51,0.38}
\definecolor{LightSalmon4}{rgb}{0.55,0.34,0.26}
\definecolor{LightSalmon}{rgb}{1.00,0.63,0.48}
\definecolor{LightSeaGreen}{rgb}{0.13,0.70,0.67}
\definecolor{LightSkyBlue1}{rgb}{0.69,0.89,1.00}
\definecolor{LightSkyBlue2}{rgb}{0.64,0.83,0.93}
\definecolor{LightSkyBlue3}{rgb}{0.55,0.71,0.80}
\definecolor{LightSkyBlue4}{rgb}{0.38,0.48,0.55}
\definecolor{LightSkyBlue}{rgb}{0.53,0.81,0.98}
\definecolor{LightSlateBlue}{rgb}{0.52,0.44,1.00}
\definecolor{LightSlateGray}{rgb}{0.47,0.53,0.60}
\definecolor{LightSlateGrey}{rgb}{0.47,0.53,0.60}
\definecolor{LightSteelBlue1}{rgb}{0.79,0.88,1.00}
\definecolor{LightSteelBlue2}{rgb}{0.74,0.82,0.93}
\definecolor{LightSteelBlue3}{rgb}{0.64,0.71,0.80}
\definecolor{LightSteelBlue4}{rgb}{0.43,0.48,0.55}
\definecolor{LightSteelBlue}{rgb}{0.69,0.77,0.87}
\definecolor{LightYellow1}{rgb}{1.00,1.00,0.88}
\definecolor{LightYellow2}{rgb}{0.93,0.93,0.82}
\definecolor{LightYellow3}{rgb}{0.80,0.80,0.71}
\definecolor{LightYellow4}{rgb}{0.55,0.55,0.48}
\definecolor{LightYellow}{rgb}{1.00,1.00,0.88}
\definecolor{LimeGreen}{rgb}{0.20,0.80,0.20}
\definecolor{MediumAquamarine}{rgb}{0.40,0.80,0.67}
\definecolor{MediumBlue}{rgb}{0.00,0.00,0.80}
\definecolor{MediumOrchid1}{rgb}{0.88,0.40,1.00}
\definecolor{MediumOrchid2}{rgb}{0.82,0.37,0.93}
\definecolor{MediumOrchid3}{rgb}{0.71,0.32,0.80}
\definecolor{MediumOrchid4}{rgb}{0.48,0.22,0.55}
\definecolor{MediumOrchid}{rgb}{0.73,0.33,0.83}
\definecolor{MediumPurple1}{rgb}{0.67,0.51,1.00}
\definecolor{MediumPurple2}{rgb}{0.62,0.47,0.93}
\definecolor{MediumPurple3}{rgb}{0.54,0.41,0.80}
\definecolor{MediumPurple4}{rgb}{0.36,0.28,0.55}
\definecolor{MediumPurple}{rgb}{0.58,0.44,0.86}
\definecolor{MediumSeaGreen}{rgb}{0.24,0.70,0.44}
\definecolor{MediumSlateBlue}{rgb}{0.48,0.41,0.93}
\definecolor{MediumSpringGreen}{rgb}{0.00,0.98,0.60}
\definecolor{MediumTurquoise}{rgb}{0.28,0.82,0.80}
\definecolor{MediumVioletRed}{rgb}{0.78,0.08,0.52}
\definecolor{MidnightBlue}{rgb}{0.10,0.10,0.44}
\definecolor{MintCream}{rgb}{0.96,1.00,0.98}
\definecolor{MistyRose1}{rgb}{1.00,0.89,0.88}
\definecolor{MistyRose2}{rgb}{0.93,0.84,0.82}
\definecolor{MistyRose3}{rgb}{0.80,0.72,0.71}
\definecolor{MistyRose4}{rgb}{0.55,0.49,0.48}
\definecolor{MistyRose}{rgb}{1.00,0.89,0.88}
\definecolor{NavajoWhite1}{rgb}{1.00,0.87,0.68}
\definecolor{NavajoWhite2}{rgb}{0.93,0.81,0.63}
\definecolor{NavajoWhite3}{rgb}{0.80,0.70,0.55}
\definecolor{NavajoWhite4}{rgb}{0.55,0.47,0.37}
\definecolor{NavajoWhite}{rgb}{1.00,0.87,0.68}
\definecolor{NavyBlue}{rgb}{0.00,0.00,0.50}
\definecolor{OldLace}{rgb}{0.99,0.96,0.90}
\definecolor{OliveDrab1}{rgb}{0.75,1.00,0.24}
\definecolor{OliveDrab2}{rgb}{0.70,0.93,0.23}
\definecolor{OliveDrab3}{rgb}{0.60,0.80,0.20}
\definecolor{OliveDrab4}{rgb}{0.41,0.55,0.13}
\definecolor{OliveDrab}{rgb}{0.42,0.56,0.14}
\definecolor{OrangeRed1}{rgb}{1.00,0.27,0.00}
\definecolor{OrangeRed2}{rgb}{0.93,0.25,0.00}
\definecolor{OrangeRed3}{rgb}{0.80,0.22,0.00}
\definecolor{OrangeRed4}{rgb}{0.55,0.15,0.00}
\definecolor{OrangeRed}{rgb}{1.00,0.27,0.00}
\definecolor{PaleGoldenrod}{rgb}{0.93,0.91,0.67}
\definecolor{PaleGreen1}{rgb}{0.60,1.00,0.60}
\definecolor{PaleGreen2}{rgb}{0.56,0.93,0.56}
\definecolor{PaleGreen3}{rgb}{0.49,0.80,0.49}
\definecolor{PaleGreen4}{rgb}{0.33,0.55,0.33}
\definecolor{PaleGreen}{rgb}{0.60,0.98,0.60}
\definecolor{PaleTurquoise1}{rgb}{0.73,1.00,1.00}
\definecolor{PaleTurquoise2}{rgb}{0.68,0.93,0.93}
\definecolor{PaleTurquoise3}{rgb}{0.59,0.80,0.80}
\definecolor{PaleTurquoise4}{rgb}{0.40,0.55,0.55}
\definecolor{PaleTurquoise}{rgb}{0.69,0.93,0.93}
\definecolor{PaleVioletRed1}{rgb}{1.00,0.51,0.67}
\definecolor{PaleVioletRed2}{rgb}{0.93,0.47,0.62}
\definecolor{PaleVioletRed3}{rgb}{0.80,0.41,0.54}
\definecolor{PaleVioletRed4}{rgb}{0.55,0.28,0.36}
\definecolor{PaleVioletRed}{rgb}{0.86,0.44,0.58}
\definecolor{PapayaWhip}{rgb}{1.00,0.94,0.84}
\definecolor{PeachPuff1}{rgb}{1.00,0.85,0.73}
\definecolor{PeachPuff2}{rgb}{0.93,0.80,0.68}
\definecolor{PeachPuff3}{rgb}{0.80,0.69,0.58}
\definecolor{PeachPuff4}{rgb}{0.55,0.47,0.40}
\definecolor{PeachPuff}{rgb}{1.00,0.85,0.73}
\definecolor{PowderBlue}{rgb}{0.69,0.88,0.90}
\definecolor{RosyBrown1}{rgb}{1.00,0.76,0.76}
\definecolor{RosyBrown2}{rgb}{0.93,0.71,0.71}
\definecolor{RosyBrown3}{rgb}{0.80,0.61,0.61}
\definecolor{RosyBrown4}{rgb}{0.55,0.41,0.41}
\definecolor{RosyBrown}{rgb}{0.74,0.56,0.56}
\definecolor{RoyalBlue1}{rgb}{0.28,0.46,1.00}
\definecolor{RoyalBlue2}{rgb}{0.26,0.43,0.93}
\definecolor{RoyalBlue3}{rgb}{0.23,0.37,0.80}
\definecolor{RoyalBlue4}{rgb}{0.15,0.25,0.55}
\definecolor{RoyalBlue}{rgb}{0.25,0.41,0.88}
\definecolor{SaddleBrown}{rgb}{0.55,0.27,0.07}
\definecolor{SandyBrown}{rgb}{0.96,0.64,0.38}
\definecolor{SeaGreen1}{rgb}{0.33,1.00,0.62}
\definecolor{SeaGreen2}{rgb}{0.31,0.93,0.58}
\definecolor{SeaGreen3}{rgb}{0.26,0.80,0.50}
\definecolor{SeaGreen4}{rgb}{0.18,0.55,0.34}
\definecolor{SeaGreen}{rgb}{0.18,0.55,0.34}
\definecolor{SkyBlue1}{rgb}{0.53,0.81,1.00}
\definecolor{SkyBlue2}{rgb}{0.49,0.75,0.93}
\definecolor{SkyBlue3}{rgb}{0.42,0.65,0.80}
\definecolor{SkyBlue4}{rgb}{0.29,0.44,0.55}
\definecolor{SkyBlue}{rgb}{0.53,0.81,0.92}
\definecolor{SlateBlue1}{rgb}{0.51,0.44,1.00}
\definecolor{SlateBlue2}{rgb}{0.48,0.40,0.93}
\definecolor{SlateBlue3}{rgb}{0.41,0.35,0.80}
\definecolor{SlateBlue4}{rgb}{0.28,0.24,0.55}
\definecolor{SlateBlue}{rgb}{0.42,0.35,0.80}
\definecolor{SlateGray1}{rgb}{0.78,0.89,1.00}
\definecolor{SlateGray2}{rgb}{0.73,0.83,0.93}
\definecolor{SlateGray3}{rgb}{0.62,0.71,0.80}
\definecolor{SlateGray4}{rgb}{0.42,0.48,0.55}
\definecolor{SlateGray}{rgb}{0.44,0.50,0.56}
\definecolor{SlateGrey}{rgb}{0.44,0.50,0.56}
\definecolor{SpringGreen1}{rgb}{0.00,1.00,0.50}
\definecolor{SpringGreen2}{rgb}{0.00,0.93,0.46}
\definecolor{SpringGreen3}{rgb}{0.00,0.80,0.40}
\definecolor{SpringGreen4}{rgb}{0.00,0.55,0.27}
\definecolor{SpringGreen}{rgb}{0.00,1.00,0.50}
\definecolor{SteelBlue1}{rgb}{0.39,0.72,1.00}
\definecolor{SteelBlue2}{rgb}{0.36,0.67,0.93}
\definecolor{SteelBlue3}{rgb}{0.31,0.58,0.80}
\definecolor{SteelBlue4}{rgb}{0.21,0.39,0.55}
\definecolor{SteelBlue}{rgb}{0.27,0.51,0.71}
\definecolor{VioletRed1}{rgb}{1.00,0.24,0.59}
\definecolor{VioletRed2}{rgb}{0.93,0.23,0.55}
\definecolor{VioletRed3}{rgb}{0.80,0.20,0.47}
\definecolor{VioletRed4}{rgb}{0.55,0.13,0.32}
\definecolor{VioletRed}{rgb}{0.82,0.13,0.56}
\definecolor{WhiteSmoke}{rgb}{0.96,0.96,0.96}
\definecolor{YellowGreen}{rgb}{0.60,0.80,0.20}
\definecolor{aliceblue}{rgb}{0.94,0.97,1.00}
\definecolor{antiquewhite}{rgb}{0.98,0.92,0.84}
\definecolor{aquamarine1}{rgb}{0.50,1.00,0.83}
\definecolor{aquamarine2}{rgb}{0.46,0.93,0.78}
\definecolor{aquamarine3}{rgb}{0.40,0.80,0.67}
\definecolor{aquamarine4}{rgb}{0.27,0.55,0.45}
\definecolor{aquamarine}{rgb}{0.50,1.00,0.83}
\definecolor{azure1}{rgb}{0.94,1.00,1.00}
\definecolor{azure2}{rgb}{0.88,0.93,0.93}
\definecolor{azure3}{rgb}{0.76,0.80,0.80}
\definecolor{azure4}{rgb}{0.51,0.55,0.55}
\definecolor{azure}{rgb}{0.94,1.00,1.00}
\definecolor{beige}{rgb}{0.96,0.96,0.86}
\definecolor{bisque1}{rgb}{1.00,0.89,0.77}
\definecolor{bisque2}{rgb}{0.93,0.84,0.72}
\definecolor{bisque3}{rgb}{0.80,0.72,0.62}
\definecolor{bisque4}{rgb}{0.55,0.49,0.42}
\definecolor{bisque}{rgb}{1.00,0.89,0.77}
\definecolor{black}{rgb}{0.00,0.00,0.00}
\definecolor{blanchedalmond}{rgb}{1.00,0.92,0.80}
\definecolor{blue1}{rgb}{0.00,0.00,1.00}
\definecolor{blue2}{rgb}{0.00,0.00,0.93}
\definecolor{blue3}{rgb}{0.00,0.00,0.80}
\definecolor{blue4}{rgb}{0.00,0.00,0.55}
\definecolor{blueviolet}{rgb}{0.54,0.17,0.89}
\definecolor{blue}{rgb}{0.00,0.00,1.00}
\definecolor{brown1}{rgb}{1.00,0.25,0.25}
\definecolor{brown2}{rgb}{0.93,0.23,0.23}
\definecolor{brown3}{rgb}{0.80,0.20,0.20}
\definecolor{brown4}{rgb}{0.55,0.14,0.14}
\definecolor{brown}{rgb}{0.65,0.16,0.16}
\definecolor{burlywood1}{rgb}{1.00,0.83,0.61}
\definecolor{burlywood2}{rgb}{0.93,0.77,0.57}
\definecolor{burlywood3}{rgb}{0.80,0.67,0.49}
\definecolor{burlywood4}{rgb}{0.55,0.45,0.33}
\definecolor{burlywood}{rgb}{0.87,0.72,0.53}
\definecolor{cadetblue}{rgb}{0.37,0.62,0.63}
\definecolor{chartreuse1}{rgb}{0.50,1.00,0.00}
\definecolor{chartreuse2}{rgb}{0.46,0.93,0.00}
\definecolor{chartreuse3}{rgb}{0.40,0.80,0.00}
\definecolor{chartreuse4}{rgb}{0.27,0.55,0.00}
\definecolor{chartreuse}{rgb}{0.50,1.00,0.00}
\definecolor{chocolate1}{rgb}{1.00,0.50,0.14}
\definecolor{chocolate2}{rgb}{0.93,0.46,0.13}
\definecolor{chocolate3}{rgb}{0.80,0.40,0.11}
\definecolor{chocolate4}{rgb}{0.55,0.27,0.07}
\definecolor{chocolate}{rgb}{0.82,0.41,0.12}
\definecolor{coral1}{rgb}{1.00,0.45,0.34}
\definecolor{coral2}{rgb}{0.93,0.42,0.31}
\definecolor{coral3}{rgb}{0.80,0.36,0.27}
\definecolor{coral4}{rgb}{0.55,0.24,0.18}
\definecolor{coral}{rgb}{1.00,0.50,0.31}
\definecolor{cornflowerblue}{rgb}{0.39,0.58,0.93}
\definecolor{cornsilk1}{rgb}{1.00,0.97,0.86}
\definecolor{cornsilk2}{rgb}{0.93,0.91,0.80}
\definecolor{cornsilk3}{rgb}{0.80,0.78,0.69}
\definecolor{cornsilk4}{rgb}{0.55,0.53,0.47}
\definecolor{cornsilk}{rgb}{1.00,0.97,0.86}
\definecolor{cyan1}{rgb}{0.00,1.00,1.00}
\definecolor{cyan2}{rgb}{0.00,0.93,0.93}
\definecolor{cyan3}{rgb}{0.00,0.80,0.80}
\definecolor{cyan4}{rgb}{0.00,0.55,0.55}
\definecolor{cyan}{rgb}{0.00,1.00,1.00}
\definecolor{darkblue}{rgb}{0.00,0.00,0.55}
\definecolor{darkcyan}{rgb}{0.00,0.55,0.55}
\definecolor{darkgoldenrod}{rgb}{0.72,0.53,0.04}
\definecolor{darkgray}{rgb}{0.66,0.66,0.66}
\definecolor{darkgreen}{rgb}{0.00,0.39,0.00}
\definecolor{darkgrey}{rgb}{0.66,0.66,0.66}
\definecolor{darkkhaki}{rgb}{0.74,0.72,0.42}
\definecolor{darkmagenta}{rgb}{0.55,0.00,0.55}
\definecolor{darkolive}{rgb}{0.33,0.42,0.18}
\definecolor{darkorange}{rgb}{1.00,0.55,0.00}
\definecolor{darkorchid}{rgb}{0.60,0.20,0.80}
\definecolor{darkred}{rgb}{0.55,0.00,0.00}
\definecolor{darksalmon}{rgb}{0.91,0.59,0.48}
\definecolor{darksea}{rgb}{0.56,0.74,0.56}
\definecolor{darkslate}{rgb}{0.18,0.31,0.31}
\definecolor{darkslate}{rgb}{0.18,0.31,0.31}
\definecolor{darkslate}{rgb}{0.28,0.24,0.55}
\definecolor{darkturquoise}{rgb}{0.00,0.81,0.82}
\definecolor{darkviolet}{rgb}{0.58,0.00,0.83}
\definecolor{deeppink}{rgb}{1.00,0.08,0.58}
\definecolor{deepsky}{rgb}{0.00,0.75,1.00}
\definecolor{dimgray}{rgb}{0.41,0.41,0.41}
\definecolor{dimgrey}{rgb}{0.41,0.41,0.41}
\definecolor{dodgerblue}{rgb}{0.12,0.56,1.00}
\definecolor{firebrick1}{rgb}{1.00,0.19,0.19}
\definecolor{firebrick2}{rgb}{0.93,0.17,0.17}
\definecolor{firebrick3}{rgb}{0.80,0.15,0.15}
\definecolor{firebrick4}{rgb}{0.55,0.10,0.10}
\definecolor{firebrick}{rgb}{0.70,0.13,0.13}
\definecolor{floralwhite}{rgb}{1.00,0.98,0.94}
\definecolor{forestgreen}{rgb}{0.13,0.55,0.13}
\definecolor{gainsboro}{rgb}{0.86,0.86,0.86}
\definecolor{ghostwhite}{rgb}{0.97,0.97,1.00}
\definecolor{gold1}{rgb}{1.00,0.84,0.00}
\definecolor{gold2}{rgb}{0.93,0.79,0.00}
\definecolor{gold3}{rgb}{0.80,0.68,0.00}
\definecolor{gold4}{rgb}{0.55,0.46,0.00}
\definecolor{goldenrod1}{rgb}{1.00,0.76,0.15}
\definecolor{goldenrod2}{rgb}{0.93,0.71,0.13}
\definecolor{goldenrod3}{rgb}{0.80,0.61,0.11}
\definecolor{goldenrod4}{rgb}{0.55,0.41,0.08}
\definecolor{goldenrod}{rgb}{0.85,0.65,0.13}
\definecolor{gold}{rgb}{1.00,0.84,0.00}
\definecolor{gray0}{rgb}{0.00,0.00,0.00}
\definecolor{gray100}{rgb}{1.00,1.00,1.00}
\definecolor{gray10}{rgb}{0.10,0.10,0.10}
\definecolor{gray11}{rgb}{0.11,0.11,0.11}
\definecolor{gray12}{rgb}{0.12,0.12,0.12}
\definecolor{gray13}{rgb}{0.13,0.13,0.13}
\definecolor{gray14}{rgb}{0.14,0.14,0.14}
\definecolor{gray15}{rgb}{0.15,0.15,0.15}
\definecolor{gray16}{rgb}{0.16,0.16,0.16}
\definecolor{gray17}{rgb}{0.17,0.17,0.17}
\definecolor{gray18}{rgb}{0.18,0.18,0.18}
\definecolor{gray19}{rgb}{0.19,0.19,0.19}
\definecolor{gray1}{rgb}{0.01,0.01,0.01}
\definecolor{gray20}{rgb}{0.20,0.20,0.20}
\definecolor{gray21}{rgb}{0.21,0.21,0.21}
\definecolor{gray22}{rgb}{0.22,0.22,0.22}
\definecolor{gray23}{rgb}{0.23,0.23,0.23}
\definecolor{gray24}{rgb}{0.24,0.24,0.24}
\definecolor{gray25}{rgb}{0.25,0.25,0.25}
\definecolor{gray26}{rgb}{0.26,0.26,0.26}
\definecolor{gray27}{rgb}{0.27,0.27,0.27}
\definecolor{gray28}{rgb}{0.28,0.28,0.28}
\definecolor{gray29}{rgb}{0.29,0.29,0.29}
\definecolor{gray2}{rgb}{0.02,0.02,0.02}
\definecolor{gray30}{rgb}{0.30,0.30,0.30}
\definecolor{gray31}{rgb}{0.31,0.31,0.31}
\definecolor{gray32}{rgb}{0.32,0.32,0.32}
\definecolor{gray33}{rgb}{0.33,0.33,0.33}
\definecolor{gray34}{rgb}{0.34,0.34,0.34}
\definecolor{gray35}{rgb}{0.35,0.35,0.35}
\definecolor{gray36}{rgb}{0.36,0.36,0.36}
\definecolor{gray37}{rgb}{0.37,0.37,0.37}
\definecolor{gray38}{rgb}{0.38,0.38,0.38}
\definecolor{gray39}{rgb}{0.39,0.39,0.39}
\definecolor{gray3}{rgb}{0.03,0.03,0.03}
\definecolor{gray40}{rgb}{0.40,0.40,0.40}
\definecolor{gray41}{rgb}{0.41,0.41,0.41}
\definecolor{gray42}{rgb}{0.42,0.42,0.42}
\definecolor{gray43}{rgb}{0.43,0.43,0.43}
\definecolor{gray44}{rgb}{0.44,0.44,0.44}
\definecolor{gray45}{rgb}{0.45,0.45,0.45}
\definecolor{gray46}{rgb}{0.46,0.46,0.46}
\definecolor{gray47}{rgb}{0.47,0.47,0.47}
\definecolor{gray48}{rgb}{0.48,0.48,0.48}
\definecolor{gray49}{rgb}{0.49,0.49,0.49}
\definecolor{gray4}{rgb}{0.04,0.04,0.04}
\definecolor{gray50}{rgb}{0.50,0.50,0.50}
\definecolor{gray51}{rgb}{0.51,0.51,0.51}
\definecolor{gray52}{rgb}{0.52,0.52,0.52}
\definecolor{gray53}{rgb}{0.53,0.53,0.53}
\definecolor{gray54}{rgb}{0.54,0.54,0.54}
\definecolor{gray55}{rgb}{0.55,0.55,0.55}
\definecolor{gray56}{rgb}{0.56,0.56,0.56}
\definecolor{gray57}{rgb}{0.57,0.57,0.57}
\definecolor{gray58}{rgb}{0.58,0.58,0.58}
\definecolor{gray59}{rgb}{0.59,0.59,0.59}
\definecolor{gray5}{rgb}{0.05,0.05,0.05}
\definecolor{gray60}{rgb}{0.60,0.60,0.60}
\definecolor{gray61}{rgb}{0.61,0.61,0.61}
\definecolor{gray62}{rgb}{0.62,0.62,0.62}
\definecolor{gray63}{rgb}{0.63,0.63,0.63}
\definecolor{gray64}{rgb}{0.64,0.64,0.64}
\definecolor{gray65}{rgb}{0.65,0.65,0.65}
\definecolor{gray66}{rgb}{0.66,0.66,0.66}
\definecolor{gray67}{rgb}{0.67,0.67,0.67}
\definecolor{gray68}{rgb}{0.68,0.68,0.68}
\definecolor{gray69}{rgb}{0.69,0.69,0.69}
\definecolor{gray6}{rgb}{0.06,0.06,0.06}
\definecolor{gray70}{rgb}{0.70,0.70,0.70}
\definecolor{gray71}{rgb}{0.71,0.71,0.71}
\definecolor{gray72}{rgb}{0.72,0.72,0.72}
\definecolor{gray73}{rgb}{0.73,0.73,0.73}
\definecolor{gray74}{rgb}{0.74,0.74,0.74}
\definecolor{gray75}{rgb}{0.75,0.75,0.75}
\definecolor{gray76}{rgb}{0.76,0.76,0.76}
\definecolor{gray77}{rgb}{0.77,0.77,0.77}
\definecolor{gray78}{rgb}{0.78,0.78,0.78}
\definecolor{gray79}{rgb}{0.79,0.79,0.79}
\definecolor{gray7}{rgb}{0.07,0.07,0.07}
\definecolor{gray80}{rgb}{0.80,0.80,0.80}
\definecolor{gray81}{rgb}{0.81,0.81,0.81}
\definecolor{gray82}{rgb}{0.82,0.82,0.82}
\definecolor{gray83}{rgb}{0.83,0.83,0.83}
\definecolor{gray84}{rgb}{0.84,0.84,0.84}
\definecolor{gray85}{rgb}{0.85,0.85,0.85}
\definecolor{gray86}{rgb}{0.86,0.86,0.86}
\definecolor{gray87}{rgb}{0.87,0.87,0.87}
\definecolor{gray88}{rgb}{0.88,0.88,0.88}
\definecolor{gray89}{rgb}{0.89,0.89,0.89}
\definecolor{gray8}{rgb}{0.08,0.08,0.08}
\definecolor{gray90}{rgb}{0.90,0.90,0.90}
\definecolor{gray91}{rgb}{0.91,0.91,0.91}
\definecolor{gray92}{rgb}{0.92,0.92,0.92}
\definecolor{gray93}{rgb}{0.93,0.93,0.93}
\definecolor{gray94}{rgb}{0.94,0.94,0.94}
\definecolor{gray95}{rgb}{0.95,0.95,0.95}
\definecolor{gray96}{rgb}{0.96,0.96,0.96}
\definecolor{gray97}{rgb}{0.97,0.97,0.97}
\definecolor{gray98}{rgb}{0.98,0.98,0.98}
\definecolor{gray99}{rgb}{0.99,0.99,0.99}
\definecolor{gray9}{rgb}{0.09,0.09,0.09}
\definecolor{gray}{rgb}{0.75,0.75,0.75}
\definecolor{green1}{rgb}{0.00,1.00,0.00}
\definecolor{green2}{rgb}{0.00,0.93,0.00}
\definecolor{green3}{rgb}{0.00,0.80,0.00}
\definecolor{green4}{rgb}{0.00,0.55,0.00}
\definecolor{greenyellow}{rgb}{0.68,1.00,0.18}
\definecolor{green}{rgb}{0.00,1.00,0.00}
\definecolor{grey0}{rgb}{0.00,0.00,0.00}
\definecolor{grey100}{rgb}{1.00,1.00,1.00}
\definecolor{grey10}{rgb}{0.10,0.10,0.10}
\definecolor{grey11}{rgb}{0.11,0.11,0.11}
\definecolor{grey12}{rgb}{0.12,0.12,0.12}
\definecolor{grey13}{rgb}{0.13,0.13,0.13}
\definecolor{grey14}{rgb}{0.14,0.14,0.14}
\definecolor{grey15}{rgb}{0.15,0.15,0.15}
\definecolor{grey16}{rgb}{0.16,0.16,0.16}
\definecolor{grey17}{rgb}{0.17,0.17,0.17}
\definecolor{grey18}{rgb}{0.18,0.18,0.18}
\definecolor{grey19}{rgb}{0.19,0.19,0.19}
\definecolor{grey1}{rgb}{0.01,0.01,0.01}
\definecolor{grey20}{rgb}{0.20,0.20,0.20}
\definecolor{grey21}{rgb}{0.21,0.21,0.21}
\definecolor{grey22}{rgb}{0.22,0.22,0.22}
\definecolor{grey23}{rgb}{0.23,0.23,0.23}
\definecolor{grey24}{rgb}{0.24,0.24,0.24}
\definecolor{grey25}{rgb}{0.25,0.25,0.25}
\definecolor{grey26}{rgb}{0.26,0.26,0.26}
\definecolor{grey27}{rgb}{0.27,0.27,0.27}
\definecolor{grey28}{rgb}{0.28,0.28,0.28}
\definecolor{grey29}{rgb}{0.29,0.29,0.29}
\definecolor{grey2}{rgb}{0.02,0.02,0.02}
\definecolor{grey30}{rgb}{0.30,0.30,0.30}
\definecolor{grey31}{rgb}{0.31,0.31,0.31}
\definecolor{grey32}{rgb}{0.32,0.32,0.32}
\definecolor{grey33}{rgb}{0.33,0.33,0.33}
\definecolor{grey34}{rgb}{0.34,0.34,0.34}
\definecolor{grey35}{rgb}{0.35,0.35,0.35}
\definecolor{grey36}{rgb}{0.36,0.36,0.36}
\definecolor{grey37}{rgb}{0.37,0.37,0.37}
\definecolor{grey38}{rgb}{0.38,0.38,0.38}
\definecolor{grey39}{rgb}{0.39,0.39,0.39}
\definecolor{grey3}{rgb}{0.03,0.03,0.03}
\definecolor{grey40}{rgb}{0.40,0.40,0.40}
\definecolor{grey41}{rgb}{0.41,0.41,0.41}
\definecolor{grey42}{rgb}{0.42,0.42,0.42}
\definecolor{grey43}{rgb}{0.43,0.43,0.43}
\definecolor{grey44}{rgb}{0.44,0.44,0.44}
\definecolor{grey45}{rgb}{0.45,0.45,0.45}
\definecolor{grey46}{rgb}{0.46,0.46,0.46}
\definecolor{grey47}{rgb}{0.47,0.47,0.47}
\definecolor{grey48}{rgb}{0.48,0.48,0.48}
\definecolor{grey49}{rgb}{0.49,0.49,0.49}
\definecolor{grey4}{rgb}{0.04,0.04,0.04}
\definecolor{grey50}{rgb}{0.50,0.50,0.50}
\definecolor{grey51}{rgb}{0.51,0.51,0.51}
\definecolor{grey52}{rgb}{0.52,0.52,0.52}
\definecolor{grey53}{rgb}{0.53,0.53,0.53}
\definecolor{grey54}{rgb}{0.54,0.54,0.54}
\definecolor{grey55}{rgb}{0.55,0.55,0.55}
\definecolor{grey56}{rgb}{0.56,0.56,0.56}
\definecolor{grey57}{rgb}{0.57,0.57,0.57}
\definecolor{grey58}{rgb}{0.58,0.58,0.58}
\definecolor{grey59}{rgb}{0.59,0.59,0.59}
\definecolor{grey5}{rgb}{0.05,0.05,0.05}
\definecolor{grey60}{rgb}{0.60,0.60,0.60}
\definecolor{grey61}{rgb}{0.61,0.61,0.61}
\definecolor{grey62}{rgb}{0.62,0.62,0.62}
\definecolor{grey63}{rgb}{0.63,0.63,0.63}
\definecolor{grey64}{rgb}{0.64,0.64,0.64}
\definecolor{grey65}{rgb}{0.65,0.65,0.65}
\definecolor{grey66}{rgb}{0.66,0.66,0.66}
\definecolor{grey67}{rgb}{0.67,0.67,0.67}
\definecolor{grey68}{rgb}{0.68,0.68,0.68}
\definecolor{grey69}{rgb}{0.69,0.69,0.69}
\definecolor{grey6}{rgb}{0.06,0.06,0.06}
\definecolor{grey70}{rgb}{0.70,0.70,0.70}
\definecolor{grey71}{rgb}{0.71,0.71,0.71}
\definecolor{grey72}{rgb}{0.72,0.72,0.72}
\definecolor{grey73}{rgb}{0.73,0.73,0.73}
\definecolor{grey74}{rgb}{0.74,0.74,0.74}
\definecolor{grey75}{rgb}{0.75,0.75,0.75}
\definecolor{grey76}{rgb}{0.76,0.76,0.76}
\definecolor{grey77}{rgb}{0.77,0.77,0.77}
\definecolor{grey78}{rgb}{0.78,0.78,0.78}
\definecolor{grey79}{rgb}{0.79,0.79,0.79}
\definecolor{grey7}{rgb}{0.07,0.07,0.07}
\definecolor{grey80}{rgb}{0.80,0.80,0.80}
\definecolor{grey81}{rgb}{0.81,0.81,0.81}
\definecolor{grey82}{rgb}{0.82,0.82,0.82}
\definecolor{grey83}{rgb}{0.83,0.83,0.83}
\definecolor{grey84}{rgb}{0.84,0.84,0.84}
\definecolor{grey85}{rgb}{0.85,0.85,0.85}
\definecolor{grey86}{rgb}{0.86,0.86,0.86}
\definecolor{grey87}{rgb}{0.87,0.87,0.87}
\definecolor{grey88}{rgb}{0.88,0.88,0.88}
\definecolor{grey89}{rgb}{0.89,0.89,0.89}
\definecolor{grey8}{rgb}{0.08,0.08,0.08}
\definecolor{grey90}{rgb}{0.90,0.90,0.90}
\definecolor{grey91}{rgb}{0.91,0.91,0.91}
\definecolor{grey92}{rgb}{0.92,0.92,0.92}
\definecolor{grey93}{rgb}{0.93,0.93,0.93}
\definecolor{grey94}{rgb}{0.94,0.94,0.94}
\definecolor{grey95}{rgb}{0.95,0.95,0.95}
\definecolor{grey96}{rgb}{0.96,0.96,0.96}
\definecolor{grey97}{rgb}{0.97,0.97,0.97}
\definecolor{grey98}{rgb}{0.98,0.98,0.98}
\definecolor{grey99}{rgb}{0.99,0.99,0.99}
\definecolor{grey9}{rgb}{0.09,0.09,0.09}
\definecolor{grey}{rgb}{0.75,0.75,0.75}
\definecolor{honeydew1}{rgb}{0.94,1.00,0.94}
\definecolor{honeydew2}{rgb}{0.88,0.93,0.88}
\definecolor{honeydew3}{rgb}{0.76,0.80,0.76}
\definecolor{honeydew4}{rgb}{0.51,0.55,0.51}
\definecolor{honeydew}{rgb}{0.94,1.00,0.94}
\definecolor{hotpink}{rgb}{1.00,0.41,0.71}
\definecolor{indianred}{rgb}{0.80,0.36,0.36}
\definecolor{ivory1}{rgb}{1.00,1.00,0.94}
\definecolor{ivory2}{rgb}{0.93,0.93,0.88}
\definecolor{ivory3}{rgb}{0.80,0.80,0.76}
\definecolor{ivory4}{rgb}{0.55,0.55,0.51}
\definecolor{ivory}{rgb}{1.00,1.00,0.94}
\definecolor{khaki1}{rgb}{1.00,0.96,0.56}
\definecolor{khaki2}{rgb}{0.93,0.90,0.52}
\definecolor{khaki3}{rgb}{0.80,0.78,0.45}
\definecolor{khaki4}{rgb}{0.55,0.53,0.31}
\definecolor{khaki}{rgb}{0.94,0.90,0.55}
\definecolor{lavenderblush}{rgb}{1.00,0.94,0.96}
\definecolor{lavender}{rgb}{0.90,0.90,0.98}
\definecolor{lawngreen}{rgb}{0.49,0.99,0.00}
\definecolor{lemonchiffon}{rgb}{1.00,0.98,0.80}
\definecolor{lightblue}{rgb}{0.68,0.85,0.90}
\definecolor{lightcoral}{rgb}{0.94,0.50,0.50}
\definecolor{lightcyan}{rgb}{0.88,1.00,1.00}
\definecolor{lightgoldenrod}{rgb}{0.93,0.87,0.51}
\definecolor{lightgoldenrod}{rgb}{0.98,0.98,0.82}
\definecolor{lightgray}{rgb}{0.83,0.83,0.83}
\definecolor{lightgreen}{rgb}{0.56,0.93,0.56}
\definecolor{lightgrey}{rgb}{0.83,0.83,0.83}
\definecolor{lightpink}{rgb}{1.00,0.71,0.76}
\definecolor{lightsalmon}{rgb}{1.00,0.63,0.48}
\definecolor{lightsea}{rgb}{0.13,0.70,0.67}
\definecolor{lightsky}{rgb}{0.53,0.81,0.98}
\definecolor{lightslate}{rgb}{0.47,0.53,0.60}
\definecolor{lightslate}{rgb}{0.47,0.53,0.60}
\definecolor{lightslate}{rgb}{0.52,0.44,1.00}
\definecolor{lightsteel}{rgb}{0.69,0.77,0.87}
\definecolor{lightyellow}{rgb}{1.00,1.00,0.88}
\definecolor{limegreen}{rgb}{0.20,0.80,0.20}
\definecolor{linen}{rgb}{0.98,0.94,0.90}
\definecolor{magenta1}{rgb}{1.00,0.00,1.00}
\definecolor{magenta2}{rgb}{0.93,0.00,0.93}
\definecolor{magenta3}{rgb}{0.80,0.00,0.80}
\definecolor{magenta4}{rgb}{0.55,0.00,0.55}
\definecolor{magenta}{rgb}{1.00,0.00,1.00}
\definecolor{maroon1}{rgb}{1.00,0.20,0.70}
\definecolor{maroon2}{rgb}{0.93,0.19,0.65}
\definecolor{maroon3}{rgb}{0.80,0.16,0.56}
\definecolor{maroon4}{rgb}{0.55,0.11,0.38}
\definecolor{maroon}{rgb}{0.69,0.19,0.38}
\definecolor{mediumaquamarine}{rgb}{0.40,0.80,0.67}
\definecolor{mediumblue}{rgb}{0.00,0.00,0.80}
\definecolor{mediumorchid}{rgb}{0.73,0.33,0.83}
\definecolor{mediumpurple}{rgb}{0.58,0.44,0.86}
\definecolor{mediumsea}{rgb}{0.24,0.70,0.44}
\definecolor{mediumslate}{rgb}{0.48,0.41,0.93}
\definecolor{mediumspring}{rgb}{0.00,0.98,0.60}
\definecolor{mediumturquoise}{rgb}{0.28,0.82,0.80}
\definecolor{mediumviolet}{rgb}{0.78,0.08,0.52}
\definecolor{midnightblue}{rgb}{0.10,0.10,0.44}
\definecolor{mintcream}{rgb}{0.96,1.00,0.98}
\definecolor{mistyrose}{rgb}{1.00,0.89,0.88}
\definecolor{moccasin}{rgb}{1.00,0.89,0.71}
\definecolor{navajowhite}{rgb}{1.00,0.87,0.68}
\definecolor{navyblue}{rgb}{0.00,0.00,0.50}
\definecolor{navy}{rgb}{0.00,0.00,0.50}
\definecolor{oldlace}{rgb}{0.99,0.96,0.90}
\definecolor{olivedrab}{rgb}{0.42,0.56,0.14}
\definecolor{orange1}{rgb}{1.00,0.65,0.00}
\definecolor{orange2}{rgb}{0.93,0.60,0.00}
\definecolor{orange3}{rgb}{0.80,0.52,0.00}
\definecolor{orange4}{rgb}{0.55,0.35,0.00}
\definecolor{orangered}{rgb}{1.00,0.27,0.00}
\definecolor{orange}{rgb}{1.00,0.65,0.00}
\definecolor{orchid1}{rgb}{1.00,0.51,0.98}
\definecolor{orchid2}{rgb}{0.93,0.48,0.91}
\definecolor{orchid3}{rgb}{0.80,0.41,0.79}
\definecolor{orchid4}{rgb}{0.55,0.28,0.54}
\definecolor{orchid}{rgb}{0.85,0.44,0.84}
\definecolor{palegoldenrod}{rgb}{0.93,0.91,0.67}
\definecolor{palegreen}{rgb}{0.60,0.98,0.60}
\definecolor{paleturquoise}{rgb}{0.69,0.93,0.93}
\definecolor{paleviolet}{rgb}{0.86,0.44,0.58}
\definecolor{papayawhip}{rgb}{1.00,0.94,0.84}
\definecolor{peachpuff}{rgb}{1.00,0.85,0.73}
\definecolor{peru}{rgb}{0.80,0.52,0.25}
\definecolor{pink1}{rgb}{1.00,0.71,0.77}
\definecolor{pink2}{rgb}{0.93,0.66,0.72}
\definecolor{pink3}{rgb}{0.80,0.57,0.62}
\definecolor{pink4}{rgb}{0.55,0.39,0.42}
\definecolor{pink}{rgb}{1.00,0.75,0.80}
\definecolor{plum1}{rgb}{1.00,0.73,1.00}
\definecolor{plum2}{rgb}{0.93,0.68,0.93}
\definecolor{plum3}{rgb}{0.80,0.59,0.80}
\definecolor{plum4}{rgb}{0.55,0.40,0.55}
\definecolor{plum}{rgb}{0.87,0.63,0.87}
\definecolor{powderblue}{rgb}{0.69,0.88,0.90}
\definecolor{purple1}{rgb}{0.61,0.19,1.00}
\definecolor{purple2}{rgb}{0.57,0.17,0.93}
\definecolor{purple3}{rgb}{0.49,0.15,0.80}
\definecolor{purple4}{rgb}{0.33,0.10,0.55}
\definecolor{purple}{rgb}{0.63,0.13,0.94}
\definecolor{red1}{rgb}{1.00,0.00,0.00}
\definecolor{red2}{rgb}{0.93,0.00,0.00}
\definecolor{red3}{rgb}{0.80,0.00,0.00}
\definecolor{red4}{rgb}{0.55,0.00,0.00}
\definecolor{red}{rgb}{1.00,0.00,0.00}
\definecolor{rosybrown}{rgb}{0.74,0.56,0.56}
\definecolor{royalblue}{rgb}{0.25,0.41,0.88}
\definecolor{saddlebrown}{rgb}{0.55,0.27,0.07}
\definecolor{salmon1}{rgb}{1.00,0.55,0.41}
\definecolor{salmon2}{rgb}{0.93,0.51,0.38}
\definecolor{salmon3}{rgb}{0.80,0.44,0.33}
\definecolor{salmon4}{rgb}{0.55,0.30,0.22}
\definecolor{salmon}{rgb}{0.98,0.50,0.45}
\definecolor{sandybrown}{rgb}{0.96,0.64,0.38}
\definecolor{seagreen}{rgb}{0.18,0.55,0.34}
\definecolor{seashell1}{rgb}{1.00,0.96,0.93}
\definecolor{seashell2}{rgb}{0.93,0.90,0.87}
\definecolor{seashell3}{rgb}{0.80,0.77,0.75}
\definecolor{seashell4}{rgb}{0.55,0.53,0.51}
\definecolor{seashell}{rgb}{1.00,0.96,0.93}
\definecolor{sienna1}{rgb}{1.00,0.51,0.28}
\definecolor{sienna2}{rgb}{0.93,0.47,0.26}
\definecolor{sienna3}{rgb}{0.80,0.41,0.22}
\definecolor{sienna4}{rgb}{0.55,0.28,0.15}
\definecolor{sienna}{rgb}{0.63,0.32,0.18}
\definecolor{skyblue}{rgb}{0.53,0.81,0.92}
\definecolor{slateblue}{rgb}{0.42,0.35,0.80}
\definecolor{slategray}{rgb}{0.44,0.50,0.56}
\definecolor{slategrey}{rgb}{0.44,0.50,0.56}
\definecolor{snow1}{rgb}{1.00,0.98,0.98}
\definecolor{snow2}{rgb}{0.93,0.91,0.91}
\definecolor{snow3}{rgb}{0.80,0.79,0.79}
\definecolor{snow4}{rgb}{0.55,0.54,0.54}
\definecolor{snow}{rgb}{1.00,0.98,0.98}
\definecolor{springgreen}{rgb}{0.00,1.00,0.50}
\definecolor{steelblue}{rgb}{0.27,0.51,0.71}
\definecolor{tan1}{rgb}{1.00,0.65,0.31}
\definecolor{tan2}{rgb}{0.93,0.60,0.29}
\definecolor{tan3}{rgb}{0.80,0.52,0.25}
\definecolor{tan4}{rgb}{0.55,0.35,0.17}
\definecolor{tan}{rgb}{0.82,0.71,0.55}
\definecolor{thistle1}{rgb}{1.00,0.88,1.00}
\definecolor{thistle2}{rgb}{0.93,0.82,0.93}
\definecolor{thistle3}{rgb}{0.80,0.71,0.80}
\definecolor{thistle4}{rgb}{0.55,0.48,0.55}
\definecolor{thistle}{rgb}{0.85,0.75,0.85}
\definecolor{tomato1}{rgb}{1.00,0.39,0.28}
\definecolor{tomato2}{rgb}{0.93,0.36,0.26}
\definecolor{tomato3}{rgb}{0.80,0.31,0.22}
\definecolor{tomato4}{rgb}{0.55,0.21,0.15}
\definecolor{tomato}{rgb}{1.00,0.39,0.28}
\definecolor{turquoise1}{rgb}{0.00,0.96,1.00}
\definecolor{turquoise2}{rgb}{0.00,0.90,0.93}
\definecolor{turquoise3}{rgb}{0.00,0.77,0.80}
\definecolor{turquoise4}{rgb}{0.00,0.53,0.55}
\definecolor{turquoise}{rgb}{0.25,0.88,0.82}
\definecolor{violetred}{rgb}{0.82,0.13,0.56}
\definecolor{violet}{rgb}{0.93,0.51,0.93}
\definecolor{wheat1}{rgb}{1.00,0.91,0.73}
\definecolor{wheat2}{rgb}{0.93,0.85,0.68}
\definecolor{wheat3}{rgb}{0.80,0.73,0.59}
\definecolor{wheat4}{rgb}{0.55,0.49,0.40}
\definecolor{wheat}{rgb}{0.96,0.87,0.70}
\definecolor{whitesmoke}{rgb}{0.96,0.96,0.96}
\definecolor{white}{rgb}{1.00,1.00,1.00}
\definecolor{yellow1}{rgb}{1.00,1.00,0.00}
\definecolor{yellow2}{rgb}{0.93,0.93,0.00}
\definecolor{yellow3}{rgb}{0.80,0.80,0.00}
\definecolor{yellow4}{rgb}{0.55,0.55,0.00}
\definecolor{yellowgreen}{rgb}{0.60,0.80,0.20}
\definecolor{yellow}{rgb}{1.00,1.00,0.00}

\definecolor{white1}{rgb}{1.00,1.00,1.00}

\definecolor{frontal1}{rgb}{0.93,0.27,0.21}
\definecolor{lateral1}{rgb}{0.68,0.66,0.80}
\definecolor{horizontal1}{rgb}{0.94,0.63,0.37}

\definecolor{tnode1}{rgb}{0.93,0.27,0.21}
\definecolor{mnode1}{rgb}{0.68,0.66,0.80}

\definecolor{unsure1}{rgb}{1.00,0.00,0.46}

\definecolor{bluegray}{rgb}{0.4, 0.6, 0.8}
\definecolor{caribbeangreen}{rgb}{0.0, 0.8, 0.6}
\definecolor{darkpastelgreen}{rgb}{0.01, 0.75, 0.24}

\definecolor{pblue2}{rgb}{0.6509803921568628, 0.807843137254902, 0.8901960784313725}
\definecolor{pblue1}{rgb}{0.12156862745098039, 0.47058823529411764, 0.7058823529411765}
\definecolor{pgreen2}{rgb}{0.6980392156862745, 0.8745098039215686, 0.5411764705882353}
\definecolor{pgreen1}{rgb}{0.2, 0.6274509803921569, 0.17254901960784313}
\definecolor{pred2}{rgb}{0.984313725490196, 0.6039215686274509, 0.6}
\definecolor{pred1}{rgb}{0.8901960784313725, 0.10196078431372549, 0.10980392156862745}
\definecolor{porange2}{rgb}{0.9921568627450981, 0.7490196078431373, 0.43529411764705883}
\definecolor{porange1}{rgb}{1.0, 0.4980392156862745, 0.0}
\definecolor{ppurple2}{rgb}{0.792156862745098, 0.6980392156862745, 0.8392156862745098}
\definecolor{ppurple1}{rgb}{0.41568627450980394, 0.23921568627450981, 0.6039215686274509}
\definecolor{pyellow}{rgb}{1.0, 1.0, 0.6}
\definecolor{pbrown}{rgb}{0.6941176470588235, 0.34901960784313724, 0.1568627450980392}

\graphicspath{{./}{./Figures/}}

\theoremstyle{definition}

\newtheorem{remark}{Remark}

\newtheorem{proposition}{Proposition}

\newtheorem{example}{Example}

\makeatletter
\patchcmd{\@maketitle}
{\addvspace{0.5\baselineskip}\egroup}
{\addvspace{-1\baselineskip}\egroup}
{}
{}
\makeatother

\begin{document}

\title{A Statistically Identifiable Model for Tensor-Valued Gaussian Random Variables}

\author{
	
Bruno Scalzo Dees, \IEEEmembership{Student Member, IEEE}, Anh-Huy Phan, \IEEEmembership{Member, IEEE}, Danilo P. Mandic, \IEEEmembership{Fellow, IEEE}
	
\thanks{B. Scalzo Dees and D. P. Mandic are with the Imperial College London, London SW7 2AZ, U.K. (e-mail: bruno.scalzo-dees12@imperial.ac.uk; d.mandic@imperial.ac.uk).}
\thanks{A.-H. Phan is with the Skolkovo Institute of Science and Technology, 143026 Moscow, Russia  (e-mail: a.phan@skolkovo.ru).}


}


\maketitle

\begin{abstract}
Real-world signals typically span across multiple dimensions, that is, they naturally reside on multi-way data structures referred to as \textit{tensors}. In contrast to standard ``flat-view'' multivariate matrix models which are agnostic to data structure and only describe linear pairwise relationships, we introduce the tensor-valued Gaussian distribution which caters for multilinear interactions -- the linear relationship between \textit{fibers} -- which is reflected by the Kronecker separable structure of the mean and covariance. By virtue of the statistical identifiability of the proposed distribution formulation, whereby different parameter values strictly generate different probability distributions, it is shown that the corresponding likelihood function can be maximised analytically to yield the maximum likelihood estimator. For rigour, the statistical consistency of the estimator is also demonstrated through numerical simulations. The probabilistic framework is then generalised to describe the joint distribution of multiple tensor-valued random variables, whereby the associated mean and covariance exhibit a Khatri-Rao separable structure. The proposed models are shown to serve as a natural basis for gridded atmospheric climate modelling.
\end{abstract}

\vspace{-0.1cm}


\IEEEpeerreviewmaketitle

\vspace{-0.2cm}

\section{Introduction}

Tensor data structures are gaining increasing prominence in modern Data Analytics, especially in relation to the Big Data paradigm where, equipped with the power of their underlying multilinear algebra, they provide a rich analysis platform for making sense from multidimensional data. In particular, \textit{tensor decompositions} have experienced a surge in popularity owing to their role as high-dimensional generalisations of the ``flat-view'' linear algebra paradigms, an example of which is the tensor-valued higher-order singular value decomposition (HOSVD), a generic extension of the ordinary matrix SVD. The tensor data domain is amenable to such generalisations of standard matrix signal processing and machine learning tools \cite{Kolda2009,Mandic2015_3,Mandic2017_2,Mandic2017_3,Sidiropoulos2017_1}. Real-world applications of tensors include those in chemometrics \cite{Smilde2005}, fluid mechanics \cite{Klus2018}, geostatistics \cite{Liu2007}, magnetic resonance imaging \cite{Basser2003}, psychometrics \cite{Kiers2001}, statistical mechanics \cite{Soize2007}, MIMO communications \cite{Sidiropoulos2000} and biomedical applications \cite{Spyrou2018}.

In addition, the analysis of tensor-valued data through multilinear algebra has been an enabling tool for solving critical information representation and storage bottlenecks, namely the curse of dimensionality. However, while multilinear algebra assumes some form of deterministic relation between data entries, the interactions between real-world observables are typically causal and probabilistic, including practical situations where even deterministic information is contaminated with random noise, missing or unreliable entries. This makes the existing multilinear models inadequate for such scenarios, since the associated \textit{probability density functions} are not yet well defined for tensor-valued data. The consideration of tensor-valued tools within a rigorous probabilistic framework would therefore offer a number of important advantages: 
\begin{enumerate}[label=(\roman*)]
	\item Possibility for statistical and hypothesis testing through the likelihood function; 
	\item Opportunity to introduce Bayesian inference methods; 
	\item Ability to employ class-conditional densities for classification tasks; 
	\item A framework to assess the \textit{degree of novelty} of a new data point, and input variable selection;
	\item A rigorous probabilistic framework to deal with missing data values;
	\item Straightforward consideration of a \textit{mixture} of probabilistic models.
\end{enumerate}
While this has naturally motivated the developments of probabilistic tensor-valued models, owing to the ambiguity in the problem formulation, a wide range of solutions have been proposed. The tensor-valued Gaussian processes were first proposed in \cite{Ghahramani2009} as a means of obtaining a probabilistic variant of the Tucker decomposition which can handle missing data entries. The work in \cite{Hoff2011} further proposed a hierarchical Bayesian extension to the Tucker decomposition. The basic properties of the distribution, such as the marginal and conditional distributions, moments, and characteristic function, were later derived in \cite{Ohlson2013}. However, all these methods are, in some sense, extensions of the matrix-valued Gaussian distribution with Kronecker separable covariance matrix \cite{Dawid1981, West1988, Gupta2000}.

Several remaining issues need to be addressed prior to a more widespread application of the class of probabilistic tensor-valued models. For example, existing parameter estimation procedures for tensor-valued Gaussian distributions are iterative, such as the \textit{expectation-maximization} algorithm \cite{Ghahramani2009,Xu2012} or the block coordinate descent \cite{Dutilleul1999,Hoff2011,Dutilleul2013,Singull2015}, also referred to as the \textit{flip-flop} algorithm. Such techniques are susceptible to local maxima and do not guarantee global optimality. A closely related topic is that of covariance matrix estimation with the Kronecker separable structure \cite{Stoica2008}. Two asymptotically efficient estimation solutions have been proposed, the first being a variant of the well-known \textit{alternating maximization} technique, while the second method is based on \textit{covariance matching} principles. However, these estimators were derived only for the Kronecker product of two matrices, and were not considered within the tensor-valued setting. 

\pagebreak

This all points out that there is a need for a general class of estimators, derived in a closed-form in the tensor domain, which would guarantee global optimality of probabilistic estimation procedures. In this way, the problem with the existing models which do not impose multilinear assumptions on the structure of the mean would be resolved, a critical issue for a complete characterisation of tensor-valued random variables. To this end, we derive a rigorous form of the tensor-valued Gaussian distribution and introduce its corresponding maximum likelihood estimator in a closed-form. This is achieved through a novel \textit{statistically identifiable} formulation of the distribution, whereby different values of the parameters generate strictly different probability distributions; this allows for the underlying likelihood function to be maximised analytically, unlike the existing formulations. Moreover, we extend the proposed probabilistic framework to account for the joint distribution of multiple tensor-valued random variables. 

The rest of this paper is organized as follows. Section \ref{section:preliminaries} provides a comprehensive introduction to multilinear algebra. Section \ref{section:kronecker} describes the underpinning Kronecker separable mean and covariance properties exhibited by tensor-valued random variables. The proposed tensor-valued Gaussian distribution and its maximum likelihood estimator are introduced in Section \ref{section:gaussian_dist}. The multivariate tensor-valued distribution is derived in Section \ref{sec:multivariate_tensor}. An intuitive example of the proposed model applied to gridded atmospheric temperature modelling is provided in Section \ref{sec:climate_modelling}.



\section{Preliminaries}

\label{section:preliminaries}

We follow the notation employed in \cite{Kolda2009}, whereby scalars are denoted by a lightface font, e.g. $x$; vectors by a lowercase boldface font, e.g. $\x$; matrices by a uppercase boldface font, e.g. $\X$; and tensors by a boldface calligraphic font, e.g. $\calX$.



The \textit{order} of a tensor defines the number of its dimensions, also referred to as \textit{modes}, i.e. the tensor $\calX \in \domR^{I_{1}\times \cdots \times I_{N}}$ has $N$ modes and $K = \prod_{n=1}^{N} I_{n}$ elements in total. 

Tensors can be reshaped into mathematically tractable lower-dimensional representations (unfoldings) which can be manipulated using standard linear algebra. The \textit{vector unfolding}, also known as vectorization, is denoted by 
\begin{align}
	\x = \vect{\calX} \quad \in \domR^{K}
\end{align}
while the \textit{mode-$n$ unfolding} (matricization), denoted by $\X_{(n)} \in \domR^{I_{n} \times \frac{K}{I_{n}}}$, is obtained by reshaping a tensor into a matrix in the form
\begin{align}
	\X_{(n)}=\left[ \begin{array}{cccc}
		\f_{1}^{(n)}, & \f_{2}^{(n)}, & \dots , & \f_{\frac{K}{I_{n}}}^{(n)}
	\end{array} \right]\label{eq:mode_n_unfolding}
\end{align}
where the column vector, $ \f_{i}^{(n)} \in \domR^{I_{n}}$, is referred to as the $i$-th \textit{mode-$n$ fiber}. Fibers are a multi-dimensional generalization of matrix rows and columns.

\begin{figure}[ht]
	\vspace{-0.2cm}
	\centering
	\includegraphics[width=0.35\textwidth, trim={2cm 0 0 0},clip]{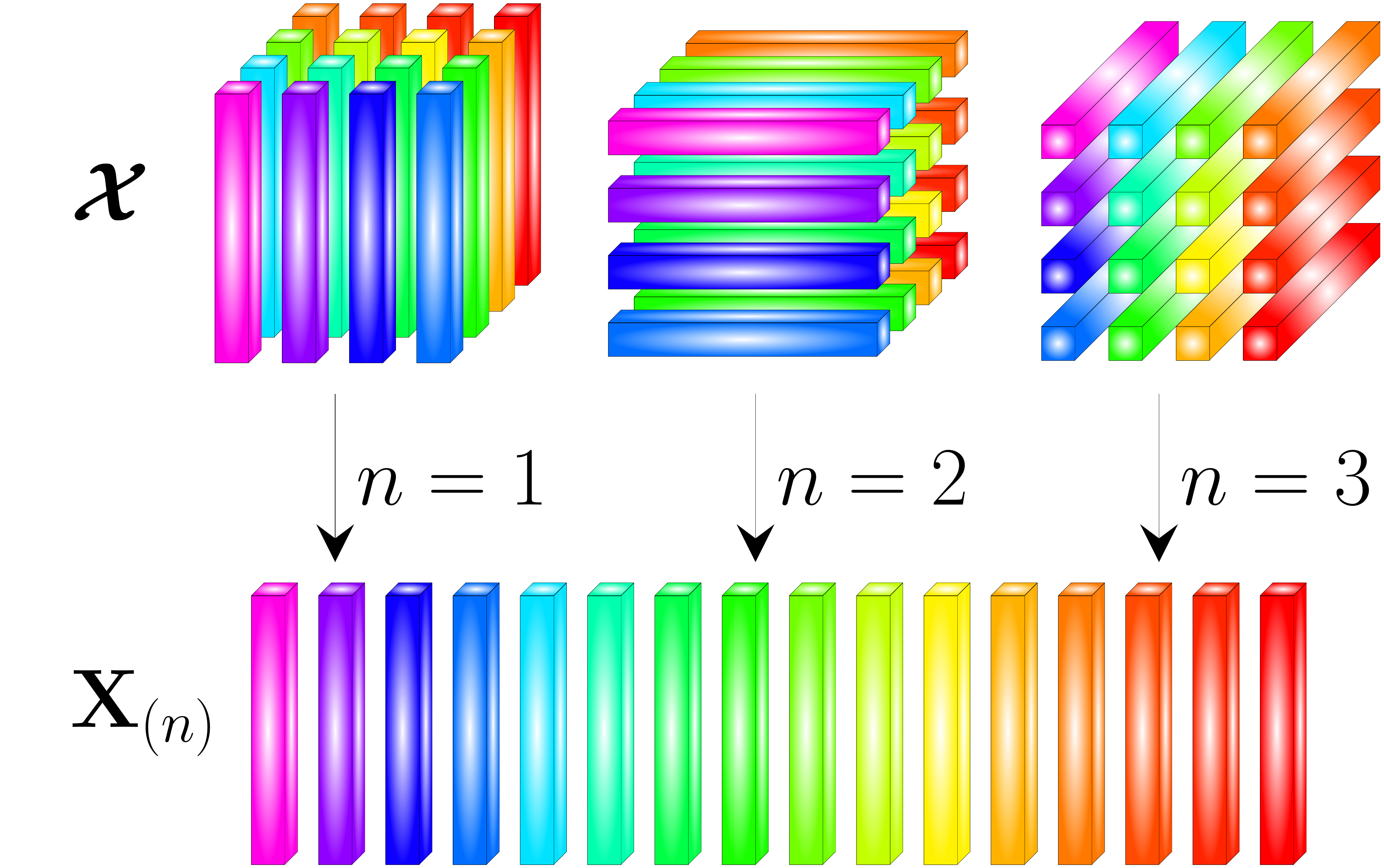}
	\caption{\label{fig:mode_n_unfold} Mode-$n$ matrix unfolding, $\X_{(n)}$, of an order-3 tensor, $\calX$, as a rearrangement of the mode-$n$ fibers, $\f_{i}^{(n)}$.}
	\vspace{-0.4cm}
\end{figure}

The operation of \textit{mode-$n$ unfolding} can be viewed as a rearrangement of the mode-$n$ fibers as column vectors of the matrix, $\X_{(n)}$, as illustrated in Figure \ref{fig:mode_n_unfold}. Notice that the considered order-$3$ tensor, $\calX$, has alternative representations in terms of mode-$1$ (left panel), mode-$2$ (middle panel) and mode-$3$ fibers (right panel), that is, its \textit{columns}, \textit{rows} and \textit{tubes}.

\subsection{Tensor products}

\begin{figure}[H]
	\centering
	\includegraphics[width=0.35\textwidth]{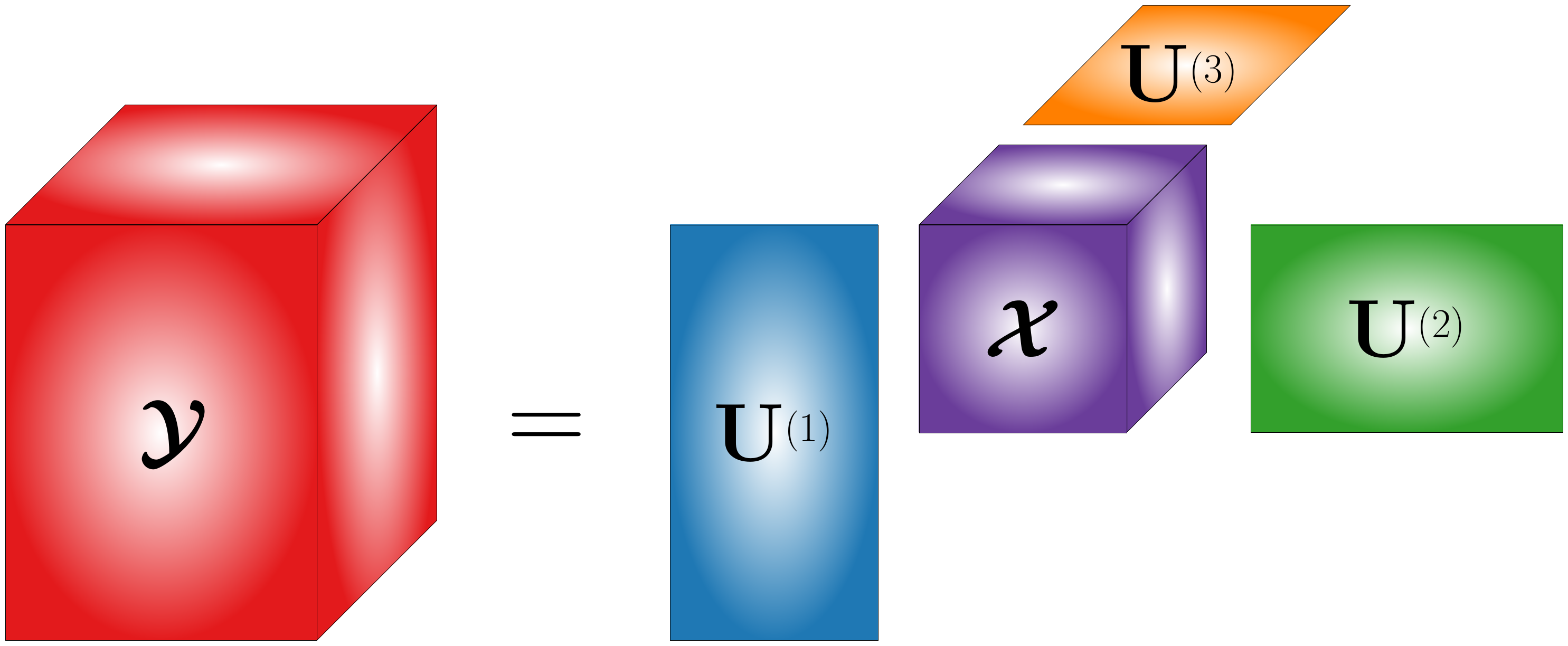}
	\caption{\label{fig:Tucker} Sequence of \textit{mode-$n$ products} for $n=1,2,3$.}
\end{figure}

The \textit{Kronecker product} between the matrices $\A \in \domR^{I\times I}$ and $\B \in \domR^{J \times J}$ yields a block matrix
\begin{align}
	\A \otimes \B = \left[\begin{array}{ccc}
		a_{11}\B & \cdots & a_{1I}\B \\
		\vdots & \ddots & \vdots \\
		a_{I1}\B & \cdots & a_{II}\B
	\end{array}\right] \quad \in \domR^{IJ \times IJ}
\end{align}
The \textit{Khatri-Rao product} between two block matrices with $M$ row and column partitions, $\A \in \domR^{MI\times MI}$ and $\B \in \domR^{MJ \times MJ}$, yields the block matrix $\A \oast \B \in \domR^{MIJ \times MIJ}$, with the $(i,j)$-th block given by
\begin{align}
	\A \oast \B & = \left[ 
	\arraycolsep=5pt
	\def\arraystretch{1.3}
	\begin{array}{ccc}
		\A_{11}\otimes\B_{11} & \cdots & \A_{1M}\otimes\B_{1M} \\
		\vdots & \ddots & \vdots \\
		\A_{M1}\otimes\B_{M1} & \cdots & \A_{MM}\otimes\B_{MM} \\
	\end{array} \right]
\end{align}
The \textit{partial trace} operator of a block matrix with $M$ row and column partitions, $\A \in \domR^{MI\times MI}$, yields the block matrix $\ptr{\A} \in \domR^{M \times M}$, with the $(i,j)$-th block given by
\begin{align}
	\ptr{\A} = \left[\begin{array}{ccc}
		\tr{\A_{11}} &\cdots & \tr{\A_{1M}} \\
		\vdots &  \ddots & \vdots \\
		\tr{\A_{M1}} & \cdots & \tr{\A_{MM}}
	\end{array}\right]
\end{align}
The \textit{mode-$n$ product} of the tensor $\calX \in \domR^{I_{1}\times \cdots \times I_{N}}$ with the matrix $\U \in \domR^{J_{n} \times I_{n}}$ is denoted by 
\begin{align}
	\calY = \calX \times_{n} \U \quad \in \domR^{I_{1} \times \cdots \times I_{n-1} \times J_{n} \times I_{n+1} \times \cdots \times I_{N}}
\end{align}
and is equivalent to performing the following steps:
\begin{spacing}{1}
	\begin{algorithmic}[1]
		\State $\X_{(n)} \leftarrow \calX $ \Comment{Mode-$n$ unfold}
		\State $\Y_{(n)} \leftarrow \U \X_{(n)}$ \Comment{Left matrix multiplication}
		\State $\calY \leftarrow \Y_{(n)}$ \Comment{Re-tensorize}
	\end{algorithmic}
\end{spacing}
\noindent For convenience, we denote the sequence of Kronecker products of the matrices $\U^{(n)} \in \domR^{I_{n} \times I_{n}}$ by
\begin{equation}
	\kronprod{n=1}{N} \U^{(n)} = \U^{(1)}\otimes\cdots\otimes\U^{(N)} \in \domR^{K \times K}
\end{equation}
and the sequence of Khatri-Rao products of the block matrices with $M$ row and column partitions, $\U^{(n)} \in \domR^{MI_{n} \times MI_{n}}$, by
\begin{equation}
\khatriprod{n=1}{N}\U^{(n)} = \U^{(1)}\oast\cdots\oast\U^{(N)} \in \domR^{MK \times MK}
\end{equation}
The sequence of outer products of the vectors $\u^{(n)} \in \domR^{I_{n}}$ is denoted by
\begin{equation}
\circprod{n=1}{N}\u^{(n)} = \u^{(1)}\circ\cdots\circ\u^{(N)} \in \domR^{I_{1} \times \cdots \times I_{N}}
\end{equation}
The sequence of \textit{mode-$n$ products} between the tensor $\calX$ and the matrices $\U^{(n)} \in \domR^{J_{n} \times I_{n}}$ is denoted by
\begin{equation}
	\calY = \calX \modeprod{n=1}{N}\U^{(n)} = \calX \times_{1}\U^{(1)}\times_{2}\cdots\times_{N}\U^{(N)} \in \domR^{J_{1} \times \cdots \times J_{N}}
\end{equation}
\noindent This operation can also be expressed in the mathematically equivalent vector and matrix representations, that is
\begin{align}
	\y & = \left( \kronprod{\subalign{n&=N}}{1} \U^{(n)} \right)\x \quad \in \domR^{L} \\
	\Y_{(n)} & = \U^{(n)}\X_{(n)}\left( \kronprod{\subalign{i&=N\\i&\neq n}}{1} \U^{(i)\Trans} \right) \in \domR^{J_{n} \times \frac{L}{J_{n}}}
\end{align}
where $L = \prod_{n=1}^{N} J_{n}$. Figure \ref{fig:Tucker} illustrates the sequence of mode-$n$ products of an order-3 tensor with matrices $\U^{(n)}$.

%


\subsection{Tensor-valued statistical operators}

For clarity, we shall first introduce the notation for the first- and second-order tensor-valued statistical operators. The operation of taking the expectation of a random tensor, $\calX$, is equal to the element-wise expectation, which yields the mean tensor $\calM = \expect{\calX}$. The \textit{variance} of $\calX$ is then defined as
\begin{align}
	\var{\calX} = \expect{\| \calX - \calM \|^{2}} = \expect{\| \calS \|^{2}}
\end{align}
that is, the expected squared Frobenius norm of the \textit{centred} tensor variable, $ \calS = \calX - \calM $. Using the mode-$n$ unfolding representation in (\ref{eq:mode_n_unfolding}) based on fibers, we can now define the \textit{mode-$n$ fiber covariance} through the total expectation theorem \cite{Weiss2005} as follows
\begin{align}
	\cov{\f^{(n)}} = E_{i}\!\left\{ \cov{ \f_{i}^{(n)} } \right\} = \expect{\S_{(n)}\S_{(n)}^{\Trans}}
\end{align}
where $E_{i}\{\cdot\}$ denotes the expectation over the indices $i$. We also denote this operation by $\cov{\X_{(n)}} \equiv \cov{\f^{(n)}}$.


\section{Kronecker Separable Statistics}

\label{section:kronecker}

In standard multivariate data analysis, multiple measurements are collected at a given trial, experiment or time instant, to form a vector-valued data sample, $\x \in \domR^{K}$. An assumption inherently adopted in statistical modeling is that the variables are described by the probability distribution, $\x \sim \mathcal{N}\left( \m, \R \right)$, which implies that the mean vector, $\m \in \domR^{K}$, and covariance matrix, $\R \in \domR^{K \times K}$, are unstructured. However, if the variables have a natural tensor representation, then it is desirable, and even necessary, to assume that the mean and covariance exhibit a more structured form motivated by physical considerations. It then naturally follows that the statistical properties of tensor-valued random variables are directly linked to those of \textit{separable Gaussian random fields} -- the continuous-space counterpart of tensors -- introduced in the next section. 

\subsection{Separable Gaussian random fields}

A Gaussian random field in an $N$-dimensional orthogonal coordinate system is given by $x : \domR^{N} \mapsto \domR$, and is described by the coordinate-dependent distribution
\begin{align}
	x(\z) \sim \Normal{m(\z),\sigma^{2}(\z)}
\end{align}
where $\z = \{ z^{(1)}, ..., z^{(N)} \} \in \domR^{N}$ is an $N$-dimensional coordinate vector, and $z^{(n)} \in \domR$ is the $n$-th axis coordinate. Furthermore, such a random variable is equipped with a covariance operator, denoted by $\sigma : \domR^{N} \times \domR^{N} \mapsto \domR$, which yields
\begin{align}
	\sigma(\z_{1},\z_{2}) = \cov{x(\z_{1}),x(\z_{2})}
\end{align}
where $\sigma(\z,\z) \equiv \sigma^{2}(\z)$. A random variable is said to exhibit a \textit{separable} mean and covariance structure if and only if the mean and covariance operators are linearly separable, that is
\begin{alignat}{2}
	m(\z) & = \prod_{n=1}^{N} m^{(n)}(z^{(n)}), \quad && \forall \z \in \domR^{N} \label{eq:separability_GRF_mean} \\
	\sigma(\z_{1},\z_{2}) & = \prod_{n=1}^{N} \sigma^{(n)}(z_{1}^{(n)},z_{2}^{(n)}), \quad && \forall \z_{1},\z_{2} \in \domR^{N} \label{eq:separability_GRF}
\end{alignat}
where $m^{(n)}:\domR \mapsto \domR$ and $\sigma^{(n)}: \domR \times \domR \mapsto \domR$ are the mean and covariance operators specific to the $n$-th coordinate axis.


\begin{remark}
	Real-world examples of fields in $N$-dimensional coordinates that are typically analysed using signal processing and machine learning techniques include: 
	\begin{enumerate}[label=(\roman*)]
		\item meteorological measurements in the \textit{longitude $\times$ latitude $\times$ altitude} space;
		\item colored pixels in the \textit{column $\times$ row $\times$ (R, G, B)} space;
		\item time-frequency multichannel signals which reside in the \textit{time $\times$ frequency $\times$ channel} space.
	\end{enumerate}
\end{remark}

\begin{remark}
	Orthogonal coordinate systems that are most commonly found in Physics and Engineering include the Cartesian, spherical polar, and cylindrical polar systems. While the reason to prefer orthogonal coordinates over general curvilinear coordinates is their \textit{simplicity}, complications typically arise when coordinates are not orthogonal, for instance, in orthogonal coordinates problems can be solved by \textit{separation of variables}, which reduces a single $N$-dimensional problem into $N$ single-dimensional problems. Tensors are naturally endowed with this powerful property.
\end{remark}




\begin{remark}
	A function that is linearly separable in a given coordinate system need not remain separable upon a change of the coordinate system. This asserts that the coordinate system used for \textit{tensorizing} a sampled field should be chosen so as to match the properties of the underlying physics. We next introduce the necessary tensorization condition to guarantee separability, which we refer to as \textit{topological coherence}.
\end{remark}


\subsection{Topologically coherent tensorization}

\begin{figure}[h!]
	\vspace{-0.3cm}
	\centering
	\includegraphics[width=0.4\textwidth, trim={0cm 0 0 0},clip]{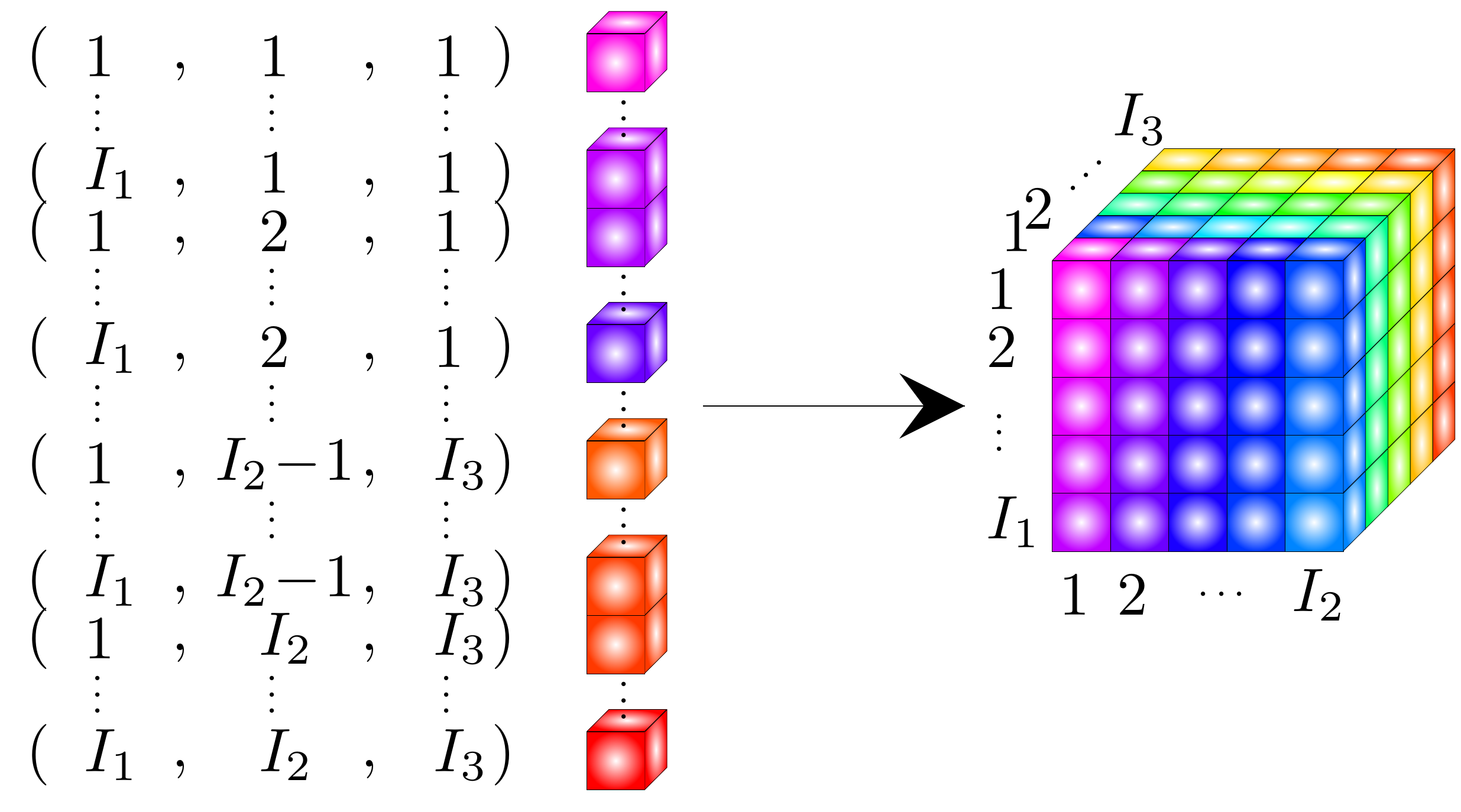}
	\vspace{-0.2cm}
	\caption{\label{fig:multi_index} Topologically coherent tensorization of samples using their $3$-dimensional coordinates.}
	\vspace{-0.3cm}
\end{figure}

The collection of $K$ samples of a separable Gaussian random field in an $N$-dimensional orthogonal coordinate system, $x : \domR^{N} \mapsto \domR$, admits the \textit{topologically coherent} tensor representation, $\calX \in \domR^{I_{1}\times \cdots \times I_{N}}$, if and only if
\begin{align}
[\calX]_{i_{1}...i_{N}} = x(z_{i_{1}}^{(1)},...,z_{i_{N}}^{(N)}), \quad i_{n} \in \domN, \quad z_{i_{n}}^{(n)} \in \domR
\end{align}
Figure \ref{fig:multi_index} illustrates the \textit{tensorization} of samples from a field on a $3$-D coordinate system to form an order-$3$ tensor.

\vspace{-0.1cm}

\begin{remark}
	Consider an order-2 tensor, $\calX \in \domR^{I_{1} \times I_{2}}$, sampled from the field, $x: \domR^{2} \mapsto \domR$, on a 2D polar coordinate system. Then, $\calX$ is topologically coherent if $[\calX]_{i_{1}i_{2}} = x(z_{i_{1}}^{(r)},z_{i_{2}}^{(\theta)})$, where $z_{i_{1}}^{(r)} $ and $z_{i_{2}}^{(\theta)}$ denote respectively the radial and angular coordinates. Notice that if, in turn, the tensor were sampled using a lattice on the 2D Cartesian coordinate system, then it would be \textit{topologically incoherent}.
\end{remark}

\vspace{-0.1cm}




\subsection{Kronecker separable statistics}

Consider a tensor, $\calX  \in \domR^{I_{1}\times \cdots \times I_{N}} $, which has been coherently sampled from a separable Gaussian field. By virtue of the statistical properties of separable Gaussian random fields in (\ref{eq:separability_GRF_mean})-(\ref{eq:separability_GRF}), it then follows that statistical properties of the scalar-valued entries of $\calX$ are also separable, that is
\begin{align}
	\expect{[\calX]_{i_{1}\cdots i_{N}}} & = \prod_{n=1}^{N} m_{i_{n}}^{(n)} \label{eq:mean_separation} \\
	\cov{[\calX]_{i_{1}\cdots i_{N}},[\calX]_{j_{1}\cdots j_{N}}} & = \prod_{n=1}^{N} \sigma_{i_{n}j_{n}}^{(n)} \label{eq:covariance_separation}
\end{align}
where $m_{i}^{(n)}$ is the mean parameter associated with the $i$-th coordinate along the $n$-th mode of $\calX$, and similarly, $\sigma_{ij}^{(n)}$ is the covariance parameter associated with the $i$-th and $j$-th coordinates along the $n$-th mode. 

By jointly considering all of the elements in $\calX$, it can be shown that the mean and covariance structures exhibit the following Kronecker separable properties \cite{Gupta2000,Hoff2011}
\begin{align}
\expect{\x} & = \kronprod{n=N}{1} \m^{(n)}  \label{eq:kron_mean} \\
\cov{\x} & = \kronprod{n=N}{1} \R^{(n)}  \label{eq:kron_covariance} \\
\cov{\X_{(n)}} & = \Biggl( \prod_{\subalign{i&=1 \\ i&\neq n}}^{N} \tr{ \R^{(i)} } \Biggr) \R^{(n)} \label{eq:mode_n_covariance}
\end{align}
where $\m^{(n)} \in \domR^{I_{n}}$ and $\R^{(n)} \in \domR^{I_{n} \times I_{n}}$ are respectively the mode-$n$ mean and covariance parameters. With reference to (\ref{eq:mean_separation})-(\ref{eq:covariance_separation}), we have that $[\m^{(n)}]_{i} = m_{i}^{(n)}$ and $[\R^{(n)}]_{ij} = \sigma_{ij}^{(n)}$.

\begin{remark}
	Intuitively, the mean structure in (\ref{eq:kron_mean}) is characterised by the parameter, $\m^{(n)}$, which describes the mean of the fibres in the $n$-th mode. This contrasts the element-wise based mean implied by the standard multivariate Gaussian distribution. Similarly, the covariance structure in (\ref{eq:kron_covariance}) is parametrized in terms of linear \textit{fiber-to-fiber} (multilinear) covariances, $\R^{(n)}$, which contrasts the linear pairwise based definition of the covariance in the multivariate Gaussian model.
\end{remark}

\begin{remark}
	A reshaping of the Kronecker separable mean in (\ref{eq:kron_mean}) reveals that the tensor-valued representation of the tensor mean exhibits a rank-$1$ canonical polyadic decomposition (CPD) structure of the form
	\begin{align}
		\expect{\calX} = \circprod{n=1}{N}  \m^{(n)} \label{eq:CPD_mean}
	\end{align}
	Figure \ref{fig:CPD_mean} shows the CPD structure of the order-$3$ tensor mean.
	\vspace{-0.4cm}
	
	\begin{figure}[H]
		\centering
		\includegraphics[width=0.3\textwidth, trim={0cm 0 0 0},clip]{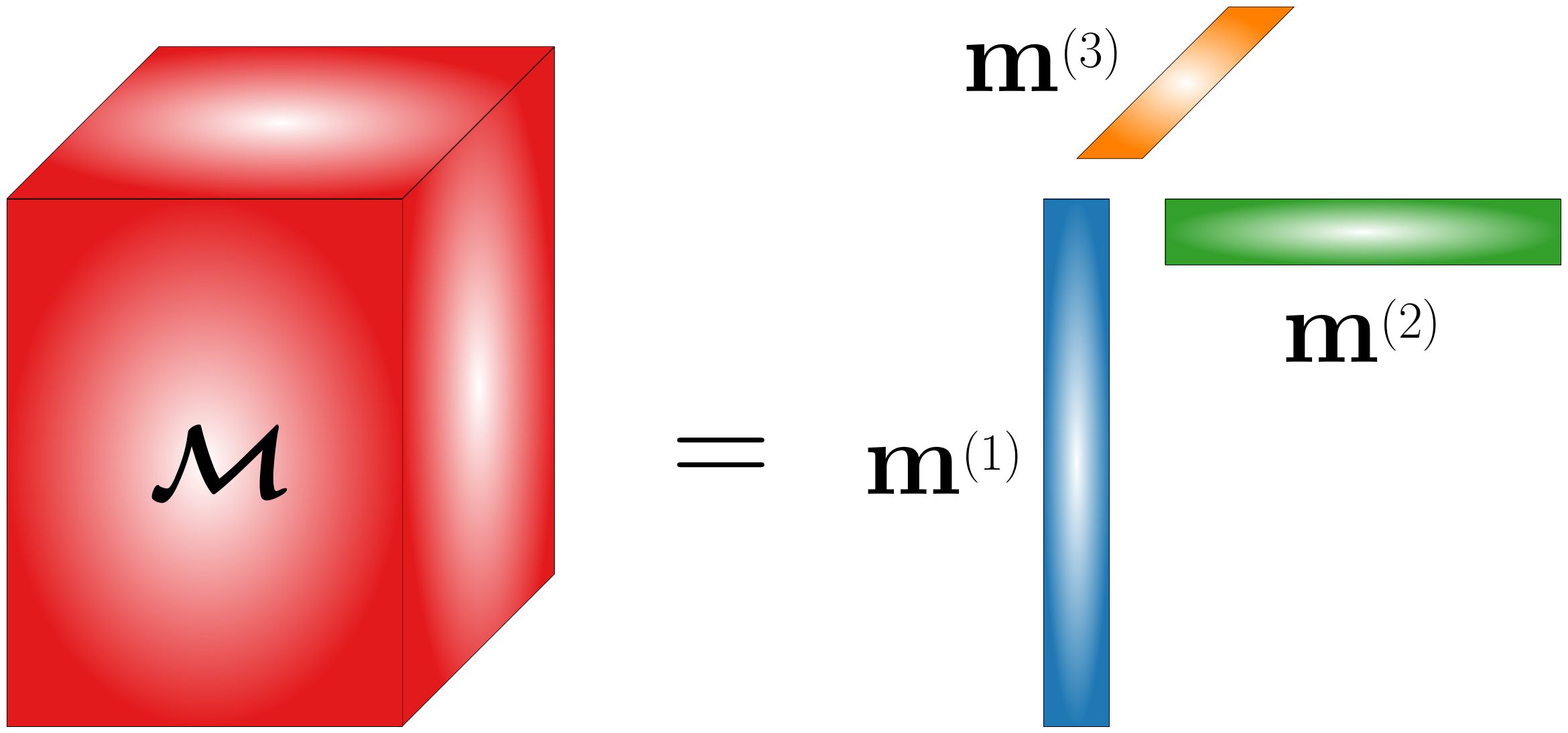}
		\caption{\label{fig:CPD_mean} CPD structure of the order-$3$ tensor mean, $\calM$.}
	\end{figure}

\end{remark}



\begin{example}
	To illustrate a physical interpretation of the Kronecker separability property of the mean in (\ref{eq:kron_mean}) and (\ref{eq:CPD_mean}), consider the order-$2$ tensor-valued random variable, $\X \in \domR^{10 \times 10}$, which exhibits a separable deterministic mean, $\M \in \domR^{10 \times 10}$, in the sense that $\M$ is given by the outer product of two single-dimensional vectors, $\m^{(1)}, \m^{(2)} \in \domR^{10}$, as illustrated in Figure \ref{fig:example_mean_2D}. The Kronecker separability arises because the vector representation of $\M =  \m^{(1)} \circ \m^{(2)} $ can be written as
	\begin{align}
		\m = \vect{\M} =  \m^{(2)} \otimes \m^{(1)} 
	\end{align}
	
	
	\begin{figure}[ht]
		\centering
		\begin{subfigure}[t]{0.49\textwidth}
			\centering
			\includegraphics[width=0.75\textwidth, trim={5cm 7cm 5cm 7cm}, clip]{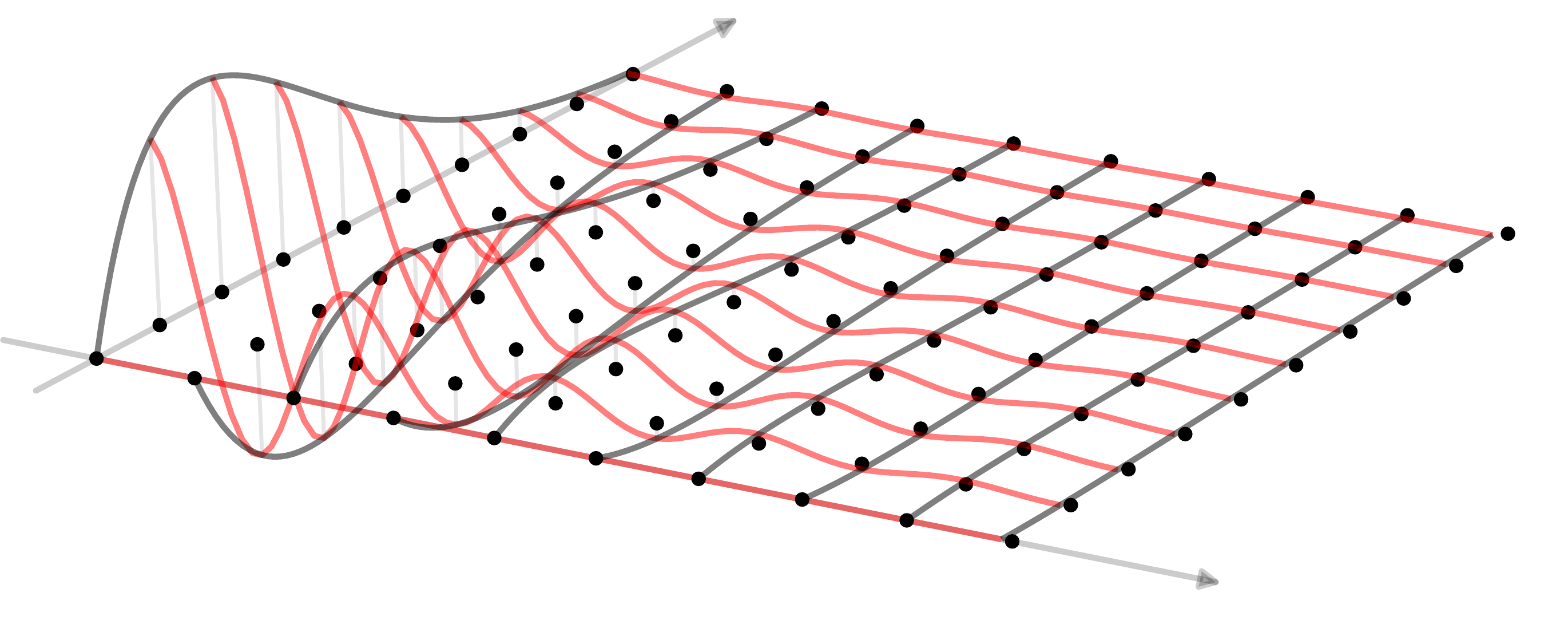}
			\caption{Tensor mean, $\M = \m^{(1)} \circ \m^{(2)}  \in \domR^{10 \times 10}$.}    
			\label{fig:mean_2D}
		\end{subfigure}
		\vspace{-0.2cm}
		\vskip\baselineskip
		\begin{subfigure}[t]{0.24\textwidth}   
			\centering 
			\includegraphics[width=0.7\textwidth, trim={13cm 9cm 12cm 8cm}, clip]{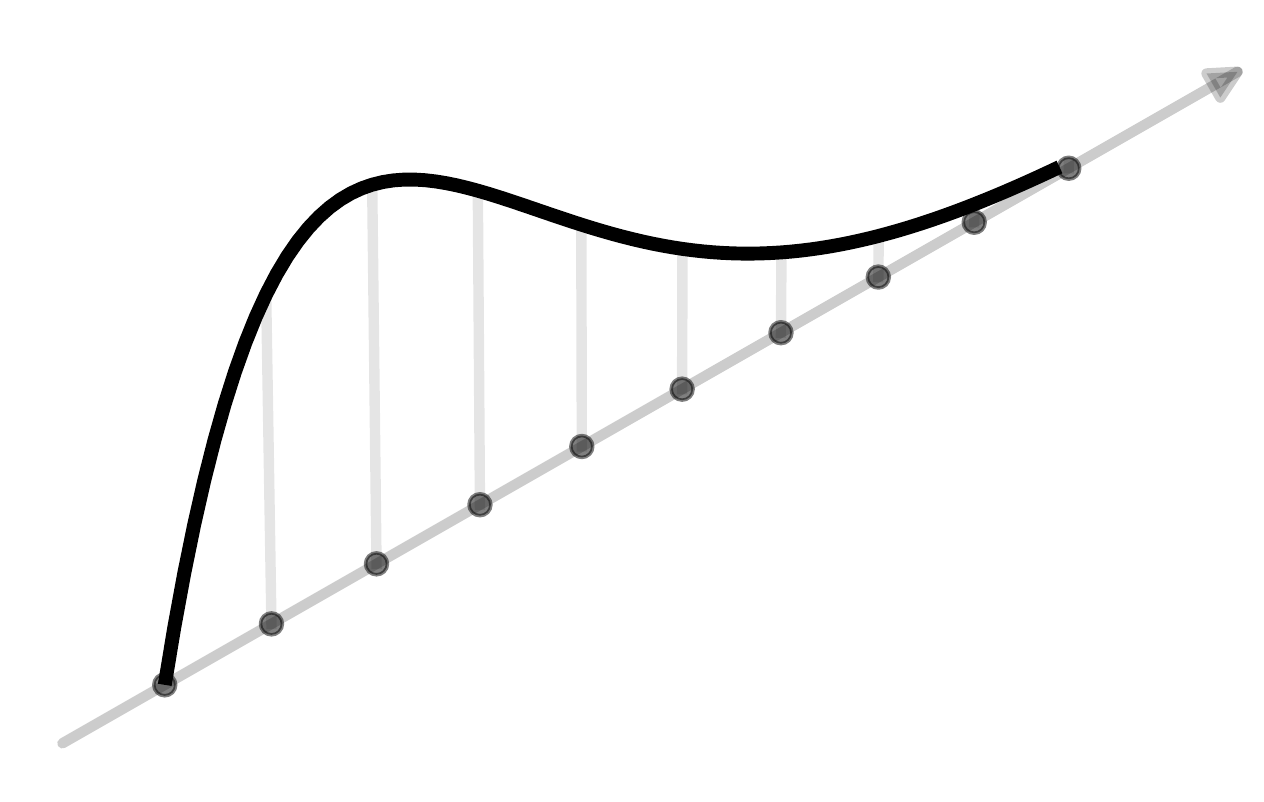} 
			\caption{{$\m^{(1)} \in \domR^{10}$.}}    
			\label{fig:mean_2D_x}
		\end{subfigure}
		\hfill
		\begin{subfigure}[t]{0.24\textwidth}   
			\centering 
			\includegraphics[width=0.9\textwidth, trim={9cm 9cm 6cm 5cm}, clip]{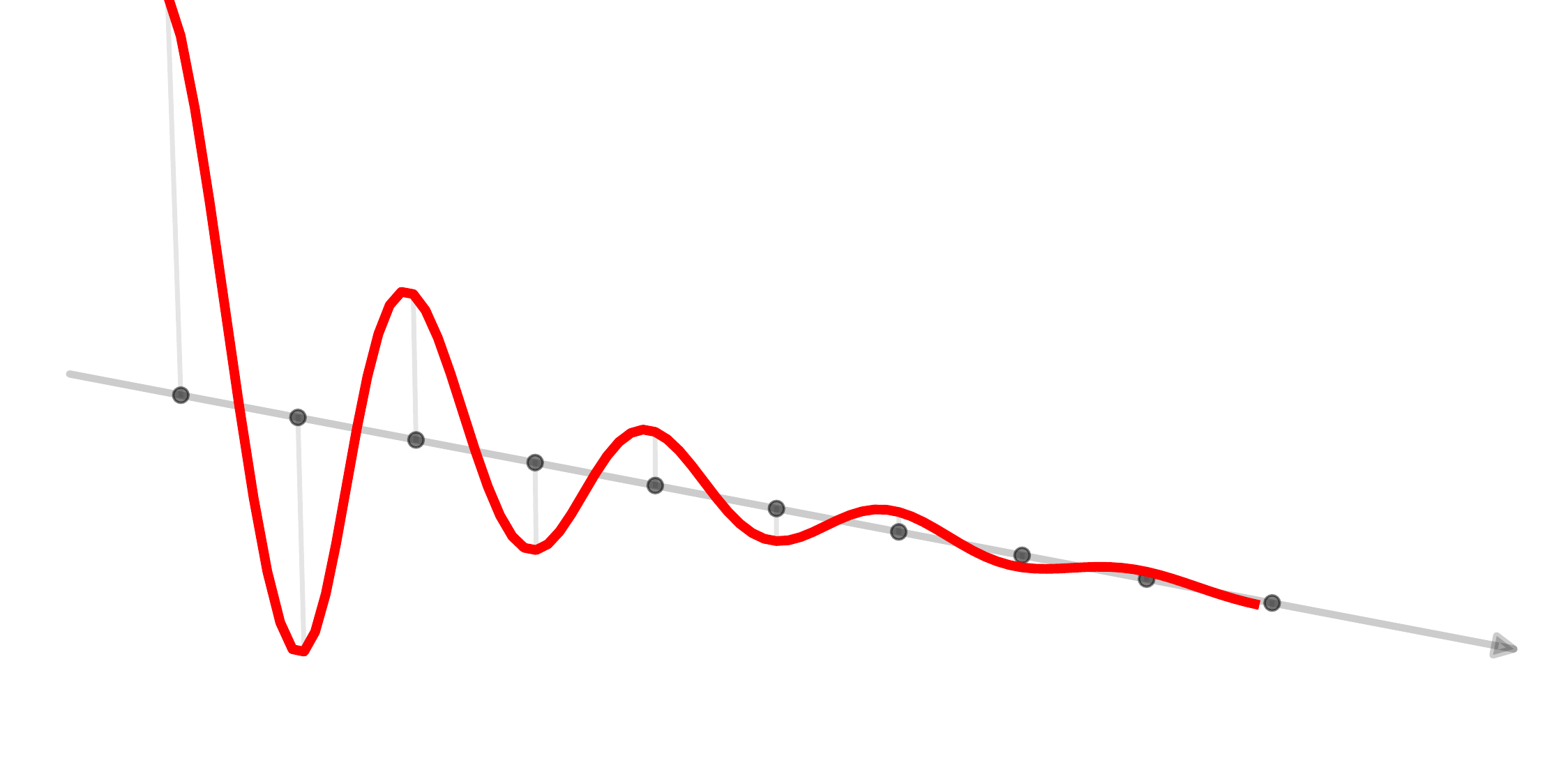} 
			\caption{$\m^{(2)} \in \domR^{10}$.}
			\label{fig:mean_2D_y}
		\end{subfigure}
		\caption{The Kronecker separable mean of an order-$2$ tensor variable, $\X \in \domR^{10 \times 10}$. (a) The tensor mean, $\M \in \domR^{10 \times 10}$. (b) The mode-$1$ mean component, $\m^{(1)} \in \domR^{10}$. (c) The mode-$2$ mean component, $\m^{(2)} \in \domR^{10}$.} 
		\label{fig:example_mean_2D}
	\end{figure}


\end{example}


\begin{example}
To provide an intuitive perspective on the formulation of the covariance structure in (\ref{eq:kron_covariance}) (which is less obvious), consider an order-$2$ tensor-valued variable, $\X \in \domR^{2 \times 2}$, which consists of 4 scalar-valued random variables, $a,b,c,d \in \domR$, as illustrated in Figure \ref{fig:order2Tensor}. Notice that $\X$ has alternative representations in terms of its mode-$1$ fibers, $\f_{i}^{(1)} \in \domR^{2}$, and mode-$2$ fibers, $\f_{i}^{(2)} \in \domR^{2}$, that is, in terms of its columns and rows.
\begin{figure}[H]
	\centering
	\includegraphics[width=0.4\textwidth, trim={0cm 0cm 0 0cm},clip]{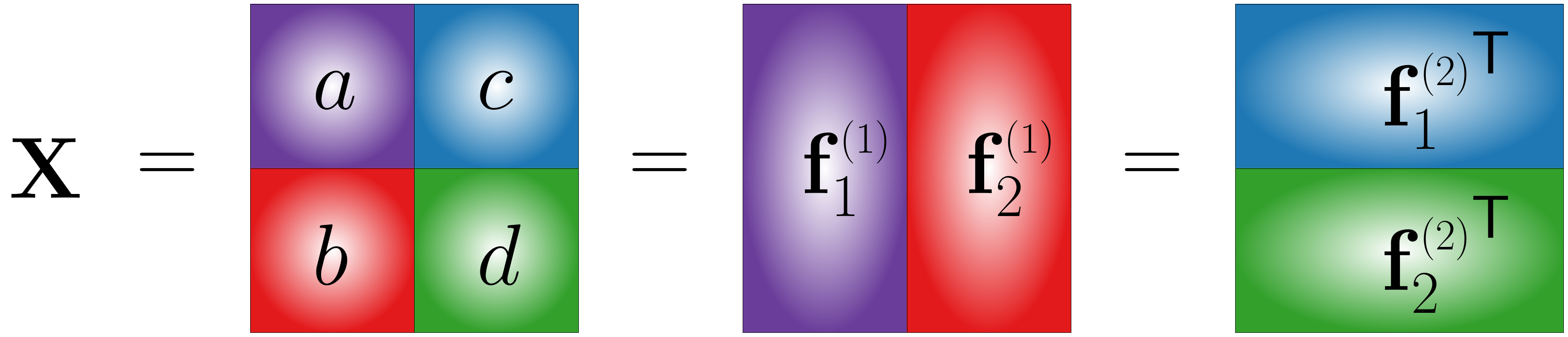}
	\caption{\label{fig:order2Tensor} The order-$2$ tensor, $\X$, represented in terms of its scalar-valued entities, $\{a,b,c,d\}$, and mode-$1$ and mode-$2$ fibers.}
\end{figure}
\noindent The tensor, $\X$, can also be described using its vector and mode-$n$ unfolded representations, as shown in Figure \ref{fig:order2TensorReps}.
\begin{figure}[H]
	\centering
	\begin{minipage}{0.09\textwidth}
		\centering
		\includegraphics[width=0.95\textwidth, trim={0cm 0cm 0 0cm},clip]{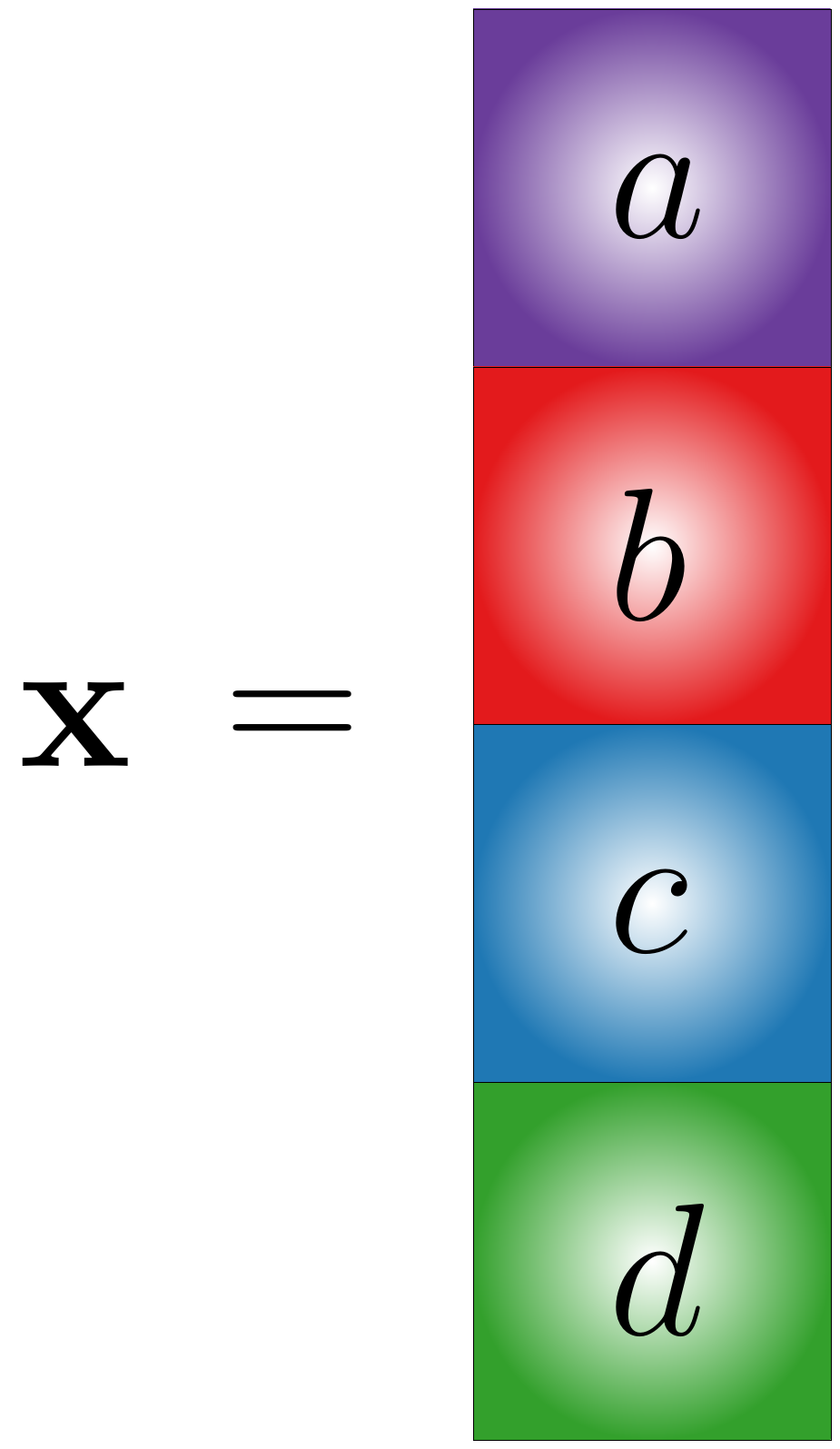}
	\end{minipage}
	\begin{minipage}{0.18\textwidth}
		\centering
		\includegraphics[width=0.85\textwidth, trim={0cm 0cm 0 0cm},clip]{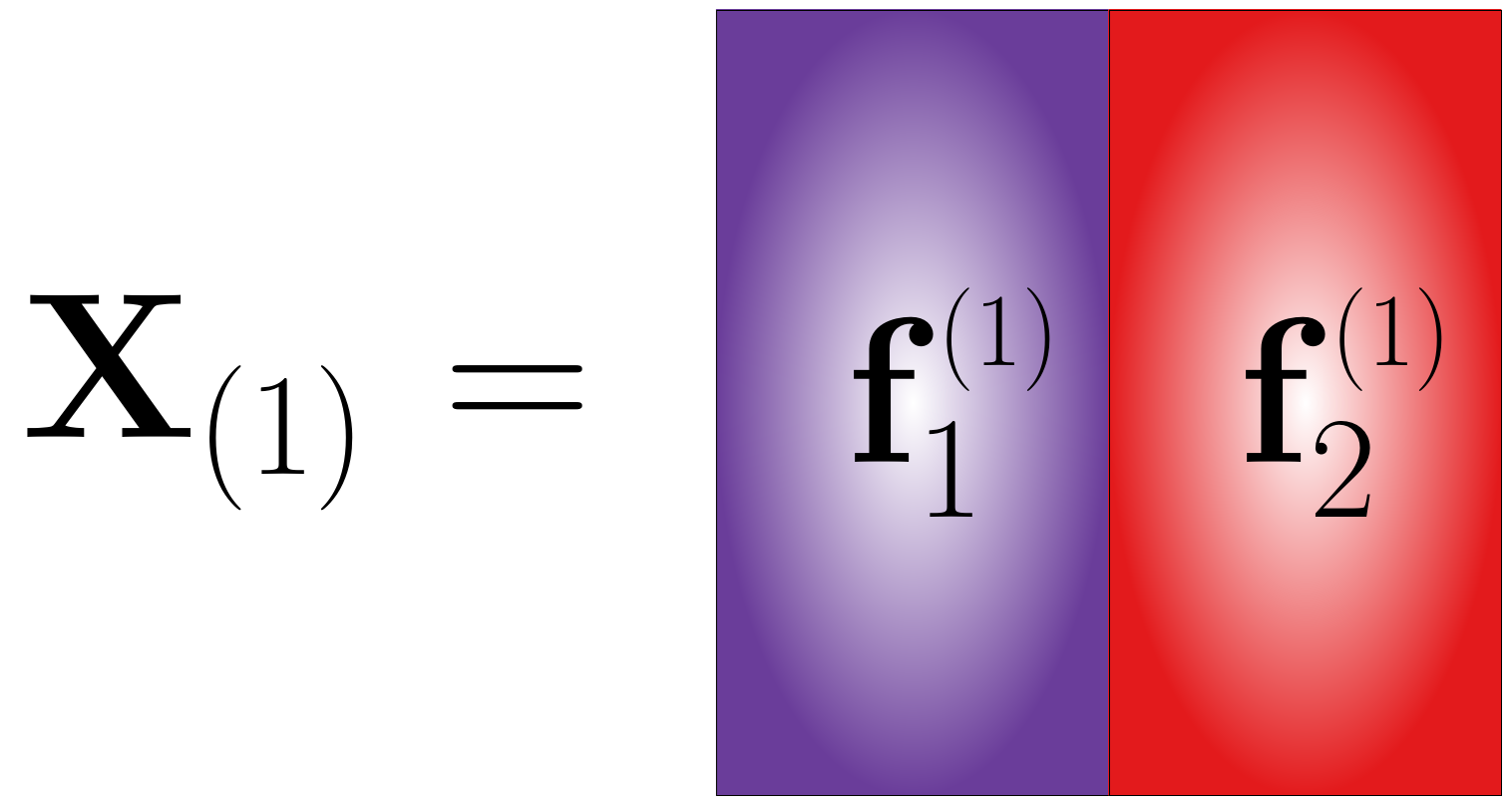}
	\end{minipage}	
	\begin{minipage}{0.18\textwidth}
		\centering
		\includegraphics[width=0.85\textwidth, trim={0cm 0cm 0 0cm},clip]{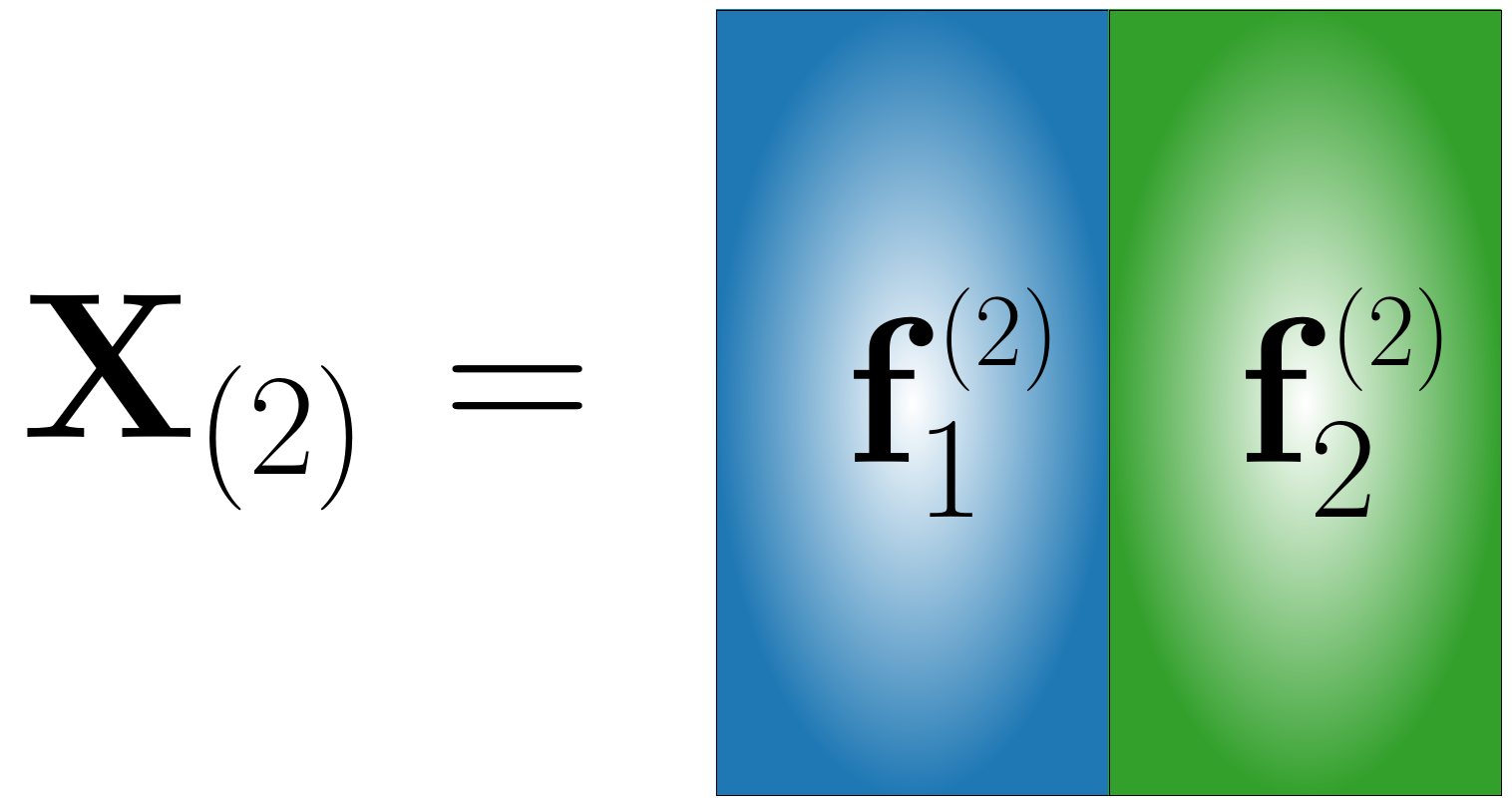}
	\end{minipage}
	\caption{\label{fig:order2TensorReps} The vector (left panel), mode-$1$ unfolding (middle panel) and mode-$2$ unfolding (right panel) representations of the order-$2$ tensor, $\X$ in Figure \ref{fig:order2Tensor}.}
\end{figure}

The mean of the vector representation of $\calX$ is given by
\begin{align}
	\expect{\x} & = \left[\begin{array}{l}
		m_{a}\\
		m_{b}\\
		m_{c}\\
		m_{d}
	\end{array}\right] 
\end{align}
with its mean parameters taking the form
\begin{align}
	\m^{(1)} & = \left[ \begin{array}{l}
		{\color{ppurple1} m_{1}^{(1)} } \\
		{\color{pred1} m_{2}^{(1)} } 
	\end{array} \right] \\
	\m^{(2)} & = \left[ \begin{array}{l}
		{\color{pblue1} m_{1}^{(2)} } \\
		{\color{pgreen1} m_{2}^{(2)} }
	\end{array} \right]
\end{align}
where, $m_{i}^{(n)} = [\m^{(n)}]_{i}$. The separability condition on the mean therefore asserts that 
\begin{align}
	\expect{\x}  =  \m^{(2)} \otimes \m^{(1)}  = \left[\begin{array}{l}
		{\color{pblue1} m_{1}^{(2)} } {\color{ppurple1} m_{1}^{(1)} }  \\
		{\color{pblue1} m_{1}^{(2)} } {\color{pred1} m_{2}^{(1)} }  \\
		{\color{pgreen1} m_{2}^{(2)} } {\color{ppurple1} m_{1}^{(1)} }  \\
		{\color{pgreen1} m_{2}^{(2)} } {\color{pred1} m_{2}^{(1)} } 
	\end{array}\right]
\end{align}
In turn, the covariance of each representation is given by
\begin{align}
\cov{\x} & = \left[ \begin{array}{llll}
\sigma_{a}^{2} & \sigma_{ab} & \sigma_{ac} & \sigma_{ad} \\
\sigma_{ab} & \sigma_{b}^{2} & \sigma_{bc} & \sigma_{bd} \\
\sigma_{ac} & \sigma_{bc} & \sigma_{c}^{2} & \sigma_{cd} \\
\sigma_{ad} & \sigma_{bd} & \sigma_{cd} & \sigma_{d}^{2}
\end{array} \right]
\end{align}
\begin{align}
	\cov{\f^{(1)}} & = \left[ \begin{array}{ll}
		{\color{ppurple1} \sigma_{11}^{(1)} } & {\color{ppurple2} \sigma_{12}^{(1)} } \\
		{\color{ppurple2} \sigma_{21}^{(1)} } & {\color{pred1} \sigma_{22}^{(1)} }
	\end{array} \right] \\
	\cov{\f^{(2)}} & = \left[ \begin{array}{ll}
		{\color{pblue1} \sigma_{11}^{(2)} } & {\color{pblue2} \sigma_{12}^{(2)} }\\
		{\color{pblue2} \sigma_{21}^{(2)} } & {\color{pgreen1} \sigma_{22}^{(2)} }
	\end{array} \right]
\end{align}
where, $\sigma_{ij}^{(n)} = [\cov{\f^{(n)}}]_{ij}$ denotes the covariance between the $i$-th and $j$-th elements of the mode-$n$ fibres, whereby $\sigma_{ii}^{(n)} \equiv \sigma_{i}^{(n)2}$. The separability condition on the covariance then asserts that 
\begin{align}
	\cov{\x} = \cov{\f^{(2)}} \otimes \cov{\f^{(1)}}
\end{align}
that is
\begin{align}
\cov{\x} \! = \! \left[ \begin{array}{llll}
{\color{pblue1} \sigma_{11}^{(2)}}{\color{ppurple1} \sigma_{11}^{(1)}} & {\color{pblue1} \sigma_{11}^{(2)}}{\color{ppurple2} \sigma_{12}^{(1)}} & {\color{pblue2} \sigma_{12}^{(2)}}{\color{ppurple1} \sigma_{11}^{(1)}}  & {\color{pblue2} \sigma_{12}^{(2)}}{\color{ppurple2} \sigma_{12}^{(1)}}\\
{\color{pblue1} \sigma_{11}^{(2)}}{\color{ppurple2} \sigma_{12}^{(1)}} & {\color{pblue1} \sigma_{11}^{(2)}}{\color{pred1} \sigma_{22}^{(1)}} & {\color{pblue2} \sigma_{12}^{(2)}}{\color{ppurple2} \sigma_{12}^{(1)}}  & {\color{pblue2} \sigma_{12}^{(2)}}{\color{pred1} \sigma_{22}^{(1)}}\\
{\color{pblue2} \sigma_{12}^{(2)}}{\color{ppurple1} \sigma_{11}^{(1)}} & {\color{pblue2} \sigma_{12}^{(2)}}{\color{ppurple2} \sigma_{12}^{(1)}} & {\color{pgreen1} \sigma_{22}^{(2)}}{\color{ppurple1} \sigma_{11}^{(1)}} & {\color{pgreen1} \sigma_{22}^{(2)}}{\color{ppurple2} \sigma_{12}^{(1)}} \\
{\color{pblue2} \sigma_{12}^{(2)}}{\color{ppurple2} \sigma_{12}^{(1)}} & {\color{pblue2} \sigma_{12}^{(2)}}{\color{pred1} \sigma_{22}^{(1)}} & {\color{pgreen1} \sigma_{22}^{(2)}}{\color{ppurple2} \sigma_{12}^{(1)}} & {\color{pgreen1} \sigma_{22}^{(2)}}{\color{pred1} \sigma_{22}^{(1)}}
\end{array} \right]
\end{align}
Figure \ref{fig:_kron_separability} further illustrates this concept, and shows that within the unstructured multivariate representation (left panel) each pairwise covariance parameter is distinct. In turn, the Kronecker separable representation (right panel) significantly reduces the number of parameters required to describe the entire covariance structure, as indicated by a distinct colour assigned to each distinct parameter. 

For instance, $\sigma_{a}^{2}=\sigma_{11}^{(2)}\sigma_{11}^{(1)}$ asserts that the variance of $a$, which resides in the first column and first row of $\X$, is equal to the product of the variance parameters associated with the first row and first column.  Similarly, $\sigma_{ac}=\sigma_{12}^{(2)}\sigma_{11}^{(1)}$ asserts that the covariance between variables $a$ and $c$ is equal to the product of the covariance parameter shared by the first and second columns, $\sigma_{12}^{(2)}$, where $a$ and $c$ respectively reside, scaled by the variance parameter associated with the first row, $\sigma_{11}^{(1)}$, where both variables reside.


\begin{figure}[H]
	\begin{minipage}{0.24\textwidth}
		\includegraphics[width=0.84\textwidth, trim={0cm 0cm 0 0cm},clip]{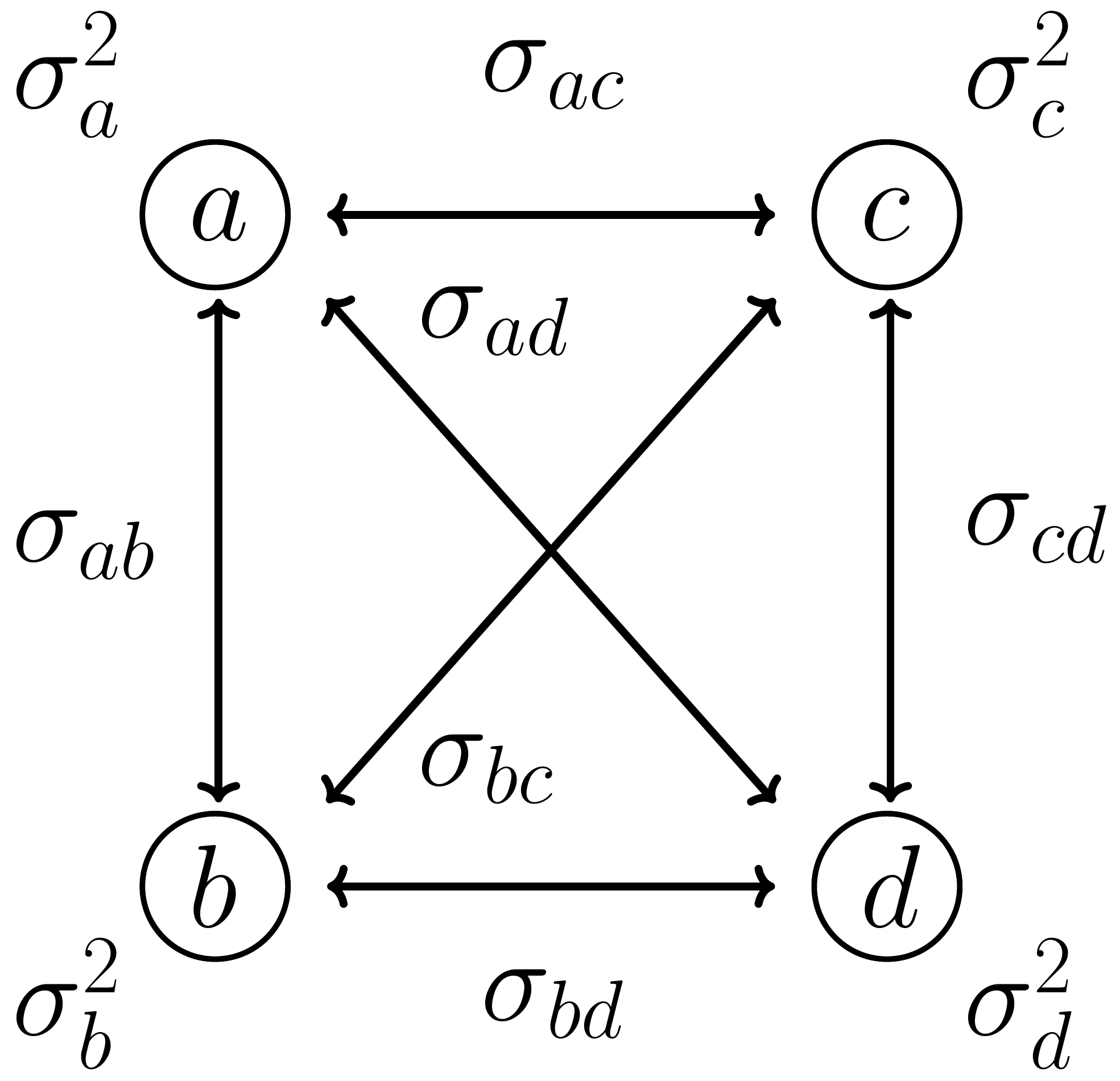}
	\end{minipage}
	\begin{minipage}{0.24\textwidth}
		\includegraphics[width=1.04\textwidth, trim={0cm 0cm 0 0cm},clip]{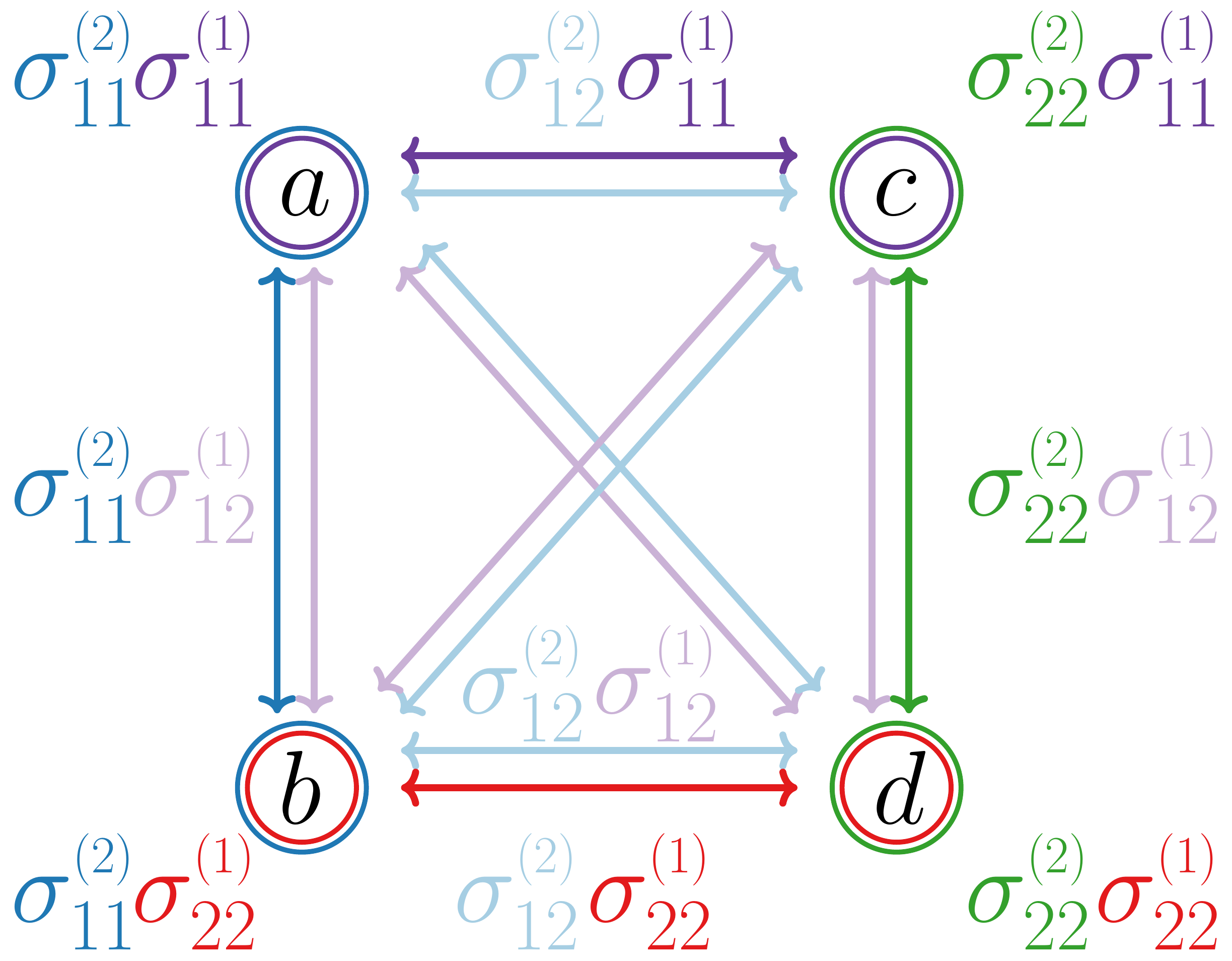}
	\end{minipage}	
	\vspace{0.2cm}
	\caption{\label{fig:_kron_separability} Illustration of the covariance of a 2D tensor, $\calX$, based on the unstructured multivariate case (left panel) and the Kronecker separable case (right panel). Each distinct parameter is highlighted in a distinct color.}
\end{figure}

\end{example}

\subsection{Parameter reduction through Kronecker separability}

Observe that the unstructured mean vector, $\m \in\domR^{K}$, contains $K$ distinct parameters, whereas its Kronecker separable counterpart, $\kronprod{n=N}{1} \m^{(n)}$, reduces to $\sum_{n=1}^{N} I_{n} < K$ parameters. 

Similarly, the unstructured covariance, $\R \in\domR^{K \times K}$, consists of $\frac{1}{2}\left( K^{2} + K \right)$ distinct parameters, whereas its Kronecker separable counterpart, $ \kronprod{n=N}{1} \R^{(n)} $, reduces to $\frac{1}{2}\sum_{n=1}^{N} \left( I_{n}^{2} + I_{n}\right) $ parameters.

\begin{remark}
	The Kronecker separability conditions in (\ref{eq:kron_mean})-(\ref{eq:kron_covariance}) provide a rigorous and parsimonious alternative to the classical unrestricted estimate of $\m$ and $\R$, which is unstable or even unavailable if the dimensions of a data tensor are large compared to the sample size. 
\end{remark}


\begin{example}
	Consider a symmetric order-$N$ tensor with all modes of the same dimension, that is, $I_{n}=I$ for all modes $n$, thereby containing $K=I^{N}$ elements in total. Then, the number of distinct parameters given by the unstructured multivariate Gaussian model and by its Kronecker separable counterpart reduce respectively to
	\begin{align}
	\eta_{\,\text{multi}}  = \frac{1}{2}\left( I^{2N} + 3 I^{N} \right), \quad \eta_{\,\text{tensor}}  = \frac{N}{2}\left(I^{2} + 3I\right)
	\end{align}
	Notice that with an increase in the order of the tensor, $N$, the ratio of the respective distinct parameters, $\frac{\eta_{\,\text{tensor}}}{\eta_{\,\text{multi}}}$, approaches zero in the limit for all $I>1$, that is
	\begin{align}
	\lim_{N  \to \infty} \,  \frac{\eta_{\,\text{tensor}}}{\eta_{\,\text{multi}}} & = 0, \quad I > 1
	\end{align}
	Figure \ref{fig:dof} illustrates the immediate reduction in parameters resulting from an increase in the order of the tensor, $N$, which demonstrates the utility of tensor-valued models for alleviating the curse of dimensionality.
	
	\begin{figure}[H]
		\vspace{-0.35cm}
		\centering
		\includegraphics[width=0.45\textwidth, trim={0.2cm 0cm 0 0cm},clip]{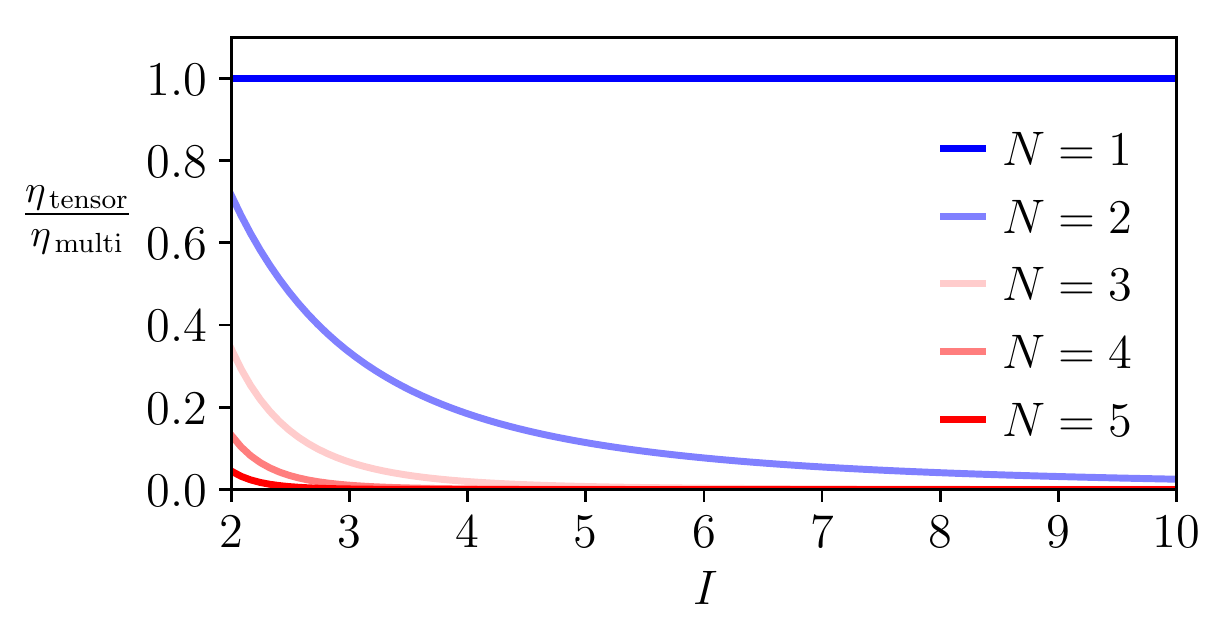}
		\vspace{-0.35cm}
		\caption{\label{fig:dof} The ratio of distinct parameters between the structured tensor model and the classical unstructured model, $\frac{\eta_{\,\text{tensor}}}{\eta_{\,\text{multi}}}$, for a varying mode dimensionality, $I$, and tensor order, $N$.}
	\end{figure}
	
\end{example}


\section{Tensor-Valued Gaussian Distribution}

\label{section:gaussian_dist}

The Gaussian distribution has become a ubiquitous statistical model for describing the mean and covariance structure of random variables observed across a broad variety of disciplines. The material in Section \ref{section:kronecker} will now serve as background to derive the tensor-valued extension of the Gaussian distribution, which can be used to describe the mean and covariance structure of multidimensional signals.

\subsection{Related work}

\label{tensorvalued_pdf_def}

It has already been established that the tensor-valued random variable, $\calX \in \domR^{I_{1} \times \cdots \times I_{N}}$,  exhibits a Gaussian distribution, defined by the tensor mean, $\calM \in \domR^{I_{1} \times \cdots \times I_{N}}$, and mode-$n$ covariance, $\R^{(n)} \in \domR^{I_{n} \times I_{n}}$, if and only if its vector representation, $\x \in \domR^{K}$, is distributed according to \cite{Hoff2011}
\begin{align}
\x \sim \mathcal{N}\left( \m, \kronprod{n=N}{1}\R^{(n)}\right) \label{eq:pdf_vector}
\end{align}
where $\m = \vect{\calM}$. With the condition in (\ref{eq:pdf_vector}), it is straightforward to show that the probability density function of $\calX$ is given by
\begin{align}
\label{eq:tensorvalued_pdf}
p(\calX) = \frac{\exp\left[ -\frac{1}{2} \left(\x - \m\right)^{\Trans} \! \left(\kronprod{n=N}{1}  \R^{(n)-1}\right)\left(\x - \m\right) \right]}{\left(2\pi\right)^{\frac{K}{2}} \det^{\frac{1}{2}}\left( \kronprod{n=N}{1}  \R^{(n)} \right) }
\end{align}
The maximum likelihood (ML) estimator of the order-$2$ tensor (matrix) Gaussian parameters was first introduced in \cite{Dutilleul1999}, and has been recently extended to the order-$N$ tensor-valued case in \cite{Hoff2011}. The estimator is obtained by maximizing the associated log-likelihood of observing $T$ samples, denoted by $\calX(t) \in \domR^{I_{1}\times \cdots \times I_{N}}$, under the distribution in (\ref{eq:tensorvalued_pdf}), that is
\begin{align}
\mathcal{L} & = \sum_{t=0}^{T-1} \ln p\left( \calX(t) \Bigl\lvert \calM,  \{ \R^{(n)} \}_{n=1}^{N} \right) \label{eq:loglikelihood_MLE}
\end{align}
The stationary points necessary to determine the ML estimator are obtained upon setting the derivatives of $\mathcal{L}$ with respect to each parameter to zero.

The expression for the stationary point with respect to the tensor mean, $\calM$, can be rearranged to yield the \textit{sample mean}
\begin{equation}
\label{eq_MLE_Solution_M}
\calM = \frac{1}{T} \sum_{t=0}^{T-1}\calX(t)
\end{equation}
In turn, the stationary point obtained with respect to each mode-$n$ covariance parameter, $\R^{(n)}$, does not lead to an estimator in closed-form. An iterative procedure based on the block coordinate descent algorithm \cite{Dutilleul1999,Tseng2001}, often referred to as the \textit{flip-flop} algorithm, is therefore employed to approach the ML estimate of $\R^{(n)}$, given by
\begin{equation}
\label{eq_MLE_Solution_Sigma}
\R^{(n)} = \frac{I_{n}}{T K} \sum_{t=0}^{T-1} \S_{(n)}(t)  \left( \kronprod{\subalign{i&=N\\i&\neq n}}{1} \R^{(i)-1} \right) \S_{(n)}^{\Trans}(t)
\end{equation}
where $\S_{(n)}$ is the mode-$n$ unfolding of the centred tensor-valued random variable, $\calS = \calX - \calM$.


It is important to notice that there exist several issues with the existing formulation of the tensor Gaussian distribution:\vspace{0.2cm}\\
\indent (i) The global optimality of the iterative scheme in (\ref{eq_MLE_Solution_Sigma}) is not guaranteed with respect to the composition $ \kronprod{n=N}{1}   \R^{(n)} $. It is well known that the general class of alternating and cascaded schemes proposed for tensor-valued estimation problems, such as the alternating least squares \cite{Uschmajew2012,Uschmajew2013} and tensor least mean square \cite{Rupp2015}, exhibit only local optimality. Global optimality can only be attained by evaluating the stationary point of $\mathcal{L}$ with respect to $ \kronprod{n=N}{1}  \R^{(n)} $, as opposed to parallel evaluations of $\R^{(n)}$ for each $n=1,...,N$;\\
\indent (ii) The estimates, $\R^{(n)}$, are \textit{non-identifiable}, whereby different values of the parameters may generate equivalent probability distributions. Since the identifiability condition is absolutely necessary for the ML estimator to be statistically consistent \cite{Newey1994}, we can immediately conclude that the estimates obtained from the \textit{flip-flop} algorithm are statistically inconsistent with respect to $\R^{(n)}$ for all $n$. With this formulation, only the composition, $\kronprod{n=N}{1} \R^{(n)}$, can be uniquely identified \cite{Dutilleul1999}. To see this, notice that an iterative solution can only estimate $\R^{(n)}$ up to a multiplicative constant, e.g. by defining $\boldTheta^{(n)} = a\R^{(n)}$ and $\boldTheta^{(m)} = \frac{1}{a}\R^{(m)}$, for any $a \neq 0$, we obtain the same composition, since $\boldTheta^{(n)} \otimes \boldTheta^{(m)} = \R^{(n)} \otimes \R^{(m)}$ and so both estimates yield the same Kronecker product. \\
\indent (iii) The tensor mean, $\calM$, does not exhibit the separability property in (\ref{eq:CPD_mean}), which is required for a rigorous definition of a tensor-valued variable which represent the discrete-space counterpart of the continuous separable Gaussian field. 


\subsection{The proposed statistically identifiable formulation}

\label{MLE_closed_form}

To resolve the aforementioned issues, we propose a new representation-invariant formulation of the tensor-valued probability distribution, based on the following rationale:\vspace{0.2cm}\\
\indent (i) The variance of the random tensor, $\calX$, is \textit{invariant} to the data representation, that is, for both the original data tensor and any of its unfoldings we have
\begin{align}
	\var{\calX} = \var{\x} = \var{\X_{(n)}}, \quad \forall n
\end{align}
since the representations contain the same data entries. Consequently, the mode-$n$ covariance parameters, $\R^{(n)}$, should exhibit the same Frobenius norm 
for all modes $n$. This contrasts the distribution formulation employed in \cite{Dutilleul1999,Hoff2011} which assumes that the variances at different modes can differ. In other words, in the existing formulation, the parameters $\R^{(n)}$ are unconstrained. 


A first step to resolving the non-identifiability issue is to dissociate the magnitude of the variance from the covariances parameters, $\{ \R^{(n)} \}_{n=1}^{N}$. This is achieved by introducing the mode-$n$ matrices, $\boldTheta^{(n)} \in \domR^{I_{n} \times I_{n}}$, and the variance parameter $\sigma^{2} \in \domR$, to yield
\begin{align}
 \kronprod{n=N}{1} \R^{(n)}  = \sigma^{2}\left( \kronprod{n=N}{1} \boldTheta^{(n)} \right) \label{eq:unit_trace_condition}
\end{align}
where
\begin{align}
	\sigma^{2} = \var{\calX} = \tr{\kronprod{n=N}{1} \R^{(n)}} \label{eq:variance_trace}
\end{align}
A physically meaningful condition is to enforce the trace of the so introduced parameters, $\{ \boldTheta^{(n)} \}_{n=1}^{N}$, to have unit value, that is, $\tr{\boldTheta^{(n)}}=1$, $\forall n$. 

Intuitively, $\boldTheta^{(n)}$ can be thought of as the \textit{covariance density} at the $n$-th mode, whereby its $(i,j)$-th element describes the percentage of the total variance, $\sigma^{2}$, assigned to the covariance between the $i$-th and $j$-th elements of the mode-$n$ fiber, as shown in the sequel. Moreover, the unit-trace condition satisfies the definition in (\ref{eq:variance_trace}). \\
\indent (ii) A rigorous definition of the tensor Gaussian variable, based on the statistical properties of separable Gaussian fields, requires the mean to be separable, as in (\ref{eq:CPD_mean}). Based on the arguments in point (i), we allow the mean to exhibit the following separable structure
\begin{align}
\m = \alpha\left( \kronprod{n=N}{1}  \boldmu^{(n)} \right)
\end{align}
where $\alpha \in \domR$ is a positive scaling factor, and the vectors $\boldmu^{(n)} \in \domR^{I_{n}}$ are constrained to be unit vectors, i.e. $\|\boldmu^{(n)}\|=1$ for all $n$. In the tensor representation, this becomes
\begin{align}
\calM = \alpha\left(\circprod{n=1}{N}  \boldmu^{(n)}\right) \label{eq:CPD_mean_identifiable}
\end{align}
In the following, we show that the distribution formulation is statistically identifiable if and only if we employ the proposed model in (\ref{eq:CPD_mean_identifiable}), as opposed to the existing model in (\ref{eq:CPD_mean}) where the vectors, $\m^{(n)}$, are unconstrained.

With the proposed distribution formulation, the Kronecker separability properties of the mean and covariance reduce to
\begin{align}
\expect{\x} & = \alpha\left( \kronprod{n=N}{1}  \boldmu^{(n)} \right) \\
	\cov{\x} & = \sigma^{2}\left( \kronprod{n=N}{1} \boldTheta^{(n)} \right) \\
	\cov{\X_{(n)}} & = \sigma^{2}\boldTheta^{(n)} \label{eq:mode_n_covariance_identif} 
\end{align}
with the conditions $\tr{\boldTheta^{(n)}}=1$ and $\|\boldmu^{(n)}\|=1$ for all $n$. Therefore, the probability density of the vector representation is described according to
\begin{align}
 \x \sim \Normal{ \alpha\left( \kronprod{n=N}{1}  \boldmu^{(n)} \right), \sigma^{2}\left( \kronprod{n=N}{1} \boldTheta^{(n)} \right) } \label{eq:tensorvalued_pdf_unit_norm}
\end{align}

\subsection{Drawing samples from the distribution}
\label{section_drawing_from_distribution}

To draw a tensor-valued sample, $\calX \in \domR^{I_{1} \times \cdots \times I_{N}}$, from the distribution in (\ref{eq:tensorvalued_pdf_unit_norm}), we must first generate a sample from the Gaussian distribution, $\w \sim \mathcal{N}\left( \0_{K\times 1}, \I_{K} \right)$, to obtain
\begin{equation}
\x = \alpha\left( \kronprod{n=N}{1}  \boldmu^{(n)} \right) + \sigma\left( \kronprod{n=N}{1} \boldTheta^{(n)\frac{1}{2}} \right) \w
\end{equation}
where $(\cdot)^{\frac{1}{2}}$ denotes the Cholesky factorization. The sample $\x$ is then reshaped into the tensor $\calX \in \domR^{I_{1} \times \cdots \times I_{N}}$.


\begin{remark}
	The authors in \cite{Singull2015} have also considered a tensor-valued Gaussian model with a structured mean of the form
	\begin{align}
		\calM = \calA \modeprod{n=1}{N} \B_{n}
	\end{align}
	However, the maximum likelihood estimates are obtained through an iterative algorithm. For rigour, our proposed model assumes that $\calM$ is a rank-$1$ tensor, which is motivated by the physical properties of Gaussian random fields, and the constituent parameters, $\alpha$ and $\{\boldmu^{(n)}\}_{n=1}^{N}$, can be obtained analytically as shown in the sequel. 
\end{remark}

\begin{remark}
	Independently, the authors in \cite{Hoff2015} proposed similar formulations for the covariance structure within the context of tensor-valued empirical Bayesian inference. The extension of \textit{Stein's loss function} was considered to derive the biased estimator of $\{ \R^{(n)} \}_{n=1}^{N}$, and the parametrization $\{  \sigma^{2}, \{ \boldTheta^{(n)} \}_{n=1}^{N} \}$, with $\det\left(\boldTheta^{(n)}\right)=1$, $\forall n$, was employed. The solution method is analogous to the \textit{flip-flip} algorithm. Owing to the non-convexity of the space of unit-determinant matrices, the authors in \cite{Hoff2015} propose a stochastic approximative solution which employs the space of unit-trace symmetric positive definite matrices, which is convex. This finding, although derived from a different perspective and employed in a different context, complements and supports the results demonstrated herein. Of particular relevance to this work is the suggestion in \cite{Hoff2015} that the unit-trace condition serves as the basis of a general solution method for tensor-valued problems.
\end{remark}


\vspace{-0.5cm}

\subsection{Maximum likelihood estimator}

To derive the maximum likelihood (ML) estimates of the proposed parameters, consider the log-likelihood of $T$ samples being distributed according to (\ref{eq:tensorvalued_pdf_unit_norm}), that is
\begin{align}
\mathcal{L} & = \sum_{t=0}^{T-1} \ln p\left( \x(t) \Bigl\lvert \alpha,\{\boldmu^{(n)}\}_{n=1}^{N}, \sigma^{2}, \{\boldTheta^{(n)}\}_{n=1}^{N} \right) \notag\\
& = -\frac{TK}{2}\ln\left( 2\pi \sigma^{2} \right) - \sum_{n=1}^{N} \frac{TK}{2I_{n}} \ln\left(\det\left( \boldTheta^{(n)} \right)\right) \notag\\
& - \frac{1}{2\sigma^{2}}\sum_{t=0}^{T-1} \left(\x(t)-\m\right)^{\Trans}\left( \kronprod{n=1}{N} \boldTheta^{(n)-1}\right)\left(\x(t)-\m\right) \label{eq:likelihood_ambiguity}
\end{align}
where $\m = \alpha\left( \kronprod{n=N}{1}  \boldmu^{(n)} \right)$. 

The stationary point of $\mathcal{L}$ with respect to the mean parameters, $\m = \alpha \left( \kronprod{n=N}{1}  \boldmu^{(n)} \right)$, yields the sample mean
\begin{align}
\alpha\left( \kronprod{n=N}{1}  \boldmu^{(n)} \right) = \frac{1}{T} \sum_{t=0}^{T-1} \x(t)
\end{align}
The optimal estimates of the constituent parameters, in the minimum mean square error sense, are given by the best rank-$1$ tensor approximation of the sample mean tensor, that is
\begin{align}
	\frac{1}{T} \sum_{t=0}^{T-1} \calX(t) = \alpha\left( \circprod{n=1}{N}  \boldmu^{(n)} \right)
\end{align}
which is, in essence, a rank-$1$ canonical polyadic decomposition (CPD). It is important to highlight that existing techniques for evaluating the best rank-$1$ CPD from a generic rank-$R$ tensor cannot guarantee global optimality of the solution. However, for rank-$1$ tensors, the best rank-$1$ CPD can be evaluated explicitly using the multilinear singular value decomposition \cite{DeLathauwer2000,VanLoan2016}, whereby $\alpha$ is the CPD core value and $\boldmu^{(n)}$ is the mode-$n$ CPD factor. Notice that in this way the unit-vector property of $\boldmu^{(n)}$ is also satisfied.


Upon introducing the centred tensor variable
\begin{align}
\calS(t) = \calX(t) - \alpha\left( \circprod{n=1}{N}  \boldmu^{(n)} \right)
\end{align}
the stationary point of $\mathcal{L}$ with respect to $\left( \kronprod{n=N}{1} \boldTheta^{(n)} \right)$ yields the globally optimum estimator
\begin{align}
\kronprod{n=N}{1} \boldTheta^{(n)} = \frac{1}{T \sigma^{2}} \sum_{t=0}^{T-1} \s(t)\s^{\Trans}(t) \label{eq:MLE_composition_covariance}
\end{align}
with $\s(t) = \vect{\calS(t)}$. By imposing the unit-trace condition, $\tr{\boldTheta^{(n)}} = 1$, $\forall n$, and using properties of the Kronecker product trace \cite{Abadir2005}, we find that $\tr{\kronprod{n=N}{1} \boldTheta^{(n)}}=1$. Therefore, by evaluating the trace of the LHS and RHS of (\ref{eq:MLE_composition_covariance}), we can directly determine the ML estimator of $\sigma^{2}$, which is of the form
\begin{align}
\sigma^{2} & = \frac{1}{T}\sum_{t=0}^{T-1}\s^{\Trans}(t)\s(t) \label{eq:MLE_sigma2}
\end{align}
Next, upon rearranging the condition in (\ref{eq:mode_n_covariance_identif}), we obtain
\begin{align}
\boldTheta^{(n)} & = \frac{1}{T\sigma^{2}} \sum_{t=0}^{T-1} \S_{(n)}(t)\S_{(n)}^{\Trans}(t) \label{eq:MLE_overline_boldsigma}
\end{align}
%



\begin{remark}
	With regard to the Kronecker separable first- and second-order moments in (\ref{eq:mean_separation})-(\ref{eq:covariance_separation}), we can show that with the proposed formulation we obtain
	\begin{align}
	\expect{[\calX]_{i_{1}...i_{N}}} & = \alpha \prod_{n=1}^{N} [\boldmu^{(n)}]_{i_{n}} \\
	\cov{[\calX]_{i_{1}...i_{N}},[\calX]_{j_{1}...j_{N}}} & = \sigma^{2} \prod_{n=1}^{N} [\boldTheta^{(n)}]_{i_{n}j_{n}}
	\end{align}
\end{remark}



\vspace{-0.5cm}

\subsection{Conditions for identifiability, uniqueness and consistency}

A \textit{consistent} estimator is one that converges in probability to the true value as the sample size, $T$, approaches infinity, for all possible values. To establish the consistency of an estimator, the following conditions are sufficient \cite{Newey1994}: (i) identifiability of the model; (ii) compactness of the parameter space; (iii) continuity of the log-likelihood function; (iv) dominance of the likelihood function. Under the assumptions that the observations, $\calX(t)$, are i.i.d. and that the law of large numbers applies, the conditions for compactness, continuity and dominance hold, and are only mild conditions. 

In turn, the condition for the identifiability is absolutely necessary for the ML estimator to be consistent, as the identifiability condition asserts that the log-likelihood has a unique global maximum \cite{Huzurbazar1947}. The importance for consistency of an ML estimator, as it reaches the global maximum, has practical implications. Namely, iterative maximization procedures may typically converge only to a local maximum, but the consistency results only apply to the global maximum solution. Therefore, under the mild conditions stated above, if the proposed estimator is unique, it is also consistent.

To this end, we show that the ML estimator for the mode-$n$ covariance density matrix, $\boldTheta^{(n)}$, is unique if and only if the number of i.i.d. random samples, $T$, drawn from (\ref{eq:tensorvalued_pdf_unit_norm}) satisfies
\begin{equation}
T > \max\left( \frac{I_{1}^{2}}{K}, \cdots, \frac{I_{N}^{2}}{K} \right) \label{eq:existence}
\end{equation}
which is consistent with Theorem 3.1 from \cite{Dutilleul2013}. Taking the rank of the LHS and RHS of (\ref{eq:MLE_overline_boldsigma}), we obtain
\begin{align}
\rank{ \boldTheta^{(n)} } \! = \! \rank{ \! \frac{1}{T\sigma^{2}} \! \sum_{t=0}^{T-1} \S_{(n)}(t) \S_{(n)}^{\Trans}(t) \! } \! = \! \left(T - 1\right) \! \frac{K}{I_{n}}
\end{align}
Since $\boldTheta^{(n)}$ is positive definite, we must have that $\rank{\boldTheta^{(n)}}=I_{n}$. It then follows that the condition  $ T > \frac{I_{n}^{2}}{K} $, $\forall n$, is necessary for the estimator to be consistent. The condition for consistency therefore reduces to (\ref{eq:existence}).

\pagebreak

\label{section:efficiency}

\begin{figure}[H]
	\vspace{-0.2cm}
	\centering
	\includegraphics[width=0.45\textwidth, trim={0cm 0cm 0 0cm},clip]{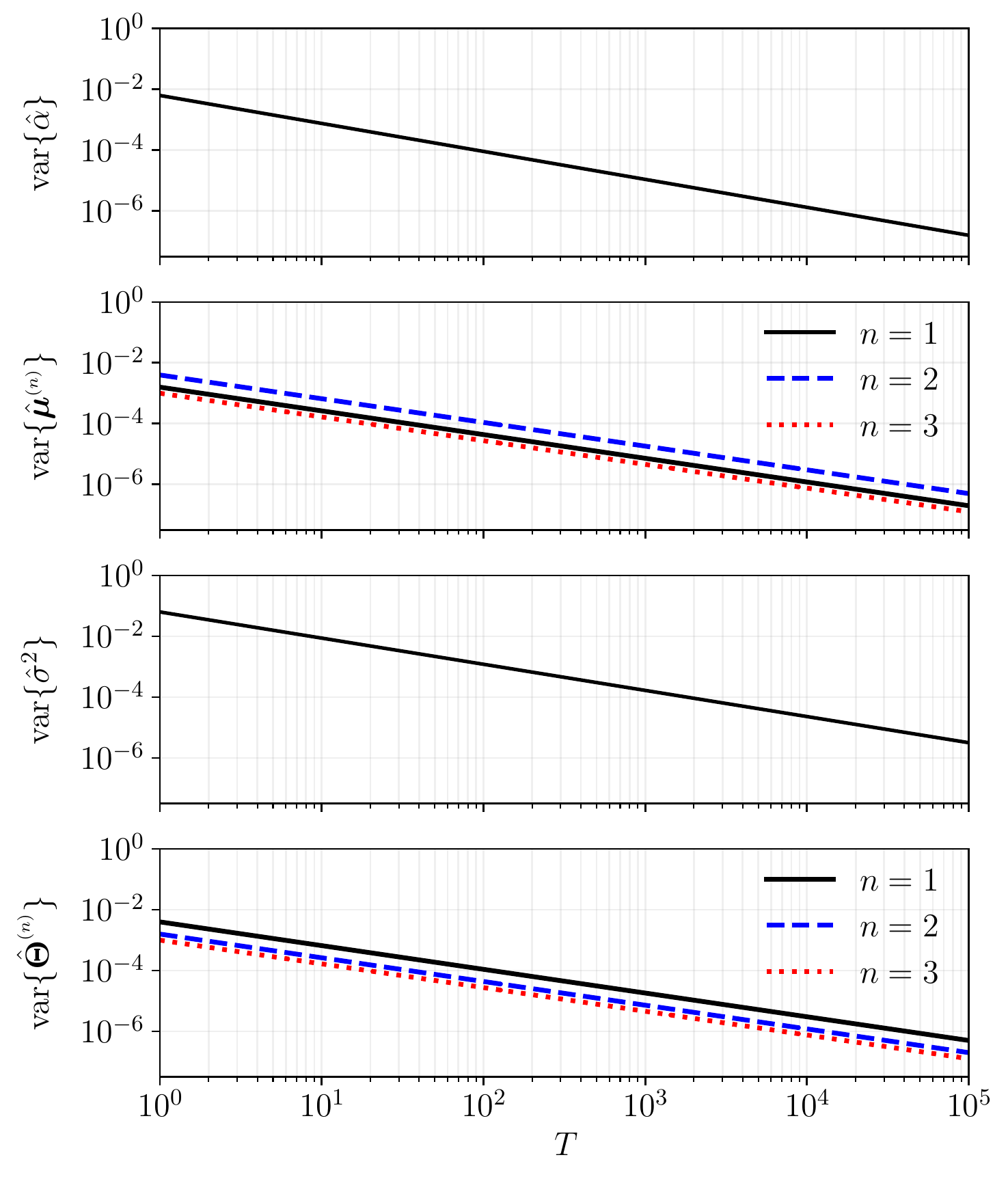}
	\vspace{-0.3cm}
	\caption{\label{fig:crlb} The variance of empirical parameter estimation as a function of the sample size, $T$, for the proposed ML estimator, computed over 1000 independent realisations.}
\end{figure}

\begin{example} \label{ex:statistical_consistency}
The statistical consistency of the proposed ML estimators were verified empirically. The sampling procedure derived in Section \ref{section_drawing_from_distribution} was employed to generate $T$ order-$3$ tensor-valued samples, $\calX(t) \in \domR^{2 \times 3 \times 4}$, from the proposed tensor-valued Gaussian distribution, parametrized as follows
\begin{align}
\x \sim \Normal{ \alpha\left( \kronprod{n=3}{1}  \boldmu^{(n)} \right), \sigma^{2}\left( \kronprod{n=3}{1} \boldTheta^{(n)} \right) }
\end{align}
The procedure was implemented using our own Python Higher-Order Tensor ToolBOX (HOTTBOX) \cite{Kisil2018}. The parameters of the distribution were chosen arbitrarily so as to satisfy the unit-vector property of $\boldmu^{(n)}$ and the unit-trace property of $\boldTheta^{(n)}$ for all $n$. The estimation variance was defined as
\begin{align}
	\var{\hat{\boldtheta}} = \expect{\|\hat{\boldtheta}-\boldtheta\|^{2}}
\end{align}
where $\hat{\boldtheta}$ is the estimate of the true parameter $\boldtheta$, and was evaluated for each parameter using sample lengths, $T$, in the range $[1,10^{5}]$. The results were averaged over 1000 independent Monte Carlo simulations and are displayed in Figure \ref{fig:crlb}. The asymptotic convergence exhibited by the proposed ML estimators to their true value with increasing sample length, $T$, therefore verifies their \textit{statistical consistency}. 
\end{example}

\vspace{-0.3cm}

\begin{remark}
	Notice that for $\calX(t) \in \domR^{2 \times 3 \times 4}$, the ML estimates can even be computed with $T=1$ samples since from (\ref{eq:existence}) we have that $T=1>\max\left( \frac{4}{24}, \frac{9}{24}, \frac{16}{24} \right)$.
\end{remark}

\vspace{-0.2cm}

\begin{remark}
	The maximum likelihood estimators of the variance and covariance parameters of the Gaussian distribution are biased, however, a multiplication of the estimates by the factor $\frac{T}{T-1}$, referred to as \textit{Bessel's correction}, yields unbiased estimators. We therefore employed Bessel's correction in Example \ref{ex:statistical_consistency} to ensure the ML estimates were unbiased.
\end{remark}

\vspace{-0.2cm}

\pagebreak

\section{Multivariate Tensor-Valued Gaussian Distribution}

\label{sec:multivariate_tensor}

After having established statistical properties of tensor-valued Gaussian random variables, a natural next step is to define the \textit{joint} probability density function of multiple univariate tensor-valued Gaussian random variables, $\calX_{1},\calX_{2},...,\calX_{M}  \in \domR^{I_{1} \times \cdots \times I_{N}}$, where the marginal distribution with respect to each variable, $\calX_{i}$, takes the form as in (\ref{eq:tensorvalued_pdf}). To this end, we first show that the density function of the joint distribution is not an obvious extension of the univariate tensor-valued case, and proceed to show that the multivariate covariance matrix does not exhibit the Kronecker separability property, but is instead \textit{Khatri-Rao} separable.


For clarity, we begin by extending the non-identifiable version of the distribution formulation in (\ref{eq:pdf_vector}) to the multivariate tensor case, and then derive its identifiable counterpart, whereby different values of the parameters strictly generate different probability distributions.

Consider an order-$N$ tensor-valued Gaussian random variables, $\calX_{i} \in \domR^{I_{1} \times \cdots \times I_{N}}$, for $i=1,...,M$, with each variate described according to the marginal distribution (in the vector form)
\begin{align}
	\x_{i} \sim \mathcal{N}\left(  \kronprod{n=N}{1} \m_{i}^{(n)} , \kronprod{n=N}{1}\R_{ii}^{(n)} \right) \label{eq:pdf_marginal_vector}
\end{align}
The tensor-valued variables, $\calX_{i}$, can be stacked together to form a multivariate tensor-valued random variable, $\calZ \in \domR^{I_{1} \times \cdots \times I_{N} \times M}$, of order $(N+1)$. In the vector representation, this amounts to forming the vector, $\z \in \domR^{MK}$, as follows
\begin{align}
	\z = \left[\begin{array}{c}
		\x_{1} \\
		\x_{2}\\
		\vdots\\
		\x_{M}
	\end{array}\right]
\end{align}
that is, $\z = \vect{\calZ}$, where each vector is given in (\ref{eq:pdf_marginal_vector}). 

Since each constituent random variable, $\x_{i}$, is Gaussian distributed, then so too is $\z$, that is, $ \z \sim \mathcal{N}\left( \m_{z}, \R_{zz} \right)$. As such, the mean vector, $\m_{z} \in \domR^{MK}$, exhibits the following block matrix structure
\begin{align}
\m_{z} & = \left[ 
\arraycolsep=1pt
\def\arraystretch{1.3}
\begin{array}{c}
\kronprod{n=N}{1}\m_{1}^{(n)} \\
\kronprod{n=N}{1}\m_{2}^{(n)}  \\
\vdots  \\
\kronprod{n=N}{1}\m_{M}^{(n)} 
\end{array} \right] \label{eq:block_matrix_kroneckers_mean}
\end{align}
However, the structure of $\R_{zz} \in \domR^{MK \times MK}$ is less obvious. From the separability property of Gaussian random fields in (\ref{eq:separability_GRF}), it follows that $\cov{\x_{i},\x_{j}} = \kronprod{n=N}{1} \R_{ij}$ is Kronecker separable for all $i,j$, which leads to the formulation in (\ref{eq:block_matrix_kroneckers})
\begin{align}
\R_{zz} & = \left[ 
\arraycolsep=2pt
\def\arraystretch{1.3}
\begin{array}{cccc}
\kronprod{n=N}{1}\R_{11}^{(n)} & \kronprod{n=N}{1}\R_{12}^{(n)} & \cdots & \kronprod{n=N}{1}\R_{1M}^{(n)} \\
\kronprod{n=N}{1}\R_{21}^{(n)} & \kronprod{n=N}{1}\R_{22}^{(n)} & \cdots & \kronprod{n=N}{1}\R_{2M}^{(n)} \\
\vdots & \vdots & \ddots & \vdots \\
\kronprod{n=N}{1}\R_{M1}^{(n)} & \kronprod{n=N}{1}\R_{M2}^{(n)} & \cdots & \kronprod{n=N}{1}\R_{MM}^{(n)} \\
\end{array} \right] \label{eq:block_matrix_kroneckers}
\end{align}

\pagebreak


\noindent The multivariate tensor mean and covariance parameters in (\ref{eq:block_matrix_kroneckers_mean})-(\ref{eq:block_matrix_kroneckers}) can be equivalently expressed through the following Khatri-Rao products
\begin{align}
	\m_{z}  = \khatriprod{n=N}{1} \m_{z}^{(n)}, \quad \R_{zz}  = \khatriprod{n=N}{1} \R_{zz}^{(n)}
\end{align}
where $\m_{z}^{(n)} \in \domR^{MI_{n}}$ and $\R_{zz}^{(n)} \in \domR^{MI_{n} \times MI_{n}}$ are respectively the multivariate tensor mode-$n$ mean and covariance parameters, defined as
\begin{align}
	\m_{z}^{(n)} \! = \! \left[
	\def\arraystretch{1.2}
	\begin{array}{c}
	\m_{1}^{(n)} \\
	\m_{2}^{(n)} \\
	\vdots \\
	\m_{M}^{(n)} \\
	\end{array} \right], \quad \!\!\! 	\R_{zz}^{(n)} \! = \! \left[
		\def\arraystretch{1.2}
		\begin{array}{cccc}
		\R_{11}^{(n)} & \R_{12}^{(n)} & \cdots & \R_{1M}^{(n)} \\
		\R_{21}^{(n)} & \R_{22}^{(n)} & \cdots & \R_{2M}^{(n)} \\
		\vdots & \vdots & \ddots & \vdots \\
		\R_{M1}^{(n)} & \R_{M2}^{(n)} & \cdots & \R_{MM}^{(n)} \\
	\end{array} \right]
\end{align}
Therefore, the vector representation, $\z \in \domR^{MK}$, is distributed according to
\begin{align}
	\z \sim \mathcal{N}\left( \khatriprod{n=N}{1}\m_{z}^{(n)}, \khatriprod{n=N}{1}\R_{zz}^{(n)} \right)
\end{align}
which asserts that the joint probability density function of the tensor-valued random variables $\calX_{1},...,\calX_{M}$ is given by
\begin{align}
	\label{eq:multi_tensorvalued_pdf}
	p(\calZ) = \frac{\exp\left[ -\frac{1}{2} \left(\z - \m_{z}\right)^{\Trans}\left( \khatriprod{n=N}{1} \R_{zz}^{(n)} \right)^{-1}\left(\z - \m_{z}\right) \right] }{\left(2\pi\right)^{\frac{MK}{2}} \det^{\frac{1}{2}}\left( \khatriprod{n=N}{1} \R_{zz}^{(n)} \right) }
\end{align}
where $\m_{z} = \left(\khatriprod{n=N}{1}\m_{z}^{(n)}\right)$.

With the above derived properties of the multivariate tensor-valued distribution we obtain the following result. 

\vspace{-0.1cm}

\begin{proposition}
	The multivariate tensor-valued random variable, $\z \in \domR^{MK}$, is said to exhibit a \textit{Khatri-Rao separable} statistics if and only if the following properties hold
	\begin{align}
		\expect{\z} & = \khatriprod{n=N}{1} \m_{z}^{(n)}  \label{eq:khatri_mean} \\
		\cov{\z} & =  \khatriprod{n=N}{1} \R_{zz}^{(n)}  \label{eq:khatri_covariance} \\
		\cov{\Z_{(n)}} & = \Biggl(\hadaprod{\subalign{i&=1\\i&\neq n}}{N}  \ptr{ \R_{zz}^{(i)} }\Biggr) \oast \R_{zz}^{(n)} \label{eq:khatri_mode_n_covariance}
	\end{align}
	where $\odot$ denotes the \textit{Hadamard} (element-wise) product. With reference to the constituent tensor-valued random variables, the separability properties in (\ref{eq:khatri_covariance})-(\ref{eq:khatri_mode_n_covariance}) reduce to the following
	\begin{align}
		\cov{\x_{i},\x_{j}} & =  \kronprod{n=N}{1} \R_{ij}^{(n)} \label{eq:multi_cov_vect} \\
		\cov{\X_{i(n)},\X_{j(n)}} & = \Biggl( \prod_{\subalign{k&=1\\k&\neq n}}^{N}  \tr{ \R_{ij}^{(k)} } \Biggr) \R_{ij}^{(n)} \label{eq:multi_cov_mode-n}
	\end{align}
\end{proposition}

\vspace{-0.2cm}

\begin{remark}
	As shown in Section \ref{tensorvalued_pdf_def} for the univariate tensor-valued case, the proposed multivariate tensor-valued Gaussian distribution described by the second-order statistics in (\ref{eq:multi_cov_vect})-(\ref{eq:multi_cov_mode-n}) is non-identifiable, whereby different values of the parameters may generate equivalent probability distributions, and therefore the parameters cannot be estimated analytically.
\end{remark}

\pagebreak

\subsection{The statistically identifiable formulation}

Following the establishment of the analytical framework for tensor-valued probability distribution in Section \ref{section:gaussian_dist}, we now introduce the identifiable counterpart of the proposed multivariate tensor-valued distribution in (\ref{eq:multi_tensorvalued_pdf}), whereby different values of the parameters strictly generate different probability distributions, based on the following rationale. 

The \textit{expected inner product} between pairwise tensor-valued random variables, $\calX_{i}$ and $\calX_{j}$, is \textit{invariant} to the data representation, that is
\begin{align}
	\expect{\inner{\calX_{i}}{\calX_{j}}} = \expect{\inner{\x_{i}}{\x_{j}}} = \expect{\inner{\X_{i(n)}}{\X_{j(n)}}}
\end{align}
for all $n$, and for both the direct data format and any vector or matrix tensor unfolding. We can therefore dissociate the expected inner product scale, $\sigma_{ij}=\expect{\inner{\calX_{i}}{\calX_{j}}}$, from the mode-$n$ covariance matrices, $ \R_{ij}^{(n)} $, by introducing the parameter $\boldTheta_{ij}^{(n)} \in \domR^{I_{n} \times I_{n}}$ to yield
\begin{align}
	\left( \kronprod{n=N}{1} \R_{ij}^{(n)} \right) = \sigma_{ij}\left( \kronprod{n=N}{1} \boldTheta_{ij}^{(n)} \right) \label{eq:unit_trace_condition_covariance}
\end{align}
where
\begin{align}
	\sigma_{ij} = \tr{\kronprod{n=N}{1} \R_{ij}^{(n)}} \label{eq:covariance_trace}
\end{align}
To obtain an identifiable parametrization we must also impose the unit-trace property on the introduced matrices, that is, $\tr{\boldTheta_{ij}^{(n)}}=1$, $\forall n$, so as to resolve the scaling ambiguity arising in Kronecker products described in Section \ref{tensorvalued_pdf_def}.

\begin{remark}
	Intuitively, $\boldTheta_{ij}^{(n)}$ can also be viewed as the \textit{covariance density} at the $n$-th mode, whereby it describes the percentage of the total covariance, $\sigma_{ij}$, assigned to each pair of mode-$n$ fibers. Moreover, the unit-trace condition satisfies the definition in (\ref{eq:covariance_trace}).
\end{remark}

Furthermore, the separability property of the mean of $\calX_{i}$ in the identifiable formulation, given by $\calM_{i} = \alpha_{i}\left( \circprod{n=1}{N}  \boldmu_{i}^{(n)} \right)$, also applies herein.

By jointly considering the parameters, $\alpha_{i}$, $\boldmu_{i}^{(n)}$, $\sigma_{ij}$ and $\boldTheta_{ij}^{(n)}$, for all $i,j=1,...,M$ and $n=1,...,N$, we can now form the matrices
\begin{align}
	\boldalpha_{z} = \left[ \begin{array}{c}
	\alpha_{1}\\
	\alpha_{2}\\
	\vdots\\
	\alpha_{M}
	\end{array} \right] \in \domR^{M}
\end{align}
\begin{align}
\boldmu_{z}^{(n)} = \left[ \begin{array}{c}
\boldmu_{1}^{(n)}\\
\boldmu_{2}^{(n)}\\
\vdots\\
\boldmu_{M}^{(n)}
\end{array} \right] \in \domR^{MI_{n}}
\end{align}
\begin{align}
	\boldSigma_{zz} = \left[\begin{array}{cccc}
		\sigma_{1}^{2} & \sigma_{12} & \cdots & \sigma_{1M} \\
		\sigma_{21} & \sigma_{2}^{2} & \cdots & \sigma_{2M} \\
		\vdots & \vdots & \ddots & \vdots \\
		\sigma_{M1} & \sigma_{M2} & \cdots & \sigma_{M}^{2}
	\end{array}\right] \in \domR^{M \times M}
\end{align}
\begin{align}
	\boldTheta_{zz}^{(n)} = \left[ 
		\def\arraystretch{1.2}
		\begin{array}{cccc}
		\boldTheta_{11}^{(n)} & \boldTheta_{12}^{(n)} & \cdots & \boldTheta_{1M}^{(n)} \\
		\boldTheta_{21}^{(n)} & \boldTheta_{22}^{(n)} & \cdots & \boldTheta_{2M}^{(n)} \\
		\vdots & \vdots & \ddots & \vdots \\
		\boldTheta_{M1}^{(n)} & \boldTheta_{M2}^{(n)} & \cdots & \boldTheta_{MM}^{(n)} \\
	\end{array} \right] \in \domR^{MI_{n} \times MI_{n}}
\end{align} 
so that the proposed distribution is formulated as follows
\begin{equation}
\z \sim \Normal{ \left( \boldalpha_{z} \!\! \khatriprod{n=N}{1}  \boldmu_{z}^{(n)} \right), \left( \boldSigma_{zz} \!\! \khatriprod{n=N}{1} \boldTheta_{zz}^{(n)} \right) }
\end{equation}
The multivariate tensor probability density function then becomes
\begin{align}
	\label{eq:multi_eq:tensorvalued_pdf_unit_norm}
	p(\z) = \frac{\exp \left[ -\frac{1}{2} \left(\z - \m_{z}\right)^{\Trans}\left( \boldSigma_{zz} \! \khatriprod{n=N}{1} \boldTheta_{zz}^{(n)}\right)^{-1}\left(\z - \m_{z}\right) \right]}{\left(2\pi\right)^{\frac{K}{2}} \det^{\frac{1}{2}}\left( \boldSigma_{zz} \! \khatriprod{n=N}{1} \boldTheta_{zz}^{(n)}\right) }
\end{align}
where $\m_{z} = \left( \boldalpha_{z} \khatriprod{n=N}{1}  \boldmu_{z}^{(n)} \right)$.

\vspace{-0.1cm}

\begin{remark}
	For $M=1$, the proposed multivariate tensor-valued probability distribution in (\ref{eq:multi_eq:tensorvalued_pdf_unit_norm}) reduces to the univariate tensor-valued probability distribution in (\ref{eq:tensorvalued_pdf_unit_norm}).
\end{remark}

\vspace{-0.1cm}

With the proposed identifiable formulation, the Khatri-Rao separability properties of multivariate tensor-valued random variable, $\z \in \domR^{MK}$, are given by
\begin{align}
 	\expect{\z} & = \left( \boldalpha_{z} \!\! \khatriprod{n=N}{1}  \boldmu_{z}^{(n)} \right)  \\
	\cov{\z} & = \left( \boldSigma_{zz} \!\! \khatriprod{n=N}{1} \boldTheta_{zz}^{(n)} \right)  \\
	\cov{\Z_{(n)}} & = \left( \boldSigma_{zz} \oast \boldTheta_{zz}^{(n)} \right)
\end{align}
With reference to the constituent tensor-valued random variables, the Khatri-Rao separability properties of the covariance simplify to
\begin{align}
	\expect{\x_{i}} & = \alpha_{i} \left( \kronprod{n=N}{1}  \boldmu_{i}^{(n)} \right)  \\
	\cov{\x_{i},\x_{j}} & = \sigma_{ij}\left( \kronprod{n=N}{1} \boldTheta_{ij}^{(n)} \right) \label{eq:khatri_cov} \\
	\cov{\X_{i(n)},\X_{j(n)}} & = \sigma_{ij}  \boldTheta_{ij}^{(n)} \label{eq:khatri_cov_n}
\end{align}


%


\subsection{Maximum likelihood tensor-valued estimator}

Since multivariate tensor-valued Gaussian probability distribution is a natural extension of the univariate tensor-valued case, this gives us the opportunity to employ the proposed relationships in (\ref{eq:khatri_cov})-(\ref{eq:khatri_cov_n}) to obtain the ML estimators
\begin{align}
	\sigma_{ij} & = \frac{1}{T}\sum_{t=0}^{T-1}\s_{i}^{\Trans}(t)\s_{j}(t) \label{eq:MLE_multi_sig} \\
	\boldTheta_{ij}^{(n)} & = \frac{1}{T\sigma_{ij}} \sum_{t=0}^{T-1} \S_{i(n)}(t)\S_{j(n)}^{\Trans}(t) \label{eq:MLE_multi_Theta}
\end{align}
where $\sigma_{i}^{2}\equiv\sigma_{ii}$. Furthermore, the ML estimation procedure for determining $\alpha_{i}$ and $\{\boldmu_{i}^{(n)}\}_{n=1}^{N}$ is equivalent to that for univariate tensor-valued variables, that is, they are obtained from the rank-$1$ multilinear SVD of the sample mean tensor $\frac{1}{T} \sum_{t=0}^{T-1} \calX_{i}(t)$. 


\pagebreak

\section{Applications to Atmospheric Climate Analysis}

\label{sec:climate_modelling}

\begin{figure}[h!]
	\vspace{-0.3cm}
	\centering
	\includegraphics[scale=0.65, trim={3.2cm 14.5cm 9cm 2cm},clip]{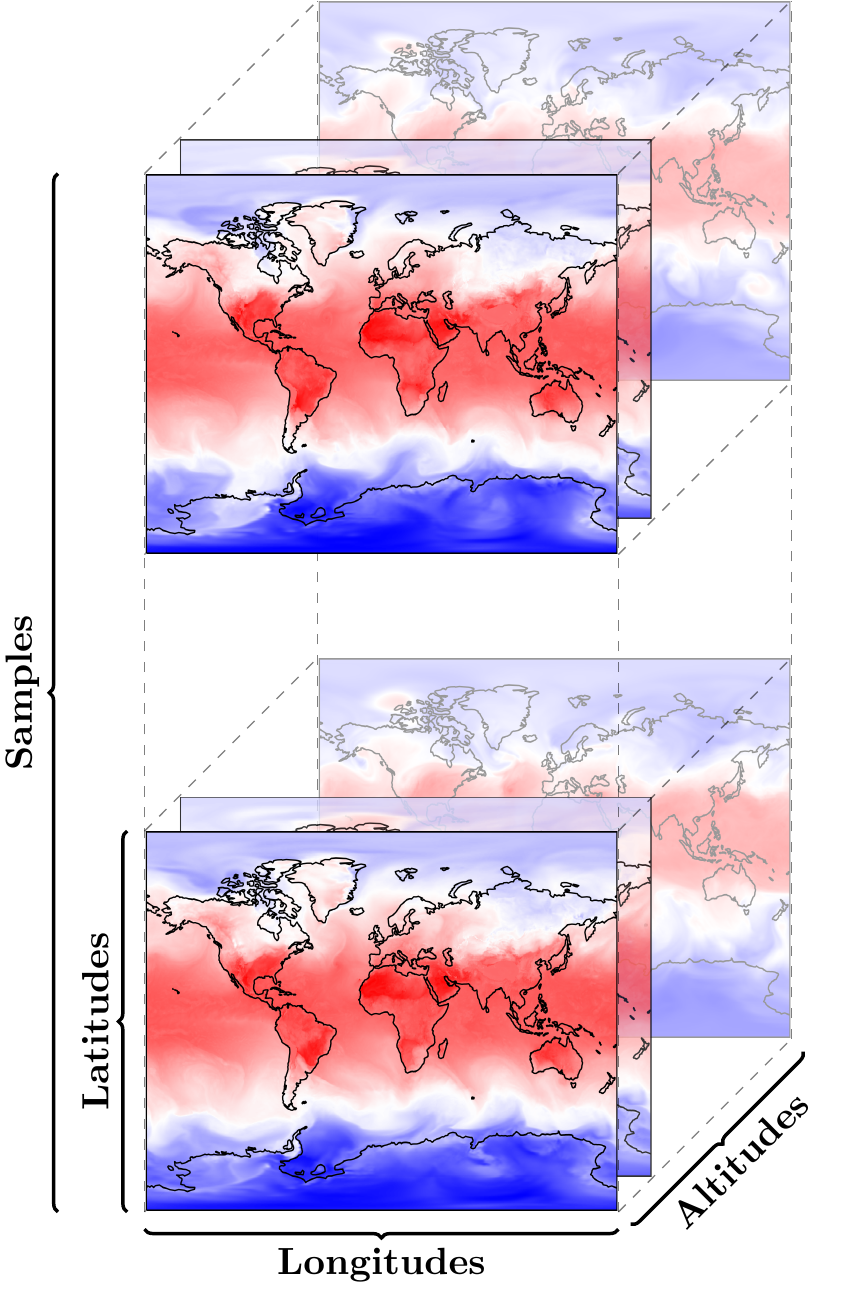}
	\caption{\label{fig:data_tensor_temperature} Construction of the order-$4$ data tensor. Each sample, $\calX(t) \in \domR^{I_{1} \times I_{2} \times I_{3}}$, consists of temperature measurements in the \textit{longitude} $\times$ \textit{latitude} $\times$ \textit{altitude} space at a time instant $t$.}
	\vspace{-0.3cm}
\end{figure}

In this section, we demonstrate the ability of the proposed probabilistic model in (\ref{eq:tensorvalued_pdf_unit_norm}) to statistically characterise atmospheric climate data. We modelled the ECMWF ReAnalysis (ERA5) dataset \cite{ERA}, which provides global estimates of atmospheric, land and oceanic climate variables in time, as a tensor-valued Gaussian random variable. We considered daily temperature measurements, each recorded at $00$:$00$ GMT, at a horizontal resolution of $31\,$km for $20$ altitude levels within the troposphere, ranging from the surface up to an altitude of $11\,$km. Each sample at a time instant $t$ naturally takes the form of an order-$3$ tensor, $\calX(t) \in \domR^{I_{1} \times I_{2} \times I_{3}}$, where $I_{1}=721$, $I_{2}=1440$ and $I_{3}=20$ denote respectively the number of latitude, longitude, and altitude grid points. We considered samples ranging in the period $2019$-$10$-$01$ to $2019$-$10$-$31$, i.e. $T=31$ daily tensor-valued samples, thereby forming of an order-$4$ tensor, $\calD \in \domR^{I_{1} \times I_{2} \times I_{3} \times T}$, as illustrated in Figure \ref{fig:data_tensor_temperature}.

\begin{figure}[t]
	\centering
	\begin{subfigure}[t]{0.49\textwidth}
		\centering
		\includegraphics[scale=0.7, trim={3.1cm 19.25cm 6.5cm 2cm},clip]{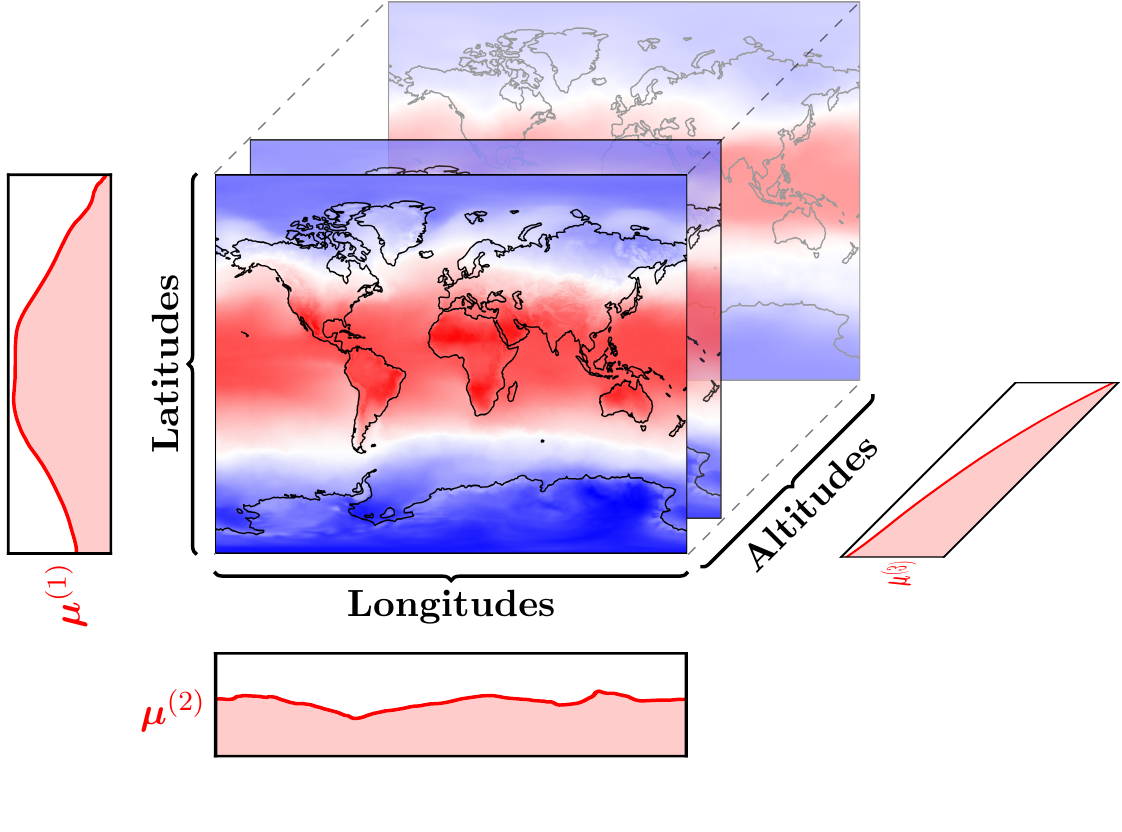}
		\vspace{-0.5cm}
		\caption{}    
		\label{fig:sample_mean_CPD}
		\vspace{0.1cm}
	\end{subfigure}
	\begin{subfigure}[t]{0.49\textwidth}   
		\centering 
		\includegraphics[scale=0.7, trim={3.2cm 21.25cm 9.5cm 2cm},clip]{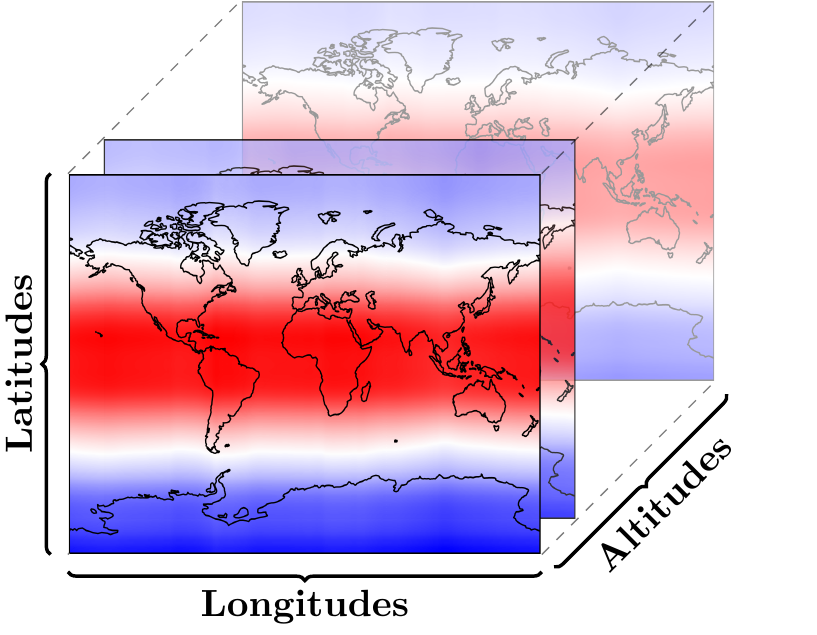} 
		\caption{}    
		\label{fig:rank1_mean_CPD}
	\end{subfigure}
	\caption{First-order statistical description of the atmospheric temperature. (a) Sample mean tensor with its rank-$1$ CPD factors, $\{ \boldmu^{(n)} \}_{n=1}^{3}$. (b) Rank-$1$ mean tensor reconstructed from the CPD factors in (a), $\alpha(\circprod{n=1}{3} \boldmu^{(n)})$.} 
	\vspace{-0.6cm}
\end{figure}

The statistical analysis was implemented using our own Python Higher-Order Tensor ToolBOX (HOTTBOX) \cite{Kisil2018}. The sample mean, $\frac{1}{T}\sum_{t=1}^{T}\calX(t)$, and its associated mean parameters, $\alpha$ and $\{ \boldmu^{(n)} \}_{n=1}^{3}$ in (\ref{eq:tensorvalued_pdf_unit_norm}), were evaluated and are shown in Figure \ref{fig:sample_mean_CPD}. Observe that the mode-$1$ CPD factor, $\boldmu^{(1)}$, conveys the expected variation of temperature in the latitude dimension, whereby the maximum temperature is observed at the equator. The mode-$2$ CPD factor associated with the variation of temperature in the longitude dimension, $\boldmu^{(2)}$, conveys the \textit{diurnal} temperature variation, that is, the difference in temperature between regions in day and night time. Similarly, the mode-$3$ CPD factor, $\boldmu^{(3)}$, depicts the expected decrease in atmospheric temperature with an increase in altitude, which arises from the radiation and convection behaviours typically observed in the troposphere. 

\pagebreak

\begin{remark}
	The rank-$1$ mean tensor $\calM = \alpha(\circprod{n=1}{3} \boldmu^{(n)})$ was constructed and is shown in Figure \ref{fig:rank1_mean_CPD}. The rank-$1$ representation explains $99.99\%$ of the variance in the sample mean, thereby demonstrating the suitability the tensor-valued model for describing the data. The main advantage of the model is the reduction of the number of parameters required to characterise the mean, which reduces $I_{1}I_{2}I_{3}=20,764,800$ parameters in $\frac{1}{T}\sum_{t=1}^{T} \calX(t)$ to $(1+I_{1}+I_{2}+I_{3}) = 2,182$ parameters in $\alpha(\circprod{n=1}{3} \boldmu^{(n)})$.
\end{remark}

\begin{figure}[ht]
	\vspace{-0.1cm}
	\centering
	\begin{subfigure}[t]{0.5\textwidth}
		\centering
		\includegraphics[scale=0.36]{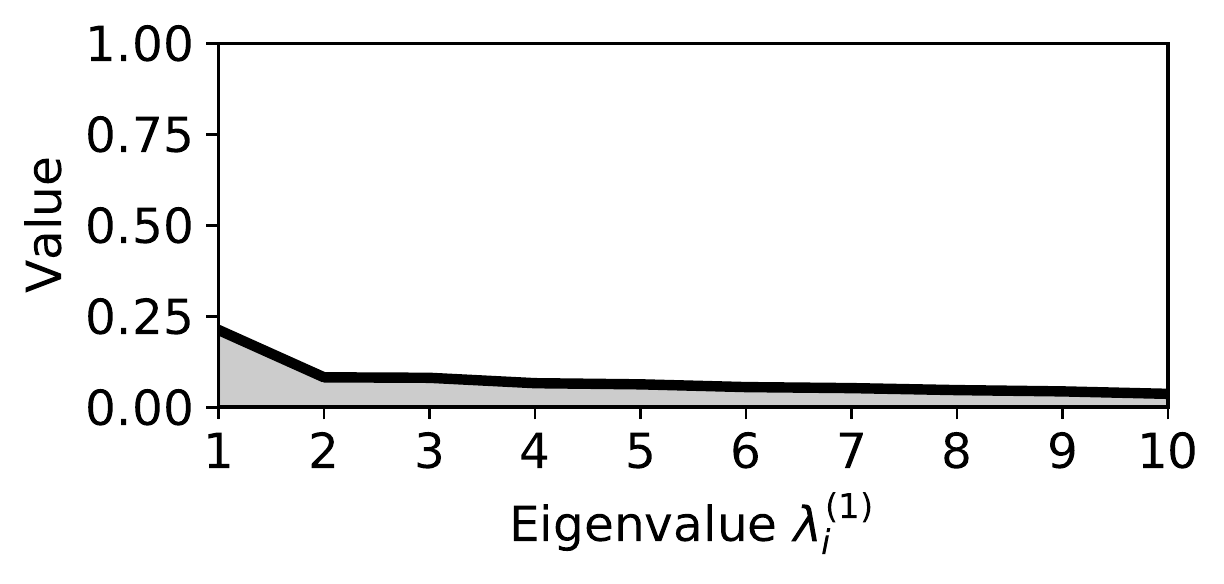}
		\includegraphics[scale=0.36]{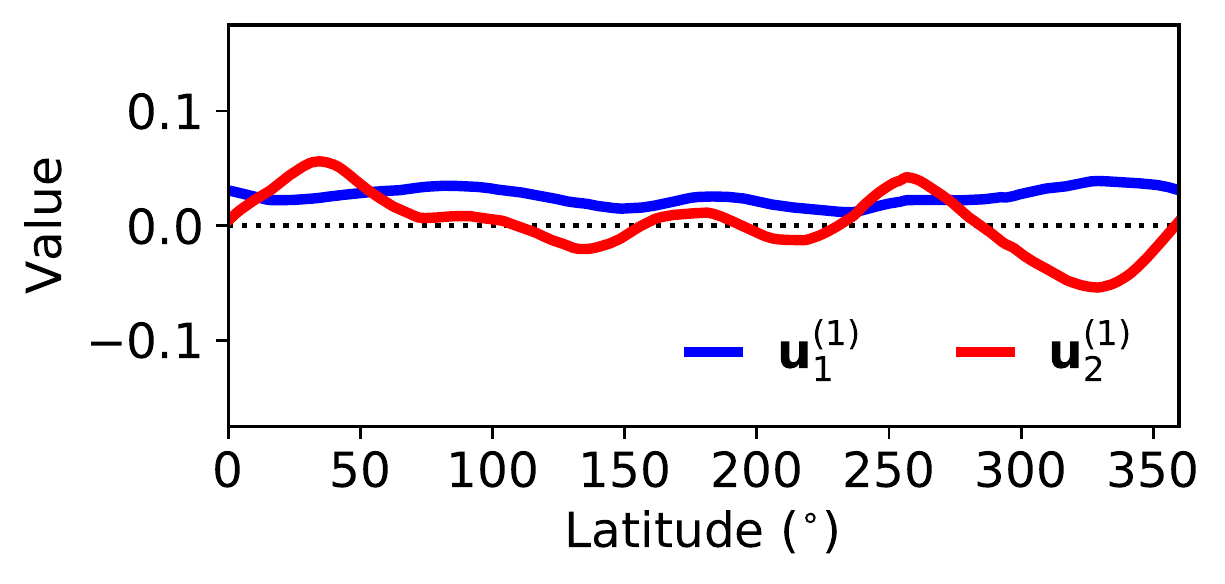}
		
		\vspace{-0.3cm}
		\caption{}    
	\end{subfigure}

	\begin{subfigure}[t]{0.5\textwidth}
		\centering
		\includegraphics[scale=0.36]{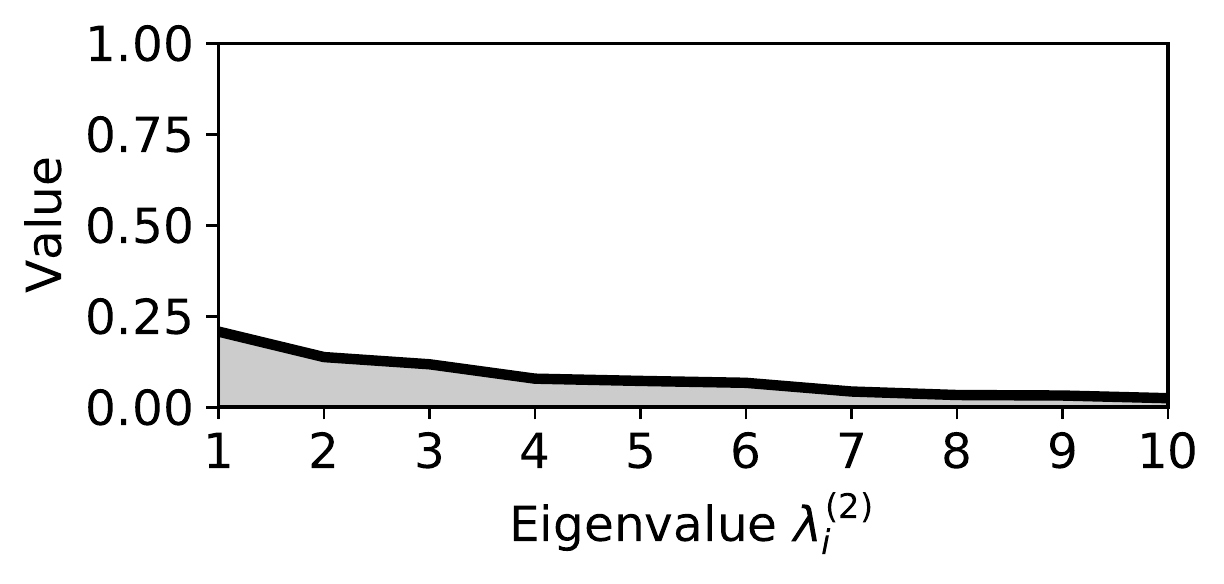}
		\includegraphics[scale=0.36]{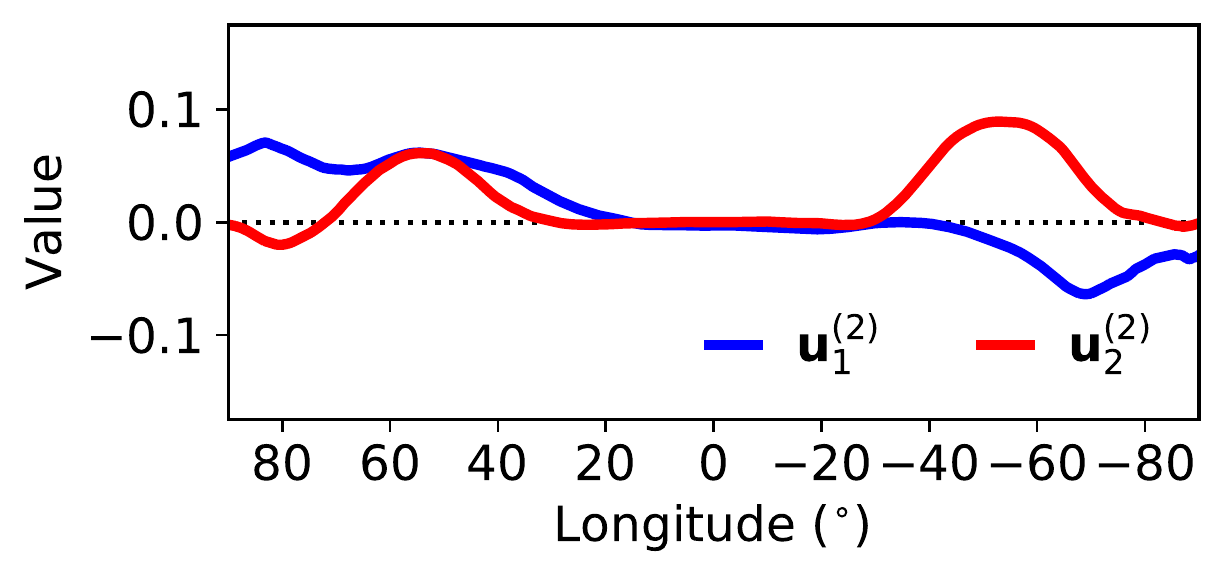}
		
		\vspace{-0.3cm}
		\caption{}
	\end{subfigure}

	\begin{subfigure}[t]{0.5\textwidth}
		\centering
		\includegraphics[scale=0.36]{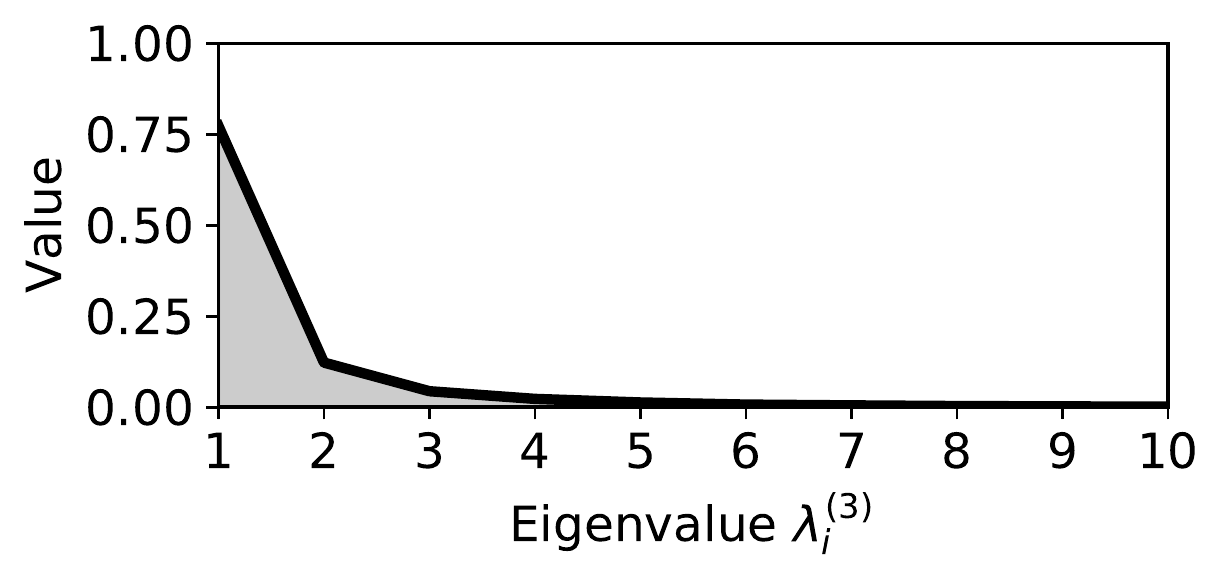}
		\includegraphics[scale=0.36]{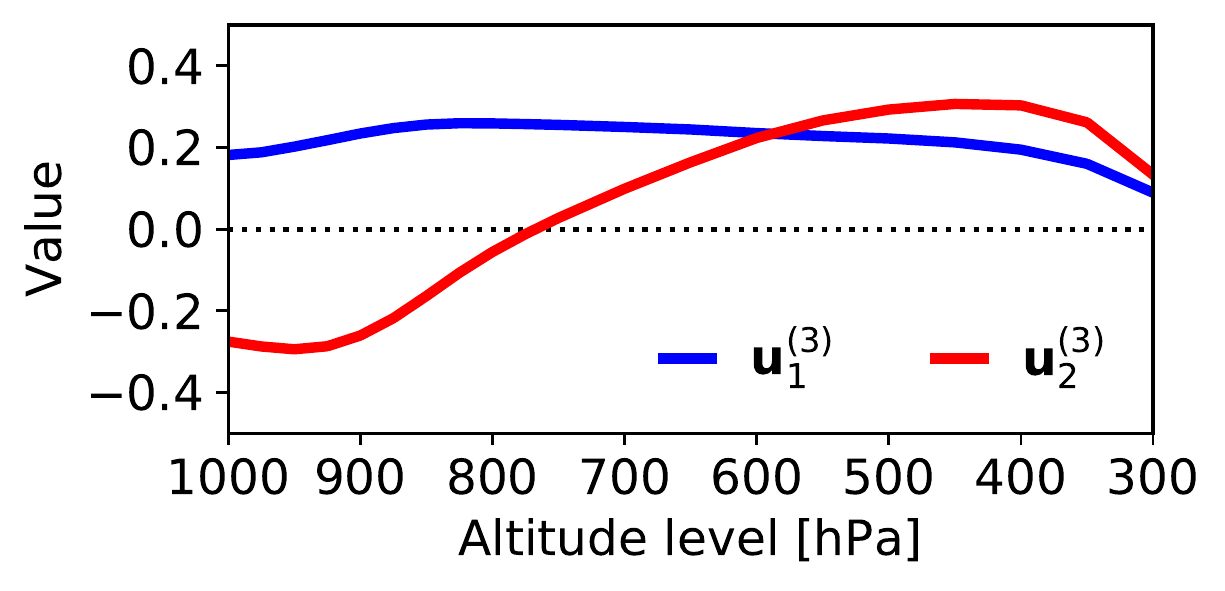}
		
		\vspace{-0.3cm}
		\caption{}
	\end{subfigure}
	\vspace{-0.3cm}
	\caption{The leading eigenvalues, $\{\lambda_{i}^{(n)}\}_{i=1}^{10}$ (left panel) and the leading two eigenvectors, $\{\u_{i}^{(n)}\}_{i=1}^{3}$, (right panel) of the mode-$n$ covariance density matrix $\boldTheta^{(n)}$. (a) Latitude ($n=1$). (b) Longitude ($n=2$). (c) Altitude ($n=3$).} 
	\label{fig:eigen_temperature}
	\vspace{-0.5cm}
\end{figure}

To describe the variation of atmospheric temperature in time with respect to the mean, we computed the mode-$n$ covariance density matrices, $\{\boldTheta^{(n)}\}_{n=1}^{N}$, and displayed their leading eigenvalues and leading two eigenvectors in Figure \ref{fig:eigen_temperature}. Observe that the sources of maximum temperature variation in time along the latitude, longitude and altitude dimensions are associated with changes in the global temperature, since all entries in the eigenvector are of the same sign. Furthermore, we can observe that temperature variation in time along the latitude and longitudes dimensions are also governed by many location-specific and idiosyncratic factors, as seen from the their flat eigenvalue spectrum. This may owe to geography-specific factors, such as the average humidity, the regime of winds or the proximity to large bodies of water. In contrast, the temperature variation in time along the altitude dimension is mostly governed by the global temperature change, since the leading eigenvalue explains more than $75\%$ of the variance.

\pagebreak

\section{Conclusions}

A statistically identifiable formulation of the tensor-valued Gaussian distribution has been proposed which, unlike existing solutions, exhibits the desired Kronecker separable statistics. For rigour, the proposed model has been used to derive the associated log-likelihood function which admits an analytical maximum and corresponds to a statistically consistent maximum likelihood estimator. The so introduced probabilistic framework has been generalised to describe the joint distribution of multiple tensor-valued random variables, which is characterised by Khatri-Rao separable statistics. It has been shown that in this way the relationships in high-dimensional data can be separated and distilled in a compact and physically meaningful manner. The results are supported by an intuitive example applied to real-world global atmospheric temperature data, computed using our own Python toolbox for tensor analysis \cite{Kisil2018}.




\footnotesize

\bibliographystyle{IEEEtran}
\bibliography{./Bibliography} 

\vspace{-1cm}

\begin{IEEEbiography}{Bruno Scalzo Dees}
	received the M.Eng. degree in aeronautical engineering from Imperial College London, U.K. He is currently working toward the Ph.D. degree at the Department of Electrical Engineering at the same institution. His research interests include statistical signal processing, maximum entropy modelling and tensor-valued random variables.
\end{IEEEbiography}

\vspace{-1cm}

\begin{IEEEbiography}{Anh-Huy Phan}
	(M'15) received the master's degree from the Ho Chi Minh City University of Technology, Ho Chi Minh City, Vietnam, in 2005, and the Ph.D. degree from the Kyushu Institute of Technology, Kitakyushu, Japan, in 2011. From October 2011 to March 2018, he was a Research Scientist with the Laboratory for Advanced Brain Signal Processing, Brain Science Institute (BSI), RIKEN, Tokyo, Japan, where he was a Visiting Research Scientist with the Toyota Collaboration Center, from April 2012 to April 2015. Since May 2018, he has been an Assistant Professor with the Center for Computational and Data-Intensive Science and Engineering, Skoltech, Moscow, Russia. He is also a Visiting Associate Professor with the Tokyo University of Agriculture and Technology (TUAT), Tokyo. He has authored three monographs. His research interests include multilinear algebra, tensor computation, tensor networks, nonlinear system, blind source separation, and brain–computer interface.
	
	Dr. Phan received best paper awards for articles in the IEEE Signal Processing Magazine in 2018 and in the International Conference on Neural Information Processing in 2016, and the Outstanding Reviewer Award for maintaining the prestige of the IEEE International Conference on Acoustics, Speech and Signal Processing 2019.
\end{IEEEbiography}

\vspace{-1cm}

\begin{IEEEbiography}{Danilo P. Mandic}
	(M'99-SM'03-F'12) received the Ph.D. degree in nonlinear adaptive signal processing from Imperial College London, U.K., in 1999.
	
	He has been a Guest Professor with Katholieke Universiteit Leuven, Leuven, Belgium, the Tokyo University of Agriculture and Technology, Tokyo, Japan, Westminster University, London, and a Frontier Researcher with RIKEN, Wako, Japan. He is currently a Professor of signal processing with Imperial College London, where he is involved in nonlinear adaptive signal processing and nonlinear dynamics. He is also the Deputy Director of the Financial Signal Processing Laboratory, Imperial College London. He has two research monographs \textit{Recurrent Neural Networks for Prediction: Learning Algorithms, Architectures and Stability} (West Sussex, U.K.: Wiley, 2001) and \textit{Complex Valued Nonlinear Adaptive Filters: Noncircularity, Widely Linear and Neural Models} (West Sussex, U.K.: Wiley, 2009), an edited book \textit{Signal Processing Techniques for Knowledge Extraction and Information Fusion} (New York, NY, USA: Springer, 2008), and more than 200 publications on signal and image processing. 
	
	Prof. Mandic has been a member of the IEEE Technical Committee on Signal Processing Theory and Methods. He has produced award winning papers and products resulting from his collaboration with the industry. He has been an Associate Editor of the \textit{IEEE Signal Processing Magazine}, the IEEE \textsc{Transactions on Circuits AND Systems II}, the IEEE \textsc{Transactions on Signal Processing}, and the IEEE \textsc{Transactions on Neural Networks}.
\end{IEEEbiography}

\vfill

\end{document}